%% file: Thesis.tex
\documentclass[12pt,a4pages,oneside,openright]{Thesis}

\input{misc/preamble}
\input{misc/information}
\graphicspath{{figures/}}

\begin{document}

\frontmatter
\pagestyle{empty}
\input{misc/titlepage}

\input{misc/copyright}

\pagestyle{thesis}
\KOMAoptions{headsepline=true}
\include{science/abstract}
\include{science/publications}

\include{misc/acknowledge}
\cleardoublepage
\phantomsection
\addcontentsline{toc}{chapter}{Contents}
\tableofcontents

\mainmatter

\include{science/ch1-Intro}

\include{science/ch2-TDSE}

\include{science/ch3-Rutherford}

\include{science/ch4-Gravity}

\include{science/ch5-MOND}

\include{science/ch6-Summary}

\include{science/appendix}

\backmatter
\cleardoublepage
\phantomsection
\addcontentsline{toc}{chapter}{Bibliography}
\bibliographystyle{Thesis}
\bibliography{Thesis}

\end{document}

%% file: misc/preamble.tex
\usepackage[utf8]{inputenc} 
\usepackage[T1]{fontenc}

\usepackage{etoolbox}
\usepackage{graphicx}
\usepackage{lmodern}
\usepackage{amsmath}
\usepackage{amssymb}
\usepackage{bm}
\usepackage{amstext}
\usepackage{lineno}
\usepackage{mathtools}
\usepackage{mathrsfs}

\allowdisplaybreaks

\usepackage{stmaryrd}
\usepackage{multicol}
\usepackage{listings}
\usepackage{float}

\emergencystretch=\maxdimen

\usepackage[italicdiff]{physics}

\usepackage{enumitem}

\usepackage{dsfont}
\newcommand{\unitop}{\hat{\mathds{1}}}

\usepackage[colorlinks=true,allcolors=blue,linktocpage=true]{hyperref}

\usepackage{setspace}
\usepackage{boxhandler}
\usepackage[caption=false]{subfig}
\usepackage{comment}
\usepackage[toc,page]{appendix}

\usepackage[sort&compress,numbers]{natbib}

\usepackage{geometry}
\newcommand{\PaperEntry}[8]{
	\noindent {#1}	\\
	\noindent {#2}	\\
	\noindent \href{https://doi.org/#7}{{#3} \textbf{#4}, #5 (#6)}	\hfill 	\href{https://arxiv.org/abs/#8}{arXiv:#8}}

\usepackage{pdfpages}
\newcommand{\nocontentsline}[3]{}
\newcommand{\tocless}[2]{\bgroup\let\addcontentsline=\nocontentsline#1{#2}\egroup}

%% file: misc/information.tex
\def\title{Entanglement Dynamics in Quantum Continuous--Variable States}

\def\titlebroken{Entanglement Dynamics in \\ Quantum Continuous--Variable States}

\def\Ankit{Ankit Kumar}

\def\RoorkeeDept{Department of Physics}
\def\IITRoorkee{Indian Institute of Technology Roorkee}

\def\Arumugam{Prof.~P.~Arumugam}

\def\Tomasz{Prof.~Tomasz Paterek}

\def\GdanskDept{Institute of Theoretical Physics and Astrophysics}
\def\GdanskFaculty{Faculty of Mathematics, Physics and Informatics}
\def\UniGdansk{University of Gda\'{n}sk}
\def\GdanskAddress{Gda\'nsk 80-308, Poland}

\def\defensetime{July, 2023}
\def\copyrightyear{2023}

\AtBeginDocument{
\hypersetup{pdftitle=\title}
\hypersetup{pdfauthor=\Ankit}
}

%% file: misc/titlepage.tex
\thispagestyle{empty}
\newgeometry{left=2.375cm,right=2.375cm,top=2.5cm,bottom=1.25cm}

\begin{center}
{
\LARGE
\textbf{\titlebroken}
}

\vspace{2cm}

{DOCTORAL THESIS}

\vfill

{
\Large
\textbf{\Ankit}
}

\IITRoorkee, India

\vfill

\emph{jointly supervised by}
\vspace{1cm}

\begin{minipage}[t]{0.46\linewidth}
\centering

{\large
\textsf{\Arumugam}}

\vspace{.5cm}
\includegraphics[width=3cm]{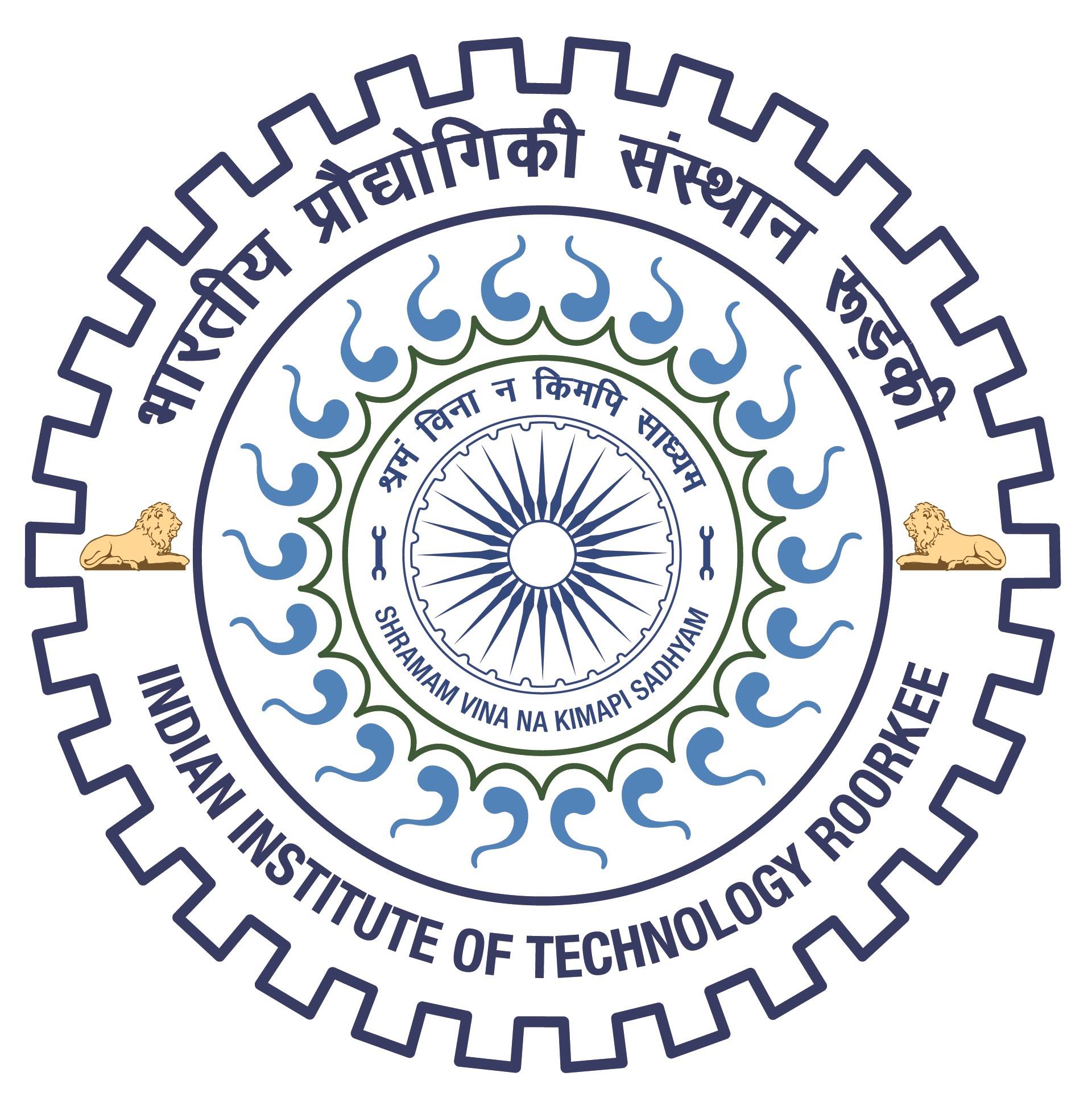}

\RoorkeeDept\\
\vspace{-1mm}
\IITRoorkee\\
\vspace{-1mm} 
Roorkee 247667, India
\end{minipage}
\hfill
\begin{minipage}[t]{0.53\linewidth}
\centering
{\large
\textsf{\Tomasz}}

\vspace{-1cm}
\includegraphics[width=5cm]{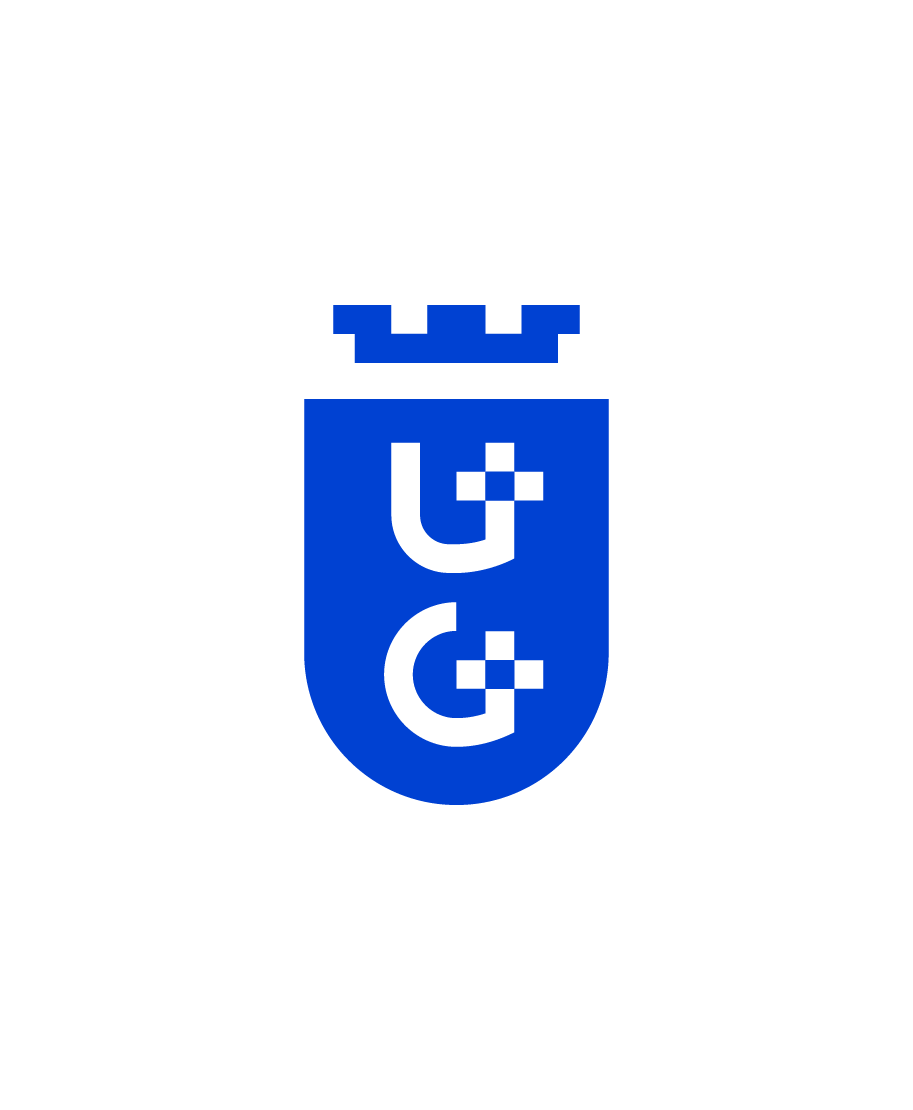}

\vspace{-1.5cm}
\GdanskDept\\
\vspace{-1mm}
\GdanskFaculty\\
\vspace{-1mm}
\UniGdansk, \GdanskAddress
\end{minipage}

\vspace{2cm}

{\large
\defensetime
}

\end{center}

%% file: misc/copyright.tex
\clearpage
\thispagestyle{empty}

\hspace{0pt}

\vfill

\begin{center}
\fontsize{12}{12}
\setstretch{1.0}
\textbf{\textcopyright \MakeUppercase{\IITRoorkee}, 
\copyrightyear \\
ALL RIGHTS RESERVED}
\end{center}

\vfill

\hspace{0pt}

%% file: science/abstract.tex
\cleardoublepage
\addchap{Abstract}

Due to the weakness of gravitational coupling, all quantum experiments up to date in which gravity plays a role utilized the field of the Earth.
Since this field undergoes practically undetectable back-action from quantum particles, it effectively admits a classical description as a fixed background Newtonian field or spacetime.
This argument strongly motivates theoretical and experimental research towards a demonstration of gravitation between two quantum masses, as this is one of the most straightforward scenarios where quantum features of gravity could be observed.
Several proposals studied the possibility of generating entanglement between two massive objects.
Along the same lines, with a particular focus on gravity, this thesis introduces general tools to tackle interaction-mediated entanglement and applies them to two particles prepared in continuous-variable states.

In order to pursue this aim systematically, this dissertation begins by introducing methods to precisely simulate the dynamics of quantum systems coupled by weak interactions.
We improve the accuracy of the numerical implementation of Cayley's operator and develop a methodology to avoid reflections from numerical infinities. 
We derive a condition under which a product state from the laboratory (LAB) perspective remains a product state in the center-of-mass (COM) frame, which reveals that only certain states are transformed into disentangled states. Even though the primary focus is on gravity, all the developed methods apply to arbitrary central interactions, and considerable parts of this thesis are devoted to explicit demonstrations of this versatility.
Accordingly, the first application is to investigate the head-on collision in the Rutherford experiment, with the projectile treated as (realistic) localized wave packets.
We observe various nonclassical effects in the average trajectories and trace them back to the convexity properties of the Coulomb potential with the help of Jensen's inequality. 
The concluding chapter also comments on the projectile-target entanglement.

Our next goal is to simplify the possible observation of weak gravitational entanglement in an inevitably noisy laboratory. 
The basic idea is to amplify correlations by pushing the particles toward each other, hoping that an ever-increasing gravitational interaction will automatically lead to a higher accumulated entanglement.
A toolbox is developed that quantifies the entanglement gain between the two particles directly in the COM frame of reference, thereby eliminating the need for inverse transformations back to the LAB frame. 
We start with the standard practice of the second-order truncation of quantum Newtonian potential, which has long kept the mathematical complexities at a minimum by forcefully constraining the system into the regime of (very well-understood) Gaussian Quantum Information. While it is known that an analytical solution exists, we utilize Ehrenfest's theorem to derive the covariance matrix in an exact closed form. The resultant entanglement is insensitive to relative motion between the two particles.

The less-understood non-Gaussian regime triggered by the cubic and higher-order potentials is considered next. We develop a hybrid analytical-numerical scheme to faithfully estimate the entanglement gain with the help of algorithms in Google TensorNetwork. The entanglement is found to be sensitive to relative motion only when the system evolves into the non-Gaussian regime. 
We prove that the position-momentum correlations originate from the force gradient in relative motion.
A derivation of closed forms for the non-Gaussian entanglement gain follows with informed guesswork. 
In experiments, it will be challenging to screen the system from all interactions but gravity.
With this in mind, we develop tools to quantify the entanglement with multiple central forces acting simultaneously.

As the final application, the thesis discusses an entanglement-based test of the Modified Newtonian Dynamics (MOND), a candidate explanation of dark matter effects which proposes to modify Newton's second law and/or the gravitational force law for accelerations smaller than $\sim 10^{-10}$ m/s$^2$. 
One verifies that the masses recently cooled by the Aspelmeyer group in Vienna, when separated by a distance of a few times their radius, are into the regime of accelerations where MOND is relevant.
Accordingly, the tools developed in this thesis offer an opportunity to test the assumptions behind MOND through entanglement between two nearby quantum masses.
We develop an experiment where departures from Newtonian gravity are certified by simply witnessing the entanglement generation starting from thermal states.

%% file: science/publications.tex
\cleardoublepage
\addchap{Publications}

\section*{Contributions in Peer-Reviewed Journals}

\begin{enumerate}[leftmargin=*,label=\arabic*)]

\item  
\PaperEntry
{Nonclassical Trajectories in Head-On Collisions}
{\textbf{\Ankit}, Tanjung Krisnanda, P. Arumugam, and Tomasz Paterek}
{Quantum}{5}{506}{2021}
{10.22331/q-2021-07-19-506}
{2011.06470}

\item  
\PaperEntry
{Continuous-Variable Entanglement through Central Forces: Application to Gravity between Quantum Masses}
{\textbf{\Ankit}, Tanjung Krisnanda, P. Arumugam, and Tomasz Paterek}
{Quantum}{7}{1008}{2023}
{10.22331/q-2023-05-15-1008}
{2206.12897}

\item
\PaperEntry
{Probing Modified Gravity with Entanglement of Microspheres}
{\textbf{\Ankit}, Yen-Kheng Lim, P. Arumugam, Tom Z\l{}o\'{s}nik, and Tomasz Paterek}
{Physical Review D}{109}{L101501}{2024}		{10.1103/PhysRevD.109.L101501}{2306.14938}

\end{enumerate}

\section*{Contributions in Peer-Reviewed Conference Proceedings}

\begin{enumerate}[leftmargin=*,label=\arabic*)]

\item
\PaperEntry
{Closest Approach of a Quantum Projectile}
{\textbf{\Ankit}, Tanjung Krisnanda, P. Arumugam, and Tomasz Paterek}
{Journal of Physics: Conference Series}{1850}{012074}{2021}
{10.1088/1742-6596/1850/1/012074}
{2112.13296}

\end{enumerate}

\section*{Contributions in Public Scientific Repositories}

\begin{enumerate}[leftmargin=*,label=\arabic*)]

\item   
{An Accurate Pentadiagonal Matrix Solution for the Time-Dependent Schr\"{o}dinger Equation} \\
\textbf{\Ankit}, and P. Arumugam    \\
\href{https://doi.org/10.5281/zenodo.7275667}{Zenodo.7275667} 
\hfill 
\href{https://github.com/vyason/Cayley-TDSE}{GitHub/vyason/Cayley-TDSE}
\hfill
\href{https://arxiv.org/abs/2205.13467}{arXiv:2205.13467}

\end{enumerate}

%% file: science/ch1-Intro.tex
\chapter{Introduction} 
\label{ch:chapter1}

Quantum mechanics originated in the 1920s when physicists such as Max Planck, Albert Einstein, and Niels Bohr, among others, sought to explain the peculiar behavior of light and atoms. 
Although its interpretation is still under debate, 
quantum mechanics is widely accepted as the theory of the atomic world. Since its inception, the quantum theory has led to numerous groundbreaking discoveries that cultivated our modern technologies. Over time, it has branched out and opened up new avenues in chemistry~\cite{QuChem-rug01:001382226}, biology~\cite{QuBio-Lambert2013}, computer science~\cite{nielsen_chuang_2010}, medicine~\cite{MRI-Machine}, etc.

The concept of entanglement was first introduced in 1935 by Einstein, Podolsky, and Rosen~\cite{paper-EPR}, when they (EPR) proposed a thought experiment to illustrate what they perceived as a paradox in quantum mechanics.
Two particles, once interacted and then separated, could admit instantaneously correlated `elements of reality', violating the principle of locality. 
Similar observations were made by Schr\" odinger in the same year~\cite{schrodinger_1935}.
Such behavior was famously known as ``spooky action at a distance'', and many physicists grappled with its implications for a long time.

Entanglement gained significant attention after 1964, when John S. Bell proposed a way to test EPR ideas~\cite{JohnBell1964}.
He derived a set of inequalities which can only be violated if two particles are truly entangled. 
Later in 1974, Clauser and Freedman provided the first evidence of such a violation in their experiments with pairs of photons~\cite{Clauser1972}, confirming the existence of entanglement.
All such inequalities are grouped under the umbrella of the Bell's theorem, which stands as one of the most important contributions to modern theoretical physics. 
With improvements in experimental techniques, researchers could demonstrate and manipulate reliable entanglement between photons, electrons, and atoms. 
Entanglement is now regarded as one of the fundamental features of quantum mechanics and sits at the heart of quantum information theory. Entanglement manipulation has resulted in various proposals for enabling tasks that are not possible classically, including quantum computing~\cite{QuantumComputing}, quantum metrology and sensing~\cite{QuantumMetrology}, quantum cryptography~\cite{QuantumCryptography}, quantum dense coding~\cite{QuantumDenseCoding}, quantum teleportation~\cite{QuantumTeleportation}, etc. 
A rapid utilisation and manipulation
of quantum entanglement in such tasks have made it a crucial resource, some argued, as real as energy~\cite{RevModPhys.81.865}.

\section{Motivation}

Unification is a very fruitful idea in physics.
Electromagnetism has been unified with the strong and weak nuclear forces into a coherent framework of the quantum field theory.  
On the other hand, gravity is described by Einstein's general theory of relativity, and its reconciliation with quantum mechanics has been one of the most challenging problems in theoretical physics.
Several approaches to a quantum theory of gravity have been put forward.
To name a few, they include 
String Theory~\cite{StringTheory}, 
Loop Quantum Gravity~\cite{LoopQuantumGravity}, Twistor Theory~\cite{TwistorTheory}, Canonical Quantum Gravity~\cite{CanonicalQuantumGravity}, etc.
Each one has its merits, demerits, and sets of distinct predictions.
However, at present, there is no experimental evidence to favor any one of these proposals.

Observing the characteristics of electromagnetism and strong and weak forces at tiny distances has been a significant factor in developing a single theory that unifies them all.
The same is not available for gravity due to its weakness, and hence problems in its consistent unification with other forces.
Accordingly, all quantum experiments in which gravity plays a role utilized the strong gravitational field of the Earth.
Milestone experiments have measured the impacts of Earth's gravity on the 
apparent frequency of photons~\cite{PhysRevLett.4.337},
time gains in Cesium clocks flying along different trajectories~\cite{doi:10.1126/science.177.4044.168}, 
phase-shift and quantum bound states of neutrons~\cite{PhysRevLett.34.1472,Nesvizhevsky2002},
gravitational acceleration of falling atoms~\cite{Peters1999}, and
phase shift in an atom interferometer~\cite{PhysRevLett.118.183602}.
Earth is massive and hence gets a practically undetectable back-action from quantum particles. 
The gravitational field effectively admits a classical description, either in terms of a fixed background Newtonian field~\cite{PhysRevLett.34.1472,Peters1999,Nesvizhevsky2002} or spacetime~\cite{PhysRevLett.4.337,doi:10.1126/science.177.4044.168,PhysRevLett.118.183602}. 

This has strongly motivated theoretical and experimental research for demonstrating gravitation between two nearby quantum masses. 
Recently, there has been an effort to utilize the concept of entanglement: given that two entities cannot be entangled without a quantum mediator~\cite{PhysRevLett.119.120402,PhysRevLett.119.240401,PhysRevLett.119.240402}, several proposals studied the possibility of the gravity-mediated gain of entanglement between massive objects~\cite{PhysRevLett.119.240401,PhysRevLett.119.240402,PhysRevA.98.043811,PhysRevLett.119.120402,npjQI_6.12,JOPB_23.235501,Rijavec_2021,PhysRevLett.128.143601,PhysRevA.102.062807,PhysRevResearch.4.023087}. This is one of the simplest scenarios where quantum features of gravity could be observed.

\subsection{Entanglement in discrete states}

\begin{figure}[H]
\centering
\includegraphics[width=0.6\linewidth]{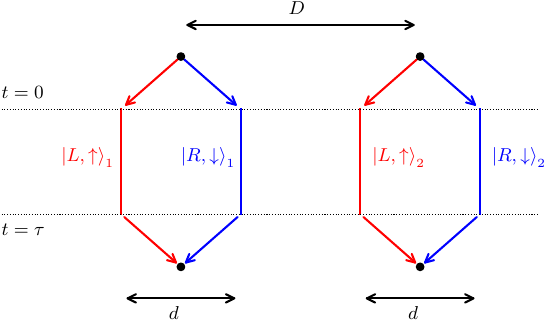}
\caption{Schematic representation of the Bose-Marletto-Vedral setup~\cite{PhysRevLett.119.240401,PhysRevLett.119.240402}. Two masses are embedded with spins and placed in adjacent interferometers.
A Stern-Gerlach magnet prepares each mass in a spatially separated superposition. The phase evolution due to the gravitational coupling generates a detectable amount of entanglement within a time $\tau$.
$L$ and $R$ are shorthand notations for left and right, and the up and down arrows represent the projections of the embedded spins.}
\label{fig:BMV}
\end{figure}
One of the most famous proposals to probe gravitational coupling in discrete superpositions is the so-called Bose-Marletto-Vedral (BMV) setup~\cite{PhysRevLett.119.240401,PhysRevLett.119.240402}, shown schematically in Fig.~\ref{fig:BMV}.
Two masses are placed in adjacent matter-wave interferometers, which prepares them in well-localised but spatially separated quantum superpositions of width $d$. The centers of the two interferometers are separated by a distance $D$.
The initial wave function is a product state given by:
\begin{equation}
\ket{\Psi(0)}_{12} = \frac{1}{\sqrt{2}}\qty(\ket{L}_1+\ket{R}_1) \ \frac{1}{\sqrt{2}}\qty(\ket{L}_2+\ket{R}_2).
\end{equation}
Given that we are in the non-relativistic regime where the gravitational field can be described by the quantum Newtonian potential, after an an evolution for time $\tau$ we get
\begin{equation}
\ket{\Psi(\tau)}_{12} = \frac{1}{\sqrt{2}} \qty{ \ket{L}_1 \frac{1}{\sqrt{2}}\qty(\ket{L}_2+e^{\Delta \phi_{LR}}\ket{R}_2)
+ \ket{R}_1 \frac{1}{\sqrt{2}}  \qty(e^{\Delta \phi_{RL}}\ket{L}_2+\ket{R}_2)
} ,
\end{equation}
where
\begin{equation}
\Delta \phi_{LR} \sim \frac{Gm_1m_2\tau}{\hbar(D-d)}, 	\hspace{1cm} \Delta \phi_{RL} \sim \frac{Gm_1m_2\tau}{\hbar(D+d)}.
\end{equation}
As long as $\Delta \phi_{LR} + \Delta \phi_{RL}$ is not an integral multiple of $2\pi$, the state cannot be factorised, generating entanglement.
Spin correlation measurements promise to witness this entanglement after the completion of interferometers. 

If the particles do get entangled, the interference pattern will be different from what is expected for non-entangled states, with the shifting of interference fringes depending on the strength of the gravitational interaction.
Much effort is being put into improving the BMV proposal to make it experimentally viable in the near future.
In particular, Ref.~\cite{PhysRevLett.128.110401} proposes to enhance the effective gravitational coupling by putting a massive mediator between two small masses, and Ref.~\cite{Sidajaya_2022} proposes to fight decoherence in noisy laboratories by freezing the quantum states through Zeno effect. Ref.~\cite{PhysRevA.105.032411} proposes a possible improvement of the original setup by considering the phase evolution in a three-qubit system. 
When compared to the earlier two-qubit setup, it leads to a higher accumulated entanglement, as well as a better resilience against environmental decoherence.

\subsection{Entanglement in continuous states}

\begin{figure}[H]
    \centering
    \includegraphics[width=0.66\linewidth]{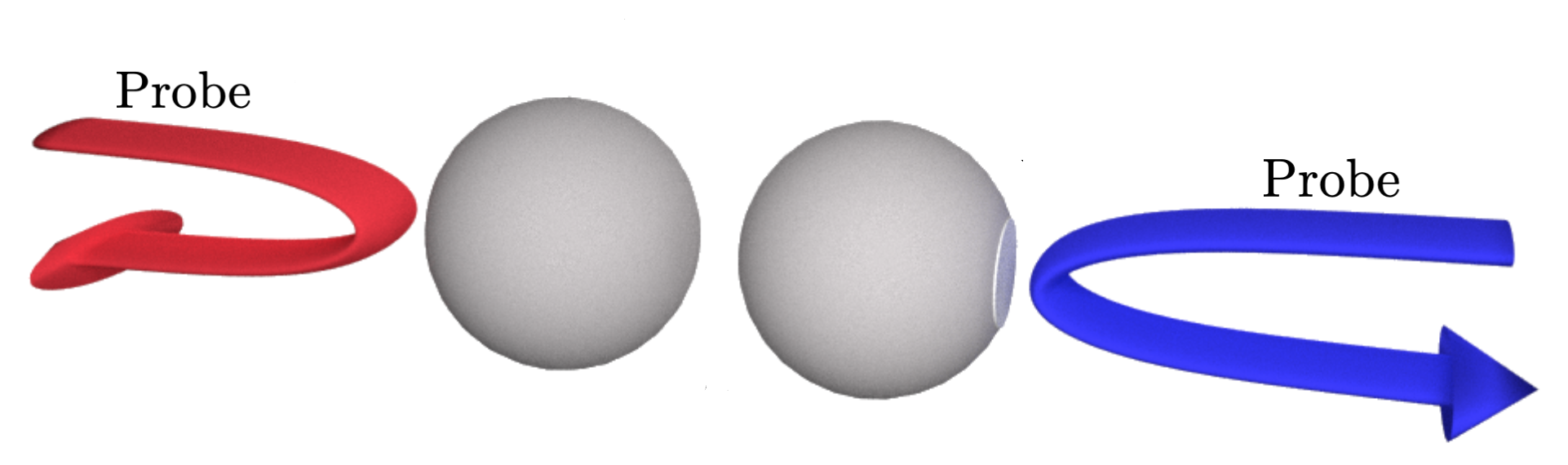}
    \caption{Two identical masses are cooled into the ground state of identical harmonic traps~\cite{npjQI_6.12,phdthesis-Tanjung}.The gravitational interaction generates position-momentum entanglement, which can be measured with weak probing lasers.}
    \label{fig:tanjung-setup}
\end{figure}

Motivated by the various advancements in optomechanics~\cite{RMP_86.1391}, in particular the cooling of macroscopic objects close to their quantum ground states~\cite{Teufel2011,Nature.478.89,NJP_11.073032}
and the measurement of bipartite entanglement~\cite{Palomaki-Science,Nature.556.473,PhysRevLett.121.220404}, another setup was proposed in Refs.~\cite{npjQI_6.12,JOPB_23.235501} where the gravitational interaction can generate position-momentum entanglement. 
The experimental setup of Ref.~\cite{npjQI_6.12} is shown schematically in Fig.~\ref{fig:tanjung-setup}. The proposal is to release two identical particles after cooling into the ground state of harmonic traps, and gravitational interaction generates a position-momentum correlation as time passes.

\begin{figure}[!b]
\centering
\subfloat[\\ Initial state.]{\includegraphics[width=0.33\linewidth]{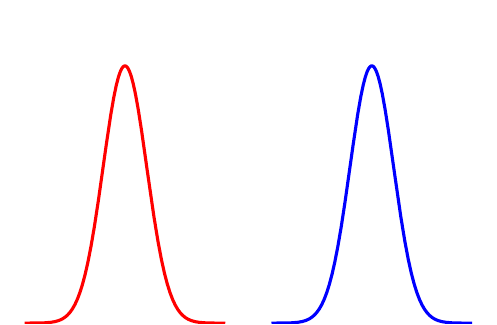}}
 \subfloat[\\ Gravitational entanglement.]{\includegraphics[width=0.33\linewidth]{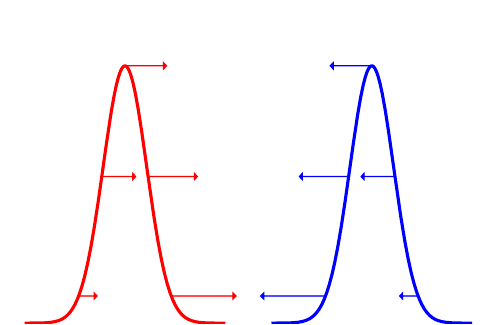}}
\subfloat[\\ Entanglement confirmation.]{\includegraphics[width=0.33\linewidth]{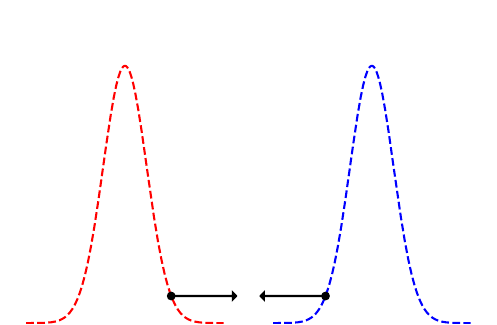}}
 
\caption{Gravity-mediated gain of position-momentum entanglement between two masses prepared in continuous-variable Gaussian states. See main text for the arguments.}
\label{fig:GenerateCorrelations}
\end{figure}

This is (crudely) explained in Fig.~\ref{fig:GenerateCorrelations}. The ground state of harmonic trapping potential is Gaussian, which means that the initial state is a product of two Gaussian wave packets separated by some distance [panel on the left].
Since the gravitational force is inversely proportional to the square of the separation, the parts of the wave packet that are closer to each other are attracted more than those that are farther away. Accordingly, different momenta are generated in different parts of the wave function [panel in the middle], leading to entanglement.
In particular, when a measurement is performed on either of the particles, the wave function collapses such that the two masses attain equal and opposite displacements and momenta, confirming generated entanglement [panel on the right].
This is of course exaggerated and can happen only in a maximally entangled state. Nevertheless, it is demonstrated in Ref.~\cite{npjQI_6.12} that entanglement keeps increasing with time. 
To be specific, with a second-order truncation of the quantum Newtonian potential and some constraints on the initial state and the time of interaction, the entanglement measured by logarithmic negativity was approximated to~\cite{npjQI_6.12}
\begin{equation}
	E \approx - \frac{1}{2} \log_2 \qty[ 1+2\qty(\frac{\hbar G t}{3\sigma^2L^3})^6 -2\qty(\frac{ \hbar G t}{3\sigma^2L^3})^3\sqrt{ 1 + \qty(\frac{ \hbar G t}{3\sigma^2L^3})^6} ],
\end{equation}
where $L$ is the initial separation between the centers of the two masses, and $\sigma$ is their inital position spread.
Note that, unlike the BMV setup, the initial states in this proposal are natural Gaussians, but the entanglement verification step is less obvious. It turns out that both the setups exhibit similar levels of resilience against environmental decoherence~\cite{Rijavec_2021}.

\section{Objectives}

This thesis aims to develop tools to resolve the interaction-mediated entanglement dynamics between two nearby quantum masses prepared in continuous-variable states.
We shall pay attention to keeping the methods generic and versatile so that they apply to arbitrary central interactions, even when many are present side by side.
As another set of objectives, we will explore the applications of the developed tools. While the (Newtonian) gravity as a coupling field is our primary focus, we will also investigate scenarios involving the Coulomb interaction, gravity and Casimir acting simultaneously, and the alternate theory of gravity (the so-called Modified Newtonian Dynamics).

\section{Organization of the Thesis}

This thesis comprises a total of six Chapters and four Appendices. 
Here in Chapter~\ref{ch:chapter1}, we described our motivations and objectives.

In Chapter~\ref{ch:chapter2}, we set up the tools for precise simulations of time evolution of (initially Gaussian) bipartite quantum states. 
In Chapter~\ref{ch:chapter3}, the introduced methods are used to study the emergence of nonclassicality during the head-on collision in the Rutherford experiment.
With a particular focus on gravity, in Chapter~\ref{ch:chapter4}, we develop a robust theoretical framework to quantify the entanglement gain between two masses prepared in natural Gaussian states (but with possibly non-Gaussian evolution).
In Chapter~\ref{ch:chapter5}, we demonstrate an exemplary utility of our methods by developing a correlation experiment to probe the Modified Newtonian Dynamics through the entanglement of microspheres.
Chapter~\ref{ch:chapter6} summarises the work reported and presents an outlook on its utility in various situations aimed at observing the interaction-mediated entanglement in quantum continuous-variable states.

Appendix~\ref{appendix:MinUncertTransf} shows the transformation of a two-mode Gaussian state between the LAB and the COM frames of reference through statistical principles.
In Appendix~\ref{appendix:Ehrenfest_COM_redmass}, we exactly resolve the Ehrenfest's dynamics for Gaussian evolution of the COM and the reduced mass wave packets.
In Appendix~\ref{appendix:EntangFormalism}, we introduce the quantifiers for entanglement between the two particles and their corresponding transformations in the COM frame of reference. 
Finally, in Appendix~\ref{appendix:ThermalStateCovMat}, we prove the Gaussianity of thermal states and relate the generation of entanglement negativity to that starting from the zero-temperature ground state.

%% file: science/ch2-TDSE.tex
\chapter{Quantum Mechanical Wave Packet Dynamics}
\label{ch:chapter2}

The quantum mechanical state of a particle is described by a wave function $\psi$. In the non-relativistic limits, this wave function evolves with time in accordance with the time-dependent Schr\"{o}dinger equation (TDSE):
\begin{equation}
 i\hbar\pdv{t}\psi(\bm{r},t) = \qty( -\frac{\hbar^2}{2m}\nabla^2 + V(\hat{\bm{r}},t) ) \psi(\bm{r},t),
 \end{equation}
where $m$ is the mass of the particle, and $V(\hat{\bm{r}},t)$ is the potential. In many physical problems where the potential is static, i.e., $V(\hat{\bm{r}},t)=V(\hat{\bm{r}})$, the resolution of TDSE is equivalent to the implementation of the time-evolution operator $\hat U$:
\begin{equation}
\psi(\bm{r},t+\Delta t) = \hat U(\Delta t) \ \psi(\bm{r},t) = \exp(-i\frac{\Delta t}{\hbar}\hat H) \psi(\bm{r},t),
\end{equation}
where $\hat H = -(\hbar^2/2m)\nabla^2 + V(\hat{\bm{r}})$ is the Hamiltonian. Note that $\hat U$ is unitary, which ensures the norm (total probability) is preserved at all times:
\begin{equation}
\braket{\psi}_{\Delta t} = \ev{\hat U^\dagger \hat U}{\psi}_{0} = \braket{\psi}_{0}.
\end{equation}
In this chapter, we develop an efficient numerical scheme for a precise resolution of single-particle TDSE, and discuss a strategy to utilize the same set of tools to handle the bipartite wave packets in the center of mass (COM) frame of reference.

\section{Introduction}

The complexities in calculating the time evolution of $\psi$ depend on the functional form of the interaction.
Even for simple Gaussians as initial states, closed analytical forms are calculable only in trivial situations, e.g., in the free space~\cite{GaussEvolFreeSpace_SMBlinder}, and the harmonic oscillator potential~\cite{GaussEvolHarmOsc_Tsuru}. Simple harmonic oscillators are regarded as the most precious tools of a theoretical physicist, but none of the fundamental forces in nature behaves so.
This demands an efficient generic numerical scheme to precisely solve the quantum evolution for arbitrary potentials which may be encountered in realistic laboratory conditions.

Along this line, the first step would be to approximate $\hat U$ up to the first order in a series expansion:
\begin{equation}
\hat U (\Delta t) =  \exp(-i\frac{\Delta t}{\hbar}\hat H)  = \sum_{n=0}^{\infty} (-i)^n \frac{\Delta t^n}{\hbar^n} \hat H^n \approx \unitop - i\frac{\Delta t}{\hbar}\hat H.
\end{equation}
However, any such truncation leads to a loss of unitarity, which in turn leads to a change in total probability over time:
\begin{eqnarray}
\braket{\psi}_{\Delta t} &=&  \ev{\hat U^\dagger \hat U}{\psi}_0   	\nonumber	\\
&=&  \ev{\qty( \unitop + i\frac{\Delta t}{\hbar}\hat H ) \qty( \unitop - i\frac{\Delta t}{\hbar}\hat H )}{\psi}_0	\nonumber \\
&=&  \ev{\qty( \unitop - i\frac{\Delta t}{\hbar}\hat H + i\frac{\Delta t}{\hbar}\hat H + \frac{\Delta t^2}{\hbar^2} \hat H^2 )}{\psi}_0	\nonumber \\
&=& \braket{\psi}_{0} + \frac{\Delta t^2}{\hbar^2} \ev{\hat H^2}{\psi}_0 > \braket{\psi}_{0} .
\end{eqnarray}
While this is acceptable for short times (the norm is preserved up to the linear order in $\Delta t$), the errors accumulate on realistic longer time scales, quickly leading to divergence. Moreover, such approximations do not respect the bidirectional numerical stability in time.
One may be tempted to include higher-order terms in the series expansion, but this would require an impractical numerical evaluation of various higher-order derivatives of the wave function. We must therefore look for alternative ways to precisely integrate the TDSE.

Various techniques have been established that are stable and mitigate errors within their capacities~\cite{FEIT1982412,Park1986,BANDRAUK1991428,Muller1999,Nurhuda1999,Watanabe2000}. In this work, we chose to utilize Cayley's form of evolution operator as it circumvents all of our problems with an \emph{unconditional stability} over long time scales~\cite{book_FiniteDiff_JWThomas}. 
We approximate the second-order derivatives with the highly accurate five-point stencil to discretise the problem onto a pentadiagonal Crank-Nicolson scheme. The resultant solutions are much more accurate when compared to the standard tridiagonal ones. This will be useful in situations where the potential is very weak, e.g., the gravitational coupling between two nearby quantum masses.

We thereafter focus on the resolution of the bipartite quantum dynamics, assuming that both the particles are initially prepared in Gaussian wave packets. The usual coordinate transformations to the COM frame of reference are discussed. At least for central interactions, the Hamiltonian decouples into the COM and the relative degrees of freedom, and the product form of a quantum state in this division is maintained at all times. However, a complete decoupling of the dynamics requires the initial quantum state to be separable in the COM frame of reference.
For a two-mode Gaussian state this happens only when the two particles are cooled in the ground state of identical harmonic traps. Note that, unlike regular problems where the COM is described by a plane wave, here it is described by a localised wave packet undergoing proper time evolution in accordance with the TDSE.
The reduced mass wave packet evolves in the interaction sourced from the COM and, based on the functional form of the potential, its time evolution can be dealt either analytically or numerically.
The methods introduced in this chapter find applications in various problems discussed throughout this thesis and beyond, which is why we make the corresponding Python implementation available publicly~\cite{Ankit_TDSE_Zenodo,Ankit_TDSE_GitHub,Ankit_2022_TDSE}.

\section{Cayley's form of evolution operator}

Cayley's form is a fractional approximation of the quantum mechanical evolution operator. The underlying idea is to evolve $\psi(\bm{r},t)$ by half of the time step forward in time, and $\psi(\bm{r},t+\Delta t)$ by half of the time step backward in time, such that they agree at time $t+\Delta t/2$~\cite{KOSLOFF198335,Ankit_2022_TDSE,Ankit_2022_Gravity,Ankit_2021_Quantum,CoP_PPuschnig}:
\begin{eqnarray}
\ket{\psi}_{t} \xrightarrow[]{\Delta t/2} &\bullet& \xleftarrow[]{\Delta t/2} \ket{\psi}_{t+\Delta t}
\nonumber	\\
\implies \hat U\qty( + \frac{\Delta t}{2} ) \ \psi(\bm{r},t) &=&  \hat U\qty( - \frac{\Delta t}{2} ) \ \psi(\bm{r},t+\Delta t)
\nonumber \\
\implies  \exp(-i\frac{\hat H\Delta t}{2\hbar}) \psi(\bm{r},t) &=&  \exp(+i\frac{\hat H\Delta t}{2\hbar}) \psi(\bm{r},t+\Delta t).
\end{eqnarray}
With a first-order approximation on both sides,
\begin{equation}
\qty(  \unitop- i\frac{ \hat H\Delta t}{2\hbar}  ) \psi(\bm{r},t)
\approx
 \qty(  \unitop+ i\frac{ \hat H\Delta t}{2\hbar}  ) \psi(\bm{r},t+\Delta t) ,
\label{eq:ImpExpExpression}
\end{equation}
we arrive at 
\begin{equation}
 \psi(\bm{r},t+\Delta t) = \qty(  \unitop+ i\frac{ \hat H\Delta t}{2\hbar}  )^{-1} \qty(  \unitop- i\frac{ \hat H\Delta t}{2\hbar}  ) \psi(\bm{r},t).
\end{equation}
Hence, Cayley's form of evolution operator is given by
\begin{equation}
\hat U(\Delta t) = \left( \unitop+ i\frac{ \hat H\Delta t}{2\hbar} \right)^{-1} \left( \unitop- i\frac{ \hat H\Delta t}{2\hbar} \right).
\label{eq:cayley_operator}
\end{equation}
An implementation of Cayley's form solves many problems at once, e.g.,
\begin{enumerate}[label=\arabic*)]

\item 
The bidirectional numerical stability in time is inbuilt into the theoretical framework [see Eq.~\eqref{eq:ImpExpExpression}].

\item 
A replacement of the second-order derivatives in Hamiltonian with finite difference formulas tells us that the wave function at different times is related by a Crank-Nicolson scheme, which is unconditionally stable for TDSE-like problems~\cite{book_FiniteDiff_JWThomas}.

\item 
The total probability is preserved over time as the resultant evolution operator in Eq.~\eqref{eq:cayley_operator} is unitary, as shown below. 

Since $\qty(\hat A \hat B)^\dagger = \hat{B}^\dagger \hat{A}^\dagger$, the hermitian conjugate of $\hat U$ is
\begin{equation}
\hat U^\dagger =  \qty[ \left( \unitop+ i\frac{ \hat H\Delta t}{2\hbar} \right)^{-1} \left( \unitop- i\frac{ \hat H\Delta t}{2\hbar} \right) ]^\dagger
= \left( \unitop- i\frac{ \hat H\Delta t}{2\hbar} \right)^\dagger \left( \unitop+ i\frac{ \hat H\Delta t}{2\hbar} \right)^{-1,\dagger}.
\end{equation}
Note that for any operator $\hat A$ we have
\begin{eqnarray}
&&	\hat{A} \hat{A}^{-1} = \unitop
\nonumber	\\
\implies 	&&	\qty( \hat{A} \hat{A}^{-1} )^\dagger = \unitop
\nonumber	\\
\implies 	&&	\qty(\hat{A}^{-1})^\dagger \hat{A} ^\dagger = \unitop,	\hspace{1.5cm}
 :\qty{ \qty(\hat A \hat B)^\dagger = \hat{B}^\dagger \hat{A}^\dagger },
\nonumber	\\
\implies 	&&	 \qty(\hat{A}^{-1})^\dagger \hat{A} ^\dagger \qty[ \qty(\hat{A} ^\dagger)^{-1} ] = \unitop  \qty[ \qty(\hat{A} ^\dagger)^{-1}  ]
\nonumber	\\
\implies 	&&	 \qty(\hat{A}^{-1})^\dagger \qty[ \hat{A} ^\dagger  \qty(\hat{A} ^\dagger)^{-1} ] =  \qty(\hat{A} ^\dagger)^{-1} 
\nonumber	\\
\implies 	&&	 \qty(\hat{A} ^{-1})^\dagger = \qty(\hat{A} ^\dagger)^{-1}.  	
\end{eqnarray}

Accordingly, we get
\begin{equation}
\hat U^\dagger	\equiv \left( \unitop - i\frac{ \hat H\Delta t}{2\hbar} \right)^\dagger \left( \unitop + i\frac{ \hat H\Delta t}{2\hbar} \right)^{\dagger,-1}	= \left( \unitop + i\frac{ \hat H\Delta t}{2\hbar} \right) \left( \unitop - i\frac{ \hat H\Delta t}{2\hbar} \right)^{-1}, 
\end{equation}
which implies
\begin{equation}
\hat U \hat U^\dagger
	= \left( \unitop+ i\frac{ \hat H\Delta t}{2\hbar} \right)^{-1} \left( \unitop- i\frac{ \hat H\Delta t}{2\hbar} \right) \left( \unitop + i\frac{ \hat H\Delta t}{2\hbar} \right) \left( \unitop - i\frac{ \hat H\Delta t}{2\hbar} \right)^{-1}.
\end{equation}
The two terms in the middle commute, and hence
\begin{equation}
	\hat U \hat U^\dagger
\equiv \left( \unitop+ i\frac{ \hat H\Delta t}{2\hbar} \right)^{-1} \left( \unitop + i\frac{ \hat H\Delta t}{2\hbar} \right)  \left( \unitop - i\frac{ \hat H\Delta t}{2\hbar} \right) \left( \unitop - i\frac{ \hat H\Delta t}{2\hbar} \right)^{-1} = \unitop.
\end{equation}

\end{enumerate}

\section{The tridiagonal discretisation}

The standard practice for calculating numerical derivatives is to implement various finite-difference approximations. In this work we only deal with one-dimensional problems, i.e.,
\begin{equation}
\hat H = - \frac{\hbar^2}{2m} \pdv[2]{x} + V(\hat x) .
\end{equation}
The easiest is to replace the second derivative in Hamiltonian with the three-point central-difference formula,
\begin{equation}
 f''(x) = \frac{f(x+\Delta x) - 2f(x) + f(x-\Delta x)}{\Delta x^2} + \mathcal{O}(\Delta x^2) ,
\end{equation}
which transforms Eq.~\eqref{eq:ImpExpExpression} to
\begin{equation}
\begin{split}
\psi_j^{n+1} + \frac{i \Delta t}{2\hbar} \left[ -\frac{\hbar^2}{2m} \left( \frac{\psi_{j+1}^{n+1} - 2\psi_{j}^{n+1} + \psi_{j-1}^{n+1}}{\Delta x^2} \right) + V_j  \psi_j^{n+1} \right] \\
= \psi_j^{n} - \frac{i \Delta t}{2\hbar} \left[ -\frac{\hbar^2}{2m} \left( \frac{\psi_{j+1}^{n} - 2\psi_{j}^{n} + \psi_{j-1}^{n}}{\Delta x^2} \right) + V_j  \psi_j^{n} \right],
\end{split}
\label{TDSE_disc_tridiag}
\end{equation}
where $f_j^n \equiv f(x_j,t_n)$, $\Delta x=x_{j+1}-x_j$ is the grid size, and $\Delta t=t_{n+1}-t_n$ is the time step. To simplify the notation, let us call:
\begin{eqnarray}
\zeta_j^n &=& \psi_j^{n} - \frac{i \Delta t}{2\hbar} \left[ -\frac{\hbar^2}{2m} \left( \frac{\psi_{j+1}^{n} - 2\psi_{j}^{n} + \psi_{j-1}^{n}}{\Delta x^2} \right) + V_j  \psi_j^{n} \right],
\nonumber	\\
a_j &=& 1 + \frac{i \Delta t}{2\hbar} \left( \frac{\hbar^2}{m\Delta x^2} + V_j \right),
\hspace{1cm}
b  =  - \frac{i\hbar\Delta t}{4m\Delta x^2}.
\end{eqnarray}
Eq.~(\ref{TDSE_disc_tridiag}) can now be re-written as
\begin{equation}
\begin{pmatrix}
a_1 & b \\
\ddots & \ddots & \ddots \\
& b & a_{j-1} & b \\
& & b & a_j & b \\
& & & b & a_{j+1} & b \\
& & & & \ddots & \ddots & \ddots \\
& & & & & b & a_{J-1}
\end{pmatrix} \\
\cdot \begin{pmatrix}
\psi_1^{n+1} \\ \vdots \\ \psi_{j-1}^{n+1} \\ \psi_{j}^{n+1} \\ \psi_{j+1}^{n+1} \\ \vdots\\ \psi_{J-1}^{n+1} \end{pmatrix} = \begin{pmatrix}
\zeta_1^{n} \\ \vdots \\ \zeta_{j-1}^{n} \\ \zeta_{j}^{n} \\ \zeta_{j+1}^{n} \\ \vdots \\ \zeta_{J-1}^{n}
\end{pmatrix},
\label{TLSE_matrix}
\end{equation}
where $J$ is the dimension of the position grid. The matrix on the left is composed entirely of constants and the old wave function is stored in the column vector on the right, $\zeta$. 
The standard practice to solve such a system of linear equations is through the Thomas algorithm, which is nothing but Gauss elimination in a tridiagonal case.

\section{The pentadiagonal discretisation}
\label{sec:pentadiagonaldiscretisation}

The three-point formula for the second derivative is accurate up to an error $\mathcal{O}(\Delta x^2)$. The corresponding tridiagonal discretisation works very well in most situations, except for astonishingly weak potentials, e.g., the gravitational field between two quantum particles. In such cases, it falls short due to the accumulation of errors over time. A numerical scheme is as good as the underlying finite-difference approximations, and hence we replace the second-order derivative with the highly accurate five-point stencil:
\begin{equation}
f''(x) = \frac{-f(x+2\Delta x) + 16f(x+\Delta x) - 30f(x) + 16f(x-\Delta x) - f(x-2\Delta x)}{12\Delta x^2} + \mathcal{O}(\Delta x^4) .
\end{equation}
In result, Eq.~\eqref{eq:ImpExpExpression} is now discretised as
\begin{equation}
\begin{split}
\psi_j^{n+1} + \frac{i \Delta t}{2\hbar} \qty[ -\frac{\hbar^2}{2m} \qty(  \frac{-\psi_{j+2}^{n+1} +16\psi_{j+1}^{n+1} - 30\psi_{j}^{n+1} +16\psi_{j-1}^{n+1} -\psi_{j-2}^{n+1}}{12\Delta x^2}  ) + V_j  \psi_j^{n+1}  ] \\ = \psi_j^{n} - \frac{i \Delta t}{2\hbar} \qty[ -\frac{\hbar^2}{2m} \qty( \frac{-\psi_{j+2}^{n} +16\psi_{j+1}^{n} - 30\psi_{j}^{n} +16\psi_{j-1}^{n} -\psi_{j-2}^{n}}{12\Delta x^2} )  + V_j  \psi_j^{n}  ].
\end{split}
\end{equation}
Following a similar approach as in the previous section, we denote
\begin{eqnarray}
\zeta_j^n &=& \psi_j^{n} - \frac{i \Delta t}{2\hbar} \qty[ -\frac{\hbar^2}{2m} \qty(  \frac{-\psi_{j+2}^{n} +16\psi_{j+1}^{n} - 30\psi_{j}^{n} +16\psi_{j-1}^{n} -\psi_{j-2}^{n}}{12\Delta x^2}  )  + V_j  \psi_j^{n} ],
\nonumber	\\
a_j &=& 1 + \frac{i\Delta t}{2\hbar} \qty( \frac{5\hbar^2}{4m\Delta x^2} + V_j  ),
\hspace{1cm}
b = -\frac{i\hbar\Delta t}{3m\Delta x^2},
\hspace{1cm}
c = \frac{i\hbar\Delta t}{48m\Delta x^2},
\end{eqnarray}
which reduces the problem to
\begin{equation}
\begin{pmatrix}
a_1 & b & c \\
\ddots & \ddots & \ddots & \ddots \\
& \ddots & \ddots & \ddots & \ddots & \ddots \\
& & c & b & a_{j-1} & b & c
\\ & & & c & b & a_j & b & c
\\ & & & & c & b & a_{j+1} & b & c
\\ & & & & & \ddots & \ddots & \ddots & \ddots & \ddots
\\ & & & & & & \ddots & \ddots & \ddots & \ddots
\\ & & & & & & & c & b & a_{J-2}
\end{pmatrix} \\
\cdot \begin{pmatrix}
\psi_1^{n+1} \\
\vdots \\
\psi_{j-2}^{n+1} \\
\psi_{j-1}^{n+1} \\
\psi_{j}^{n+1} \\
\psi_{j+1}^{n+1} \\
\psi_{j+2}^{n+1} \\
\vdots\\
\psi_{J-2}^{n+1} \end{pmatrix} = \begin{pmatrix}
\zeta_1^{n} \\
\vdots \\
\zeta_{j-2}^{n} \\
\zeta_{j-1}^{n} \\
\zeta_{j}^{n} \\
\zeta_{j+1}^{n} \\
\zeta_{j+2}^{n} \\
\vdots \\
\zeta_{J-2}^{n}
\end{pmatrix}.
\label{eq:PDiagEqn}
\end{equation}
We now have a pentadiagonal system of linear equations for $J-2$ unknown wave function values at time $t_{n+1}$. Note that the Thomas algorithm is not applicable anymore, as it works only for tridiagonal matrices. Accordingly, we implement the LU-factorisation techniques as they are versatile enough to solve both the tridiagonal and the pentadiagonal system of equations. We perform an LU-factorisation of the constant matrix on the left, followed by forward and backward substitutions of the $\zeta$ vector on the right~\cite{CoP_PPuschnig,Ankit_2021_Quantum}. A Python implementation is publicly available at Zenodo and GitHub~\cite{Ankit_TDSE_Zenodo,Ankit_TDSE_GitHub}, with the corresponding documentation in Ref.~\cite{Ankit_2022_TDSE}. Note that these methods work only for well-localised square-integrable wave functions.
Otherwise, the reflections from numerical boundaries lead to unwanted interference.

\section{Comparison of numerical errors}

In this section, we calculate the evolution of a Gaussian wave packet with the (standard) tridiagonal and the (improved) pentadiagonal methods. For demonstration, we assume that the wave packet is initially centered around $x_0$ with a position spread $\sigma$ and a momentum $p_0$:
\begin{equation}
\psi(x,t=0) = \frac{1}{\sqrt{\sigma\sqrt{2\pi}}}
\ \exp\qty( -\frac{(x-x_0)^2}{4\sigma^2} + i \frac{p_0}{\hbar}(x-x_0) ),
\end{equation}
and compare the numerical errors accumulated in both methods for the cases of evolution in free space and the harmonic oscillator potential.
As a parameter of interest, we chose Heisenberg's uncertainty product $\bm{\Delta} x \bm{\Delta} p$, since it involves both the statistical moments of the position and momentum variables.

\begin{enumerate}[label=\alph*)]

\item Evolution in the free space, i.e., $V = 0$, can be solved analytically using Fourier transformation techniques~\cite{GaussEvolFreeSpace_SMBlinder}, which imply
\begin{equation}
\psi(x,t) = \frac{1}{\sqrt{\sigma(1+i\omega_0 t)\sqrt{2\pi}}} \ \exp[ - \frac{1}{4\sigma^2(1 + i \omega_0 t)} \qty( x-x_0-2i \sigma^2 \frac{p_0}{\hbar} )^2 - \sigma^2\frac{p_0^2}{\hbar^2}  ],
\end{equation}
where $\omega_0 = \hbar/2m\sigma^2$. One can calculate the position and momentum spreads to see that the uncertainty product is given by
\begin{equation}
\bm{\Delta}x  \bm{\Delta}p = \frac{\hbar}{2}\sqrt{1+\omega_0^2t^2}.
\label{eq:uncp_FS}
\end{equation}

\item Evolution in the harmonic oscillator potential, i.e., $V = \frac{1}{2}m\omega^2x^2$, can also be solved analytically~\cite{GaussEvolHarmOsc_Tsuru}. The closed form is rather complicated, but one can use the Ehrenfest's theorem to calculate the time evolution of the average statistical moments.
The corresponding uncertainty product is given by
\begin{equation}
\bm{\Delta} x \bm{\Delta} p  = \frac{\hbar}{2}\sqrt{ \cos^4(\omega t) + \sin^4(\omega t) + \frac{1}{4} \qty(  \frac{\omega_0^2}{\omega^2} + \frac{\omega^2}{\omega_0^2}  ) \sin^2(2\omega t) } .
\label{eq:uncp_HO}
\end{equation}

\end{enumerate}

\begin{figure}[!t] 
\centering

\subfloat[Evolution in the free space, $V = 0$. Initial wave packet is centered at $x = -50$ with a width of $2$ units and a momentum of $+1$ unit.]{\includegraphics[width=\linewidth]{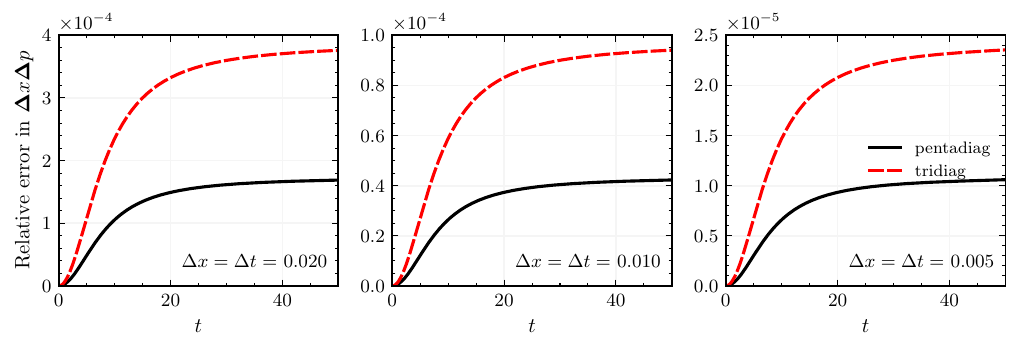}\label{fig:errors_cayley_FS}}

\subfloat[Evolution in the harmonic oscillator potential, $V=\frac{1}{2}m\omega^2x^2$ with $\omega = 0.1$. Initial wave packet is centered at $x = -10$, with a width of $2$ units.]{\includegraphics[width=\linewidth]{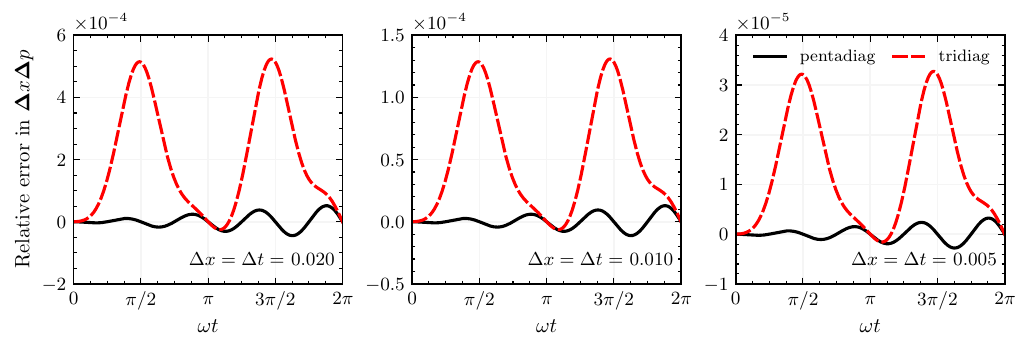}\label{fig:errors_cayley_HO}}

\caption{Comparison of errors in the tridiagonal and the pentadiagonal solutions of the time-dependent Schr\"{o}dinger equation. $\bm{\Delta} x \bm{\Delta} p$ is the Heisenberg's uncertainty product. We assume $\hbar = 1$, $m = 1$, and the relative errors are calculated w.r.t. the analytical results discussed in the main text. $\Delta x$ denotes the grid size, and $\Delta t$ is the time step. Note different vertical scales in each panel.}

\label{fig:errors_cayley}

\end{figure}

Assuming $\hbar = 1$, $m = 1$, in Fig.~\ref{fig:errors_cayley} we show the errors in the uncertainty product, computed relative to the closed forms in Eqs.~\eqref{eq:uncp_FS} and~\eqref{eq:uncp_HO}. It can be easily seen that our pentadiagonal solutions are far more accurate than the standard ones. Accordingly, they will be used for simulating quantum evolution in extremely weak fields. In the following chapters, we use the standard tridiagonal solutions for studying the head-on collision of charged particles~\cite{Ankit_2021_Quantum}, and the highly accurate pentadiagonal solutions for the astonishingly weak gravitational coupling between two nearby quantum objects~\cite{Ankit_2022_Gravity}.

Note that Heisenberg's uncertainty product requires the first two statistical moments of position and momentum operators, which can be evaluated with:
\begin{equation}
    \ev{\hat x^n}  =  \int_{-\infty}^{+\infty} dx \ \psi^* \ x^n \ \psi,
\hspace{1cm}
    \ev{\hat p^n}  =  (-i\hbar)^n \int_{-\infty}^{+\infty} dx \ \psi^* \ \pdv[n]{\psi}{x}.
\end{equation}
It is worth mentioning that, for a well localised problem, $\ev{\hat p^2}$ can be calculated with the first-order derivative only.
In such cases the wave functions are square-integrable: $\lim_{x\to\pm \infty} \psi = 0$ and $\lim_{x\to\pm \infty} d\psi/dx = 0$, and integration by parts implies
\begin{equation}
\ev{\hat p^2} 
=
 -\hbar^2 \int_{-\infty}^{+\infty} dx \ \psi^* \pdv[2]{\psi}{x}
=
-\hbar^2 \qty[ \psi^* \pdv{\psi}{x} - \int dx \ \pdv{\psi^*}{x} \pdv{\psi}{x} ]_{-\infty}^{+\infty}
=
\hbar^2 \int_{-\infty}^{+\infty} dx \ \abs{\pdv{\psi}{x}}^2.
\end{equation}
Furthermore, we go one step further and utilise the law of conservation of energy to calculate $\ev{\hat p^2}$ without evaluating any numerical derivative whatsoever. This is explained as follows.
A unitary evolution implies that the total energy, $\ev{\hat H}$, is a constant of motion. At $t=0$ we start with a minimum uncertainty Gaussian wave packet characterized by $\bm{\Delta} x \bm{\Delta} p (0) = \hbar/2$, which implies
\begin{equation}
\ev{\hat p^2(0)} = \ev{\hat p(0)}^2 + \bm{\Delta} p^2 (0) = p_0^2 + \frac{\hbar^2}{4\sigma^2}.
\end{equation}
On equating $\ev{\hat H(0)}$ with $\ev{\hat H}$ we arrive at
\begin{equation}
    \ev{\hat p^2} = p_0^2 + \frac{\hbar^2}{4\sigma^2} + 2m \Big(  \ev{V(0)} - \ev{V} \Big) ,
\end{equation}
where $\ev{V(0)}$ is readily available in closed formulas, e.g., in the free space $\ev{V(0)} = 0$, and in the harmonic oscillator potential $\ev{V(0)} = \frac{1}{2}m\omega_0^2\ev{\hat x^2(0)} = \frac{1}{2}m\omega_0^2(x_0^2+\sigma^2)$.

\section{The heartbeating inside a box}

\begin{figure}[!t]
\centering
\includegraphics[width=\linewidth]{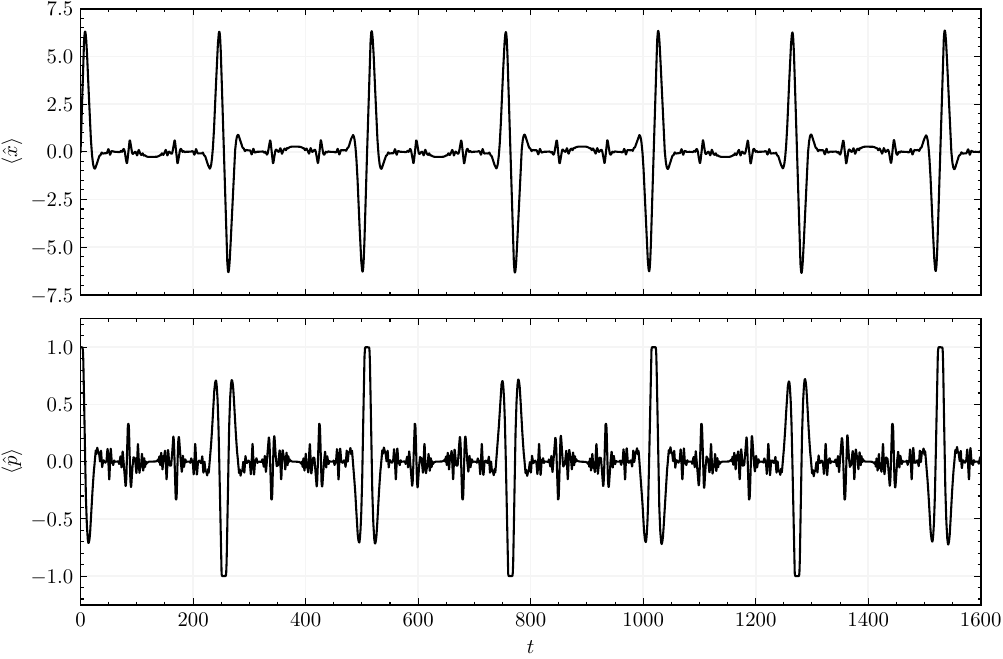}
\caption{Expected position and momentum of a Gaussian evolving inside a box extending from $x=-10$ to $+10$. Assuming $\hbar = m = 1$, the initial wave packet is centered at the origin with a width of 1 unit and a momentum of +1 unit. The size of the position grid is $\Delta x = 0.01$, and the time step is $\Delta t = 0.01$.}
\label{fig:heartbeat_in_a_box}
\end{figure}

Now that we have a numerical scheme for resolving the quantum dynamics of localised wave packets, in this section we play around and demonstrate the fascinating dance of an (initially Gaussian) wave packet evolving inside a box. Assuming $\hbar  = 1$, we consider a particle of mass $m = 1$ trapped inside a box extending between $x = \pm 10$. As a sanity check, it is first confirmed that the probability distribution does not change with time when the initial state corresponds to any one of the eigenstates:
\begin{equation}
\psi_n(x,0) = \begin{cases}
\sqrt{\frac{2}{L}} \sin \qty( \frac{n\pi x}{L} ) ,	\hspace{1cm}	n = 0,2,4,\dots 
\\
\sqrt{\frac{2}{L}} \cos \qty( \frac{n\pi x}{L} ) ,	\hspace{1cm}	n = 1,3,5,\dots 
\end{cases}
\end{equation}
We then consider the initial state as a Gaussian wave packet with a width of $1$ unit centered at the origin with $1$ unit of momentum to the right. This way, the forward part of the wave packet hits the boundary at $x = +10$ and is reflected back. These reflections interfere with the rest of the wave packet to create a beautiful dancing pattern.
In Fig.~\ref{fig:heartbeat_in_a_box}, we show the expected position and momentum as a function of time which looks like a periodically repeating heartbeat pattern. In the following chapters we shall discuss the practical applications of Cayley's propagator in the Rutherford experiment and the gravitational entanglement dynamics.

\section{Bipartite wave packet dynamics}	
\label{appendix:TDSE-Transf2COM}

Till now we have discussed the case of a single-particle wave packet evolving in a classical background potential.
It turns out that the same methods can be utilised to solve the bipartite dynamics after a careful change of coordinates. The TDSE for a system of two particles $A$ and $B$ is given by
\begin{eqnarray}
	\Bigg( -\frac{\hbar^2}{2m_A} \pdv[2]{x_A}  -\frac{\hbar^2}{2m_B} \pdv[2]{x_B} + V(\hat x_A,\hat x_B) \Bigg) \Psi(x_A,x_B,t) 
	= i\hbar\pdv{t} \Psi(x_A,x_B,t),
\end{eqnarray}
where $x_A$ and $x_B$ are the positions/displacements of the masses $m_A$ and $m_B$, and $\hat p_A = -i\hbar \partial/\partial x_A$ and $\hat p_B = -i\hbar \partial/\partial x_B$ are their respective momenta. In an attempt to decouple this two-body problem, we make a coordinate transformation to the COM frame of reference:
\begin{equation}
R = \frac{m_Ax_A+m_Bx_B}{m_A+m_B},
\hspace{1cm}
r = x_B-x_A,
\end{equation}
where $R$ and $r$ are the positions/displacements of the COM [mass $M = m_A+m_B$] and the reduced mass [mass $\mu = m_Am_B/(m_A+m_B)$], respectively. One can take their time derivatives to write their respective momenta as
\begin{eqnarray}
P = M\dv{R}{t} 	= \frac{m_A+m_B}{m_A+m_B} \qty( m_A \dv{x_A}{t} + m_B \dv{x_B}{t} ) 	= p_A+p_B,
\nonumber	\\
p = \mu\dv{r}{t} = \frac{m_Am_B}{m_A+m_B} 	\qty( \dv{x_B}{t} - \dv{x_A}{t} )	 = \frac{m_Ap_B-m_Bp_A}{m_A+m_B} ,
\end{eqnarray}
which implies that the inverse transformations are
\begin{equation}
	x_A = R  -  \frac{m_B}{M}r,
\hspace{1cm}
	x_B = R + \frac{m_A}{M}r,
\hspace{1cm}
	p_A = \frac{m_A}{M}P - p,
\hspace{1cm}
	p_B = \frac{m_B}{M}P + p.
\end{equation}
Given that $x_A$ and $x_B$ are functions of $R$ and $r$, the rules of partial differentiation imply
\begin{eqnarray}
\pdv{x_A} &=& \qty( \pdv{R}{x_A} ) \pdv{R}		+	\qty( \pdv{r}{x_A} ) \pdv{r}	= \frac{m_A}{M} \pdv{R} - \pdv{r},
\\
\pdv{x_B} &=& \qty( \pdv{R}{x_B} ) \pdv{R}		+	\qty( \pdv{r}{x_B} ) \pdv{r}	= \frac{m_B}{M} \pdv{R} + \pdv{r}.
\end{eqnarray}
Similarly, the second-order derivatives can be calculated as
\begin{eqnarray}
\pdv[2]{x_A} &=& \qty( \pdv{R}{x_A} ) \pdv{R}	 \qty( \frac{m_A}{M} \pdv{R} - \pdv{r} ) + \qty( \pdv{r}{x_A} ) \pdv{r}	 \qty( \frac{m_A}{M} \pdv{R} - \pdv{r} )	\nonumber \\
&& =	\frac{m_A}{M} \qty( \frac{m_A}{M} \pdv[2]{R} - \pdv{}{R}{r} )	- 	\qty( \frac{m_A}{M} \pdv{}{r}{R} - \pdv[2]{r} )	\nonumber	\\
&& =  \frac{m_A^2}{M^2} \pdv[2]{R} + \pdv[2]{r} - 2\frac{m_A}{M}\pdv{}{R}{r},
\\	\nonumber \\
\pdv[2]{x_B} &=& \qty( \pdv{R}{x_B} ) \pdv{R}	 \qty( \frac{m_B}{M} \pdv{R} + \pdv{r} ) + \qty( \pdv{r}{x_B} ) \pdv{r}	 \qty( \frac{m_B}{M} \pdv{R} + \pdv{r} )	\nonumber \\
&& =	\frac{m_B}{M} \qty( \frac{m_B}{M} \pdv[2]{R} + \pdv{}{R}{r} )	+ 	\qty( \frac{m_B}{M} \pdv{}{r}{R} + \pdv[2]{r} )	\nonumber	\\
&& =  \frac{m_B^2}{M^2} \pdv[2]{R} + \pdv[2]{r} + 2\frac{m_B}{M}\pdv{}{R}{r},
\end{eqnarray}
and hence the kinetic energy part of the Hamiltonian is
\begin{equation}
- \frac{\hbar^2}{2m_A} \pdv[2]{x_A}  - \frac{\hbar^2}{2m_B} \pdv[2]{x_B} = -\frac{\hbar^2}{2M} \pdv[2]{R}  -\frac{\hbar^2}{2\mu} \pdv[2]{r}.
\end{equation}
Within the scope of this thesis, we only deal with central interactions, $V(x_A,x_B) = V(x_B-x_A)=V(r)$. The Schr\"odinger equation now becomes
\begin{equation}
	\Bigg( -\frac{\hbar^2}{2M} \pdv[2]{R}   -\frac{\hbar^2}{2\mu} \pdv[2]{r} + V(\hat r) \Bigg) \Psi(x_A,x_B,t)
	=  i\hbar\pdv{t}\Psi(x_A,x_B,t).
	\label{eq:TDSE_in_COM_general}
\end{equation}
Given that the initial wave function transforms to the COM frame as $\Psi(x_A,x_B,t=0) = \phi(R,t=0) \ \psi(r,t=0)$, the separation of variables in Eq.~\eqref{eq:TDSE_in_COM_general} will ensure that the product form is maintained at all times. Accordingly, the bipartite problem decouples as
\begin{eqnarray}
	-\frac{\hbar^2}{2M} \pdv[2]{R} \phi(R,t)  &=&  i\hbar\pdv{t}\phi(R,t),	
\label{eq:COMSE-free}
\\
	\qty( -\frac{\hbar^2}{2\mu} \pdv[2]{r} + V(\hat r) ) \psi(r,t) &=& i\hbar\pdv{t}\psi(r,t),
\end{eqnarray}
where $\hat P = -i\hbar\partial/\partial R$ and $\hat p = -i\hbar\partial/\partial r$ can now be identified as the momentum operators for the COM and the reduced mass, respectively. Here the COM evolves in the free space, which can be easily solved with analytical techniques~\cite{GaussEvolFreeSpace_SMBlinder}. The reduced mass evolves under the influence of the interaction $V(r)$ and, depending on its functional form, one can use either the analytical or the numerical method to calculate the corresponding time evolution. The two-body wave function is given by the product
\begin{equation}
	\Psi(x_A,x_B,t) = \phi \qty( \frac{m_Ax_A+m_Bx_B}{m_A+m_B},t  ) \ \psi \qty( x_B-x_A,t ).
	\label{eq:TBWF_LAB2COM_relation}
\end{equation}

\subsection{Transformation of a two-mode Gaussian state}
\label{sec:transform_TMGS}

\begin{figure}[!t]
	\centering
	\includegraphics[width=\linewidth]{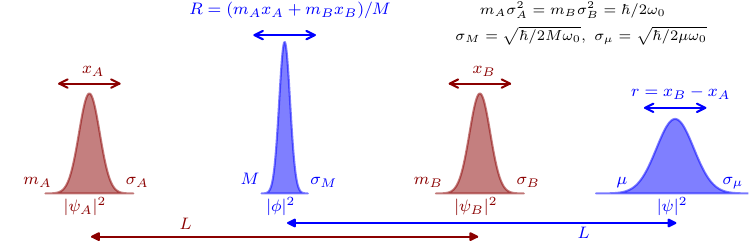}
	\caption{From LAB frame to COM frame. Gaussianity of the initial state is preserved as well as the product form. The widths, however, are different in different frames.}
	\label{fig:TMGS_genreal}
\end{figure}

Recall that in the previous section the bipartite TDSE decouples to two independent TDSEs only when the initial state can be written as a product form in the COM frame of reference. In this section we derive the conditions under which this happens for a two-mode Gaussian state $\Psi(x_A,x_B,t=0) = \psi_A(x_A) \ \psi_B(x_B)$. In the displacement space,
\begin{eqnarray}
\psi_A(x_A) &=& \qty( \frac{1}{2\pi\sigma_A^2} )^{1/4} \exp(-\frac{x_A^2}{4\sigma_A^2} + i\frac{p_{A0} }{\hbar}x_A ),
\\
\psi_B(x_B) &=& \qty( \frac{1}{2\pi\sigma_B^2} )^{1/4} \exp(-\frac{x_B^2}{4\sigma_B^2} + i\frac{p_{B0} }{\hbar}x_B ) ,
\end{eqnarray}
where $p_{A0}$ and $p_{B0}$ denote the initial momenta of the two particles. With simple algebra, we can rearrange the initial wave function as
\begin{eqnarray}
\Psi(t=0) &=& \qty( \frac{1}{2\pi\sigma_A^2} )^{1/4} \exp(-\frac{x_A^2}{4\sigma_A^2} + i\frac{p_{A0} }{\hbar}x_A ) \ \qty( \frac{1}{2\pi\sigma_B^2} )^{1/4} \exp(-\frac{x_B^2}{4\sigma_B^2} + i\frac{p_{B0} }{\hbar}x_B )	\nonumber	\\
&=&   \qty( \frac{1}{2\pi\sigma_A^2})^{1/4} \qty( \frac{1}{2\pi\sigma_B^2} )^{1/4} \exp(-\frac{x_A^2}{4\sigma_A^2}  -\frac{x_B^2}{4\sigma_B^2}) \exp\qty( i\frac{ p_{A0}x_A+p_{B0}x_B }{\hbar} ) .	
\end{eqnarray}
We shall now make use of the inverse transformations to express this in the COM frame. To start with, the Gaussian part is given by
\begin{eqnarray}
\frac{x_A^2}{\sigma_A^2} + \frac{x_B^2}{\sigma_B^2} &=& \frac{1}{\sigma_A^2} \qty( R - \frac{m_B}{M}r )^2 + \frac{1}{\sigma_B^2} \qty( R + \frac{m_A}{M}r )^2
\nonumber	\\
&=& \frac{ 1 }{\sigma_A^2} \qty( R^2 + \frac{m_B^2}{M^2}r^2 - 2\frac{m_B}{M}Rr  ) + \frac{ 1 }{\sigma_B^2} \qty( R^2 + \frac{m_A^2}{M^2}r^2 + 2\frac{m_A}{M}Rr )
\nonumber	\\
&=&	\qty( \frac{ \sigma_A^2+\sigma_B^2 }{\sigma_A^2\sigma_B^2} ) R^2 + \qty( \frac{m_A^2\sigma_A^2+m_B^2\sigma_B^2}{M^2\sigma_A^2\sigma_B^2}) r^2 + \qty( \frac{m_A\sigma_A^2-m_B\sigma_B^2}{M\sigma_A^2\sigma_B^2}) Rr.
\end{eqnarray}
The last term needs to vanish for the state to decouple into independent Gaussians in $R$ and $r$, which happens only when $m_A\sigma_A^2=m_B\sigma_B^2$. Note that
\begin{equation}
m_A \sigma_A^2 = m_A \times \frac{\hbar}{2m_A\omega_A} = \frac{\hbar}{2\omega_A},
\hspace{1cm}
m_B \sigma_B^2 = m_B \times \frac{\hbar}{2m_B\omega_B} = \frac{\hbar}{2\omega_B},
\end{equation}
and hence the conditionality $m_A\sigma_A^2=m_B\sigma_B^2$ essentially implies $\omega_A = \omega_B \equiv \omega_0$. The wave packet dynamics decouples only when the two particles are prepared in the ground state of identical harmonic traps of frequency $\omega_0$.
 Under this assumption,
\begin{eqnarray}
\frac{x_A^2}{\sigma_A^2} + \frac{x_B^2}{\sigma_B^2} &=&	\qty[ \frac{ \qty(\frac{\hbar}{2m_A\omega_0}) + \qty(\frac{\hbar}{2m_B\omega_0}) } {  \qty(\frac{\hbar}{2m_A\omega_0}) \qty(\frac{\hbar}{2m_B\omega_0}) } ] R^2 + \qty[ \frac{m_A^2\qty(\frac{\hbar}{2m_A\omega_0})+m_B^2\qty(\frac{\hbar}{2m_B\omega_0})}{M^2\qty(\frac{\hbar}{2m_A\omega_0})\qty(\frac{\hbar}{2m_B\omega_0})} ] r^2 
\nonumber	\\
&=& \qty[ \frac{2(m_A+m_B)\omega_0}{\hbar} ] R^2 +  \qty[ \frac{2m_Am_B\omega_0}{(m_A+m_B)\hbar} ] r^2
\nonumber	\\
&=& \qty( \frac{2M\omega_0}{\hbar} ) R^2 +  \qty( \frac{2\mu\omega_0}{\hbar} ) r^2
\nonumber	\\
&=&	\frac{R^2}{\sigma_M^2} + \frac{r^2}{\sigma_\mu^2},
\end{eqnarray}
where $\sigma_M^2 = \hbar/2M\omega_0$ and $\sigma_\mu^2 = \hbar/2\mu\omega_0$. For the plane wave part of $\Psi(t=0)$ we have
\begin{eqnarray}
p_{A0}x_A+p_{B0}x_B &=& p_{A0}\qty( R - \frac{m_B}{M}r ) + p_{B0}\qty( R + \frac{m_A}{M}r )	\nonumber	\\
&=&	\qty(p_{A0}+p_{B0})R + \qty(\frac{m_Ap_{B0}-m_Bp_{A0}}{m_A+m_B}) r	\nonumber	\\
&=& p_{M0}R + p_{\mu 0}r ,
\end{eqnarray}
where $p_{M0}$ and $p_{\mu 0}$ correspond to the total initial momenta for the COM and the reduced mass, respectively.
At last, in the normalisation constant we can put
\begin{eqnarray}
\sigma_A^2 \sigma_B^2 	&=&	\frac{\hbar}{2m_A\omega_0} \times \frac{\hbar}{2m_B\omega_0}
\nonumber	\\
&=&		\frac{\hbar}{2\omega_0} \times \frac{m_A+m_B}{m_A m_B} \times \frac{1}{m_A+m_B} \times \frac{\hbar}{2\omega_0} 
\nonumber	\\
&=&		\frac{\hbar}{2 M \omega_0} \times \frac{\hbar}{2 \mu \omega_0}
\nonumber	\\
&\equiv&		\sigma_M^2 \sigma_\mu^2 .
\end{eqnarray}
With all these transformations, the initial wave function nicely separates as
\begin{eqnarray}
\Psi(t=0) &=& \qty( \frac{1}{2\pi\sigma_M^2} )^{1/4} \qty( \frac{1}{2\pi\sigma_\mu^2} )^{1/4} \exp( -\frac{R^2}{\sigma_M^2} - \frac{r^2}{\sigma_\mu^2} ) \exp(i\frac{p_{M0}R + p_{\mu 0}r}{\hbar})	\nonumber	\\
&=& \qty( \frac{1}{2\pi\sigma_M^2} )^{1/4} \exp(-\frac{R^2}{4\sigma_M^2} + i \frac{p_{M0}}{\hbar}R) \ \qty( \frac{1}{4\pi\sigma_\mu^2} )^{1/4} \exp(-\frac{r^2}{4\sigma_\mu^2} + i\frac{p_{\mu 0}}{\hbar}r )	\nonumber	\\
&=& \phi(R,t=0) \ \psi(r,t=0) ,
\label{eq:InitialState_inCOMframe_generalised}
\end{eqnarray}
where $\phi(R,t=0)$ and $\psi(r,t=0)$ describe the initial states for the COM and the reduced mass, respectively:
\begin{eqnarray}
\phi(R,t=0) &=& \qty( \frac{1}{2\pi\sigma_M^2} )^{1/4} \exp(-\frac{R^2}{4\sigma_M^2} + i \frac{p_{M0}}{\hbar}R) ,
\label{eq:RMwavefunction_t0_general}
\\
\psi(r,t=0) &=& \qty( \frac{1}{2\pi\sigma_\mu^2} )^{1/4} \exp(-\frac{r^2}{4\sigma_\mu^2} + i\frac{p_{\mu 0}}{\hbar}r ) .
\label{eq:COMwavefunction_t0_general}
\end{eqnarray}

The COM wave packet admits a width of $\sigma_M = \sqrt{\hbar/2M\omega_0}$, and the reduced mass wave packet has a width of $\sigma_\mu = \sqrt{\hbar/2\mu\omega_0}$. For two identical masses $m$ prepared in Gaussians of width $\sigma$, the COM would have a smaller width of $\sigma/\sqrt{2}$, and the reduced mass will have a larger width of $\sigma\sqrt{2}$. The corresponding relations are illustrated in Fig.~\ref{fig:TMGS_genreal}. A separable Hamiltonian implies that the two-body wave function retains its product form at all times, i.e., $\Psi(x_A,x_B,t) = \phi(R,t) \ \psi(r,t)$. The condition for the separability of a two-mode Gaussian state is also implied from statistical principles as shown in Appendix~\ref{appendix:MinUncertTransf}.
Note that the time-dependence of $\phi$ is governed by Eq.~\eqref{eq:COMSE-free}, which is solvable analytically using Fourier techniques~\cite{GaussEvolFreeSpace_SMBlinder}.

\subsection{The case of optomechanically held masses}

\begin{figure}
\centering
\includegraphics[width=0.66\linewidth]{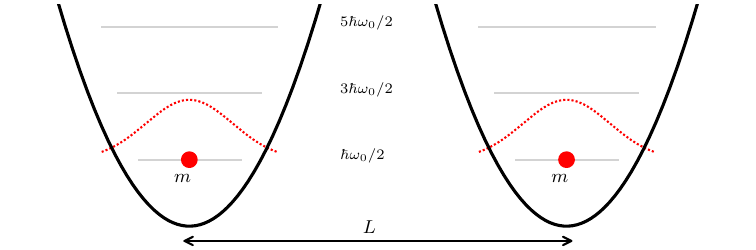}
\caption{The case of two masses $m_A$ and $m_B$ interacting with each other while levitated in the ground state of identical optomechanical harmonic traps with frequency $\omega_0$.}
\label{fig:system-optomech}
\end{figure}

In most of the problems we will discuss the harmonic traps are opened after the particles have been cooled into their ground states. We now highlight the salient differences in quantum evolution when the traps are not opened. In such a case, the Hamiltonian describing two quantum particles interacting via a central potential is given by
\begin{equation}
\hat H = -\frac{\hbar^2}{2m_A} \pdv[2]{x_A} + \frac{1}{2}m_A\omega_0^2\hat x_A^2 -\frac{\hbar^2}{2m_B} \pdv[2]{x_B} + \frac{1}{2}m_B\omega_0^2\hat x_B^2 + V(\hat x_B-\hat x_A).
\end{equation}
We can use inverse coordinate transformations to prove that
\begin{eqnarray}
\frac{1}{2}m_A\omega_0^2x_A^2 &+& \frac{1}{2}m_B\omega_0^2x_B^2 
\nonumber	\\
&=&  \frac{1}{2}m_A\omega_0^2 \qty(R - \frac{m_B}{M}r)^2  +   \frac{1}{2}m_B\omega_0^2 \qty(R + \frac{m_A}{M}r)^2
\nonumber	\\
&=&		\frac{1}{2}m_A\omega_0^2 \qty( R^2 + \frac{m_B^2}{M^2}r^2 - 2\frac{m_B}{M}Rr )
	+ \frac{1}{2}m_B\omega_0^2 \qty( R^2 + \frac{m_A^2}{M^2}r^2 + 2\frac{m_A}{M}Rr )
\nonumber	\\
&=& 	\frac{1}{2} (m_A+m_B) \omega_0^2 R^2 + \frac{1}{2} \qty(\frac{m_Am_B}{M}) \omega_0^2 r^2	\nonumber	\\
&=& 	\frac{1}{2} M \omega_0^2 R^2 + \frac{1}{2} \mu \omega_0^2 r^2.
\end{eqnarray}
Similar to the previous section, the bipartite TDSE now decouples into
\begin{eqnarray}
	\qty(	-\frac{\hbar^2}{2M} \pdv[2]{R} + \frac{1}{2} M \omega_0^2 \hat R^2 ) \phi(R,t)  &=&  i\hbar\pdv{t}\phi(R,t),	
\label{eq:COMSE-trap}
\\
	\qty( -\frac{\hbar^2}{2\mu} \pdv[2]{r} + \frac{1}{2} \mu \omega_0^2 \hat r^2 +  V(\hat r) ) \psi(r,t) &=& i\hbar\pdv{t}\psi(r,t).
\end{eqnarray}
Note that here the COM is trapped in a virtual harmonic potential with a trap frequency equal to that for the two particles.
Also, since the initial state of the COM [Eq.~\eqref{eq:RMwavefunction_t0_general}] is the ground state of this Hamiltonian [\eqref{eq:COMSE-trap}], the COM wave function does not undergo any time evolution whatsoever.

\section{Summary}

We demonstrated the utility of Cayley's form of evolution operator in the numerical resolution of continuous-variable quantum dynamics. The highly accurate five-point stencil was utilized to discretise the problem as an implicit-explicit pentadiagonal Crank-Nicolson scheme, which is unconditionally stable on realistic time scales. Given the same grid size and time step, the resultant numerical solutions achieve much higher accuracy than the standard ones.
We also discussed the coordinate transformations to the COM frame of reference and the situations when the bipartite TDSE decouples into two single-particle TDSEs. For a two-mode Gaussian state, this happens only when the two particles are prepared in the ground state of identical harmonic traps. The theory works for arbitrary central interactions, and for multiple central forces acting at the same time.

%% file: science/ch3-Rutherford.tex
\chapter{Head-On Collision in Rutherford Experiment}
\label{ch:chapter3}

The unavoidable existence of a finite momentum variance implies that quantum mechanical wave packets cannot be stopped completely. 
Therefore, the situations where classical particles stop, like head-on collisions, are natural candidates to probe the emergence of nonclassicality. 
We demonstrate this phenomenon in the paradigmatic Rutherford experiment. 
Taking a leap over the traditional practice of assuming quantum mechanical plane waves, we treat the projectiles as localised wave packets and study their head-on collisions with the stationary target nuclei. 
We simulate the quantum dynamics of this one-dimensional system and study deviations of the average quantum solution from the classical one. 
These deviations are traced back to the convexity properties of Coulomb potential. Finally, we sketch how these theoretical findings could be tested in experiments looking for the onset of nuclear reactions.

\section{Introduction}

The seminal theoretical discussion of scattering angles in the Rutherford experiment is based on the Coulomb interaction between point charges modeling the alpha projectiles and the stationary gold nuclei~\cite{RutherfordExp_Students,RutherfordExp_Prof}.
In the conventional quantum approach, the projectiles are described as incident plane waves, and the collision is studied in asymptotic limits under suitable approximations~\cite{book_CJJoachain}. A fuller approach that we pursue here is to compute the time dependence of the quantum evolution. With this in mind we study the simplest case in Rutherford experiment, i.e., the head-on collision.

As in the original discussion, we model the nuclei as stationary sources of the Coulomb potential, but in contradistinction we describe the alpha particles by incident Gaussian wave packets. This leads to several predictions that differ from their classical counterparts. We focus on the average quantum behavior and show that it does not recover the classical solutions. In particular, the quantum projectile does not approach the target as close as its classical counterpart, the quantum dynamics is not symmetric about the time of collision, and the expected quantum trajectories do not overlap with the classical ones even in the asymptotic limits after the collision.

We trace back these discrepancies to the Ehrenfest theorem and emphasize that the average quantum dynamics recovers classical solutions only for potentials that are at most quadratic in the position operator~\cite{Ehrenfest1927,book-QMech-BCHall,book-QMech-MaxJammer}.
We derive inequalities between average quantum and classical forces that hold for any cubic potential as well as for potentials with fixed convexity properties.
This is directly applicable to Coulomb or gravitational forces and clearly shows that average quantum dynamics are different from their classical counterparts in a plethora of physically interesting scenarios.

Finally, we briefly discuss tunneling through the Coulomb barrier to infer the distance of closest approach and its dependence on the initial spread of the incident Gaussian wave packet.
The tunneling probability is an experimentally accessible parameter as
the particles that have crossed the barrier give rise to nuclear reactions~\cite{ZP-54.656,Nature-106.14}. In addition to computing the probability in the dynamical quantum model we also show the conditions under which the Wentzel-Kramers-Brillouin (WKB) formula accurately approximates it~\cite{PR-40.621}.
In particular, we note that the latter may be orders of magnitude off for low energy projectiles even for a negligible momentum dispersion (see also Refs.~\cite{PLA-220.41,PLA-225.303,JPA-37.2423,PLA-378.1071}).

\section{Ehrenfest's dynamics and classical limit}

Before moving to the collisions we would like to present a general discussion on a comparison between the average Schr\"odinger dynamics and the classical one, especially that some textbooks give an incorrect statement that the average quantum trajectory recovers the classical motion~\cite{book_QM_Eisberg}.
This comes in relation to the Ehrenfest theorem showing that the quantum equations of motion for average position and momentum,
\begin{equation}
\dv{t}\ev{\hat x} = \frac{\ev{\hat p}}{m}, 
\hspace{1cm} 
\dv{t}\ev{\hat p} = -\ev{V'(\hat x)},
\label{eq:ehrenfesteqs}
\end{equation}
have the same general form as the Hamilton's equations of classical mechanics:
\begin{equation}
\dv{t}x = \frac{p}{m}, 
\hspace{1cm}
 \dv{t}p = -V'(x),
\label{eq:hamiltoneqs}
\end{equation}
where $x$ is the position, $p$ is the momentum, $m$ is the mass of the particle and $V'$ stands for the gradient of potential to which we will refer as (negative) force. In what follows, we only consider the one-dimensional motion.

While the general form of Eqs.~\eqref{eq:ehrenfesteqs} and~\eqref{eq:hamiltoneqs} is the same, they become identical only if the average of the force equals the force for the average position~\cite{book-QMech-BCHall}:
\begin{equation}
\ev{V'(\hat x)} = V'(\ev{\hat x}).
\label{eq:EeqH}
\end{equation}
For arbitrary wave functions this condition is satisfied only for potentials that are at most quadratic in $x$.
Already a cubic potential shows that there is a consistent difference between the quantum and the classical forces.
Namely, the derivative of the potential has a form $V'(x) = a_1 + a_2 \, x + a_3 \, x^2$.
The non-negativity of variance, $\ev{\hat x^2} \geq \ev{\hat x}^2$, then implies the inequality $\ev{V'(\hat x)} \geq V'(\ev{\hat x})$. Similar inequalities follow for derivatives of convex or concave potentials using Jensen's inequalities. Many potentials of natural interactions have well-defined convexity properties, e.g., the repulsive Coulomb potential that we study here. In this case, the wave packet's average quantum force can be very different from what a classical particle experiences in the same potential. This leads to consistent differences between the average quantum trajectory and the classical one.

\section{Comparison of classical and quantum trajectories}
\label{sec:formalism}

For comparison, we first show how to compute the classical trajectories.
Consider a projectile, initially at $x = -L$ (negative), is shot towards the target, located at $x = 0$.
We follow the tradition that projectiles are propagating from left to right.
If the particle's initial kinetic energy is $T_0$, the law of conservation of energy implies $T(x) + V(x) = T_0 + V(L) \equiv E_0$,
with the Coulomb potential,
\begin{equation}
V(x) = Z_{P}Z_{T} \alpha \hbar c / | x | \equiv \frac{d_\text{cl}}{|x|}E_0,
\label{CoulombPotential}
\end{equation}
where $Z_{P}$ and $Z_{T}$ are the atomic numbers of the projectile and the target, respectively, $\alpha$ is the fine structure constant, and $d_{\mathrm{cl}} = Z_P Z_T \alpha \hbar c / E_0$ is the classical distance of closest approach. Accordingly, the classical equation of motion can be derived as
\begin{eqnarray}
&& T(x) + V(x) = E_0
\nonumber	\\
\implies &&	\frac{1}{2}m\qty(\dv{x}{t})^2 + \frac{d_\text{cl}}{|x|}E_0 = E_0
\nonumber	\\
\implies && \dv{x}{t} = \pm \sqrt{\frac{2E_0}{m}} \sqrt{1-\frac{d_\text{cl}}{|x|}}
\nonumber	\\
\implies &&	\dv{x}{t} = \pm \sqrt{\frac{2E_0}{m}} \sqrt{1+\frac{d_\text{cl}}{x}},	\hspace{1cm} : \qty{ x < 0 }.
\label{cleom}
\end{eqnarray}
 The velocity is positive (with `+' sign) for the projectile approaching the target, and negative (`-') when it is reflected back.
The resultant trajectories are integrated as follows.
\begin{itemize}
\item For the projectile approaching the target we have
	\begin{eqnarray}
	&&  \dv{x}{t} = \sqrt{\frac{2E_0}{m}} \sqrt{1+\frac{d_\text{cl}}{x}} 
\nonumber	\\
\implies	&& 	\int_{-L}^{x} \frac{dx'}{\sqrt{1+\frac{d_\text{cl}}{x'}}} = \sqrt{\frac{2E_0}{m}} \int_{0}^{t} dt'
\nonumber	\\
\implies	&&  	d_\text{cl} \int_{-L/d_\text{cl}}^{x/d_\text{cl}} \frac{du}{\sqrt{1+\frac{1}{u}}} = \sqrt{\frac{2E_0}{m}} t	,	\hspace{1cm} : \qty{ x'/d_\text{cl} = u },
\nonumber	\\
\implies	&&  d_\text{cl} \qty[ u\sqrt{1+\frac{1}{u}} -  \frac{1}{2} \ln \qty( \frac{ 1 + \sqrt{1+\frac{1}{u}} }{ \left|1-\sqrt{1+\frac{1}{u}}\right| } ) ]_{-L/d_\text{cl}}^{x/d_\text{cl}} = \sqrt{\frac{2E_0}{m}} t	
\nonumber	\\
\implies	&&  d_\text{cl} \qty[ u\sqrt{1+\frac{1}{u}} -  \frac{1}{2} \ln \qty( \frac{ 1 + \sqrt{1+\frac{1}{u}} }{ 1-\sqrt{1+\frac{1}{u}} } ) ]_{-L/d_\text{cl}}^{x/d_\text{cl}} = \sqrt{\frac{2E_0}{m}} t	,	\hspace{1cm} : \qty{ u < -1 },	
\nonumber	\\
\implies	&&  d_\text{cl} \Bigg[ \frac{x}{d_\text{cl}} \sqrt{1+\frac{d_\text{cl}}{x}}	-   \frac{1}{2} \ln \qty( \frac{ 1 + \sqrt{1+\frac{d_\text{cl}}{x}} }{ 1-\sqrt{1+\frac{d_\text{cl}}{x}} } ) 
+ \frac{L}{d_\text{cl}} \sqrt{1-\frac{d_\text{cl}}{L}} 
\nonumber   \\ 
&& \hspace{5cm}   + \frac{1}{2} \ln \qty( \frac{ 1 + \sqrt{1-\frac{d_\text{cl}}{L}} }{ 1-\sqrt{1-\frac{d_\text{cl}}{L}} } ) \Bigg] = \sqrt{\frac{2E_0}{m}} t	
\nonumber	\\
\implies	&&   x \sqrt{1+\frac{d_\text{cl}}{x}}	-   \frac{d_\text{cl}}{2} \ln \qty( \frac{ 1 + \sqrt{1+\frac{d_\text{cl}}{x}} }{ 1-\sqrt{1+\frac{d_\text{cl}}{x}} } ) + L \sqrt{1-\frac{d_\text{cl}}{L}} 
\nonumber   \\ 
&& \hspace{5cm} + \frac{d_\text{cl}}{2} \ln \qty( \frac{ 1 + \sqrt{1-\frac{d_\text{cl}}{L}} }{ 1-\sqrt{1-\frac{d_\text{cl}}{L}} } )  = \sqrt{\frac{2E_0}{m}} t,	
\end{eqnarray}
At $t = \tau_\text{cl}$ (time of collision), $x = -d_\text{cl}$, which implies
\begin{equation}
\tau_\text{cl} = \sqrt{\frac{m}{2E_0}} \qty[  L \sqrt{1-\frac{d_\text{cl}}{L}}  + \frac{d_\text{cl}}{2} \ln \qty( \frac{ 1 + \sqrt{1-\frac{d_\text{cl}}{L}} }{ 1-\sqrt{1-\frac{d_\text{cl}}{L}} } )  ].
\end{equation}

\item For the projectile reflected back, the equation is to be integrated from the time of the classical collision:
\begin{eqnarray}
	&&  \dv{x}{t} = -\sqrt{\frac{2E_0}{m}} \sqrt{1+\frac{d_\text{cl}}{x}}
\nonumber	\\
\implies	&&   \int_{-d_\text{cl}}^{x} \frac{dx}{\sqrt{1+\frac{d_\text{cl}}{x}}} = -\sqrt{\frac{2E_0}{m}} \int_{\tau_{cl}}^{t} dt'
\nonumber	\\
\implies	&&  d_\text{cl} \qty[ u\sqrt{1+\frac{1}{u}} -  \frac{1}{2} \ln \qty( \frac{ 1 + \sqrt{1+\frac{1}{u}} }{ \abs{1-\sqrt{1+\frac{1}{u}}} } ) ]_{-1}^{x/d_\text{cl}} = -\sqrt{\frac{2E_0}{m}} \qty(t - \tau_\text{cl} )	
\nonumber	\\
\implies	&&   d_\text{cl} \qty [ \frac{x}{d_\text{cl}} \sqrt{1+\frac{d_\text{cl}}{x}}	-   \frac{1}{2} \ln \qty( \frac{ 1 + \sqrt{1+\frac{d_\text{cl}}{x}} }{ 1-\sqrt{1+\frac{d_\text{cl}}{x}} } ) ] = -\sqrt{\frac{2E_0}{m}} \qty(t - \tau_\text{cl} )	
\nonumber	\\
\implies	&&  x \sqrt{1+\frac{d_\text{cl}}{x}}	-   \frac{d_\text{cl}}{2} \ln \qty( \frac{ 1 + \sqrt{1+\frac{d_\text{cl}}{x}} }{ 1-\sqrt{1+\frac{d_\text{cl}}{x}} } ) = -\sqrt{\frac{2E_0}{m}} \qty(t - \tau_\text{cl} ).
\end{eqnarray}
\end{itemize}
The above trajectories can be concisely written as
\begin{equation}
x\sqrt{1+\frac{d_\text{cl}}{x}} - \frac{d_\mathrm{cl}}{2} \ln \left( \frac{1+\sqrt{1+\frac{d_\text{cl}}{x}}}{1-\sqrt{1+\frac{d_\text{cl}}{x}}} \right)  + \sqrt{\frac{2E_0}{m}}(t-\tau_{\mathrm{cl}}) \ \text{sign}(t-\tau_{\mathrm{cl}}) = 0.
\label{xcl}
\end{equation}
Since this is a transcendental equation, we solve for $x(t)$ numerically. The momentum and force are thereafter calculated using Eqs.~(\ref{cleom}) and (\ref{CoulombPotential}), respectively.

\begin{figure}
\centering
\includegraphics[width=0.55\linewidth]{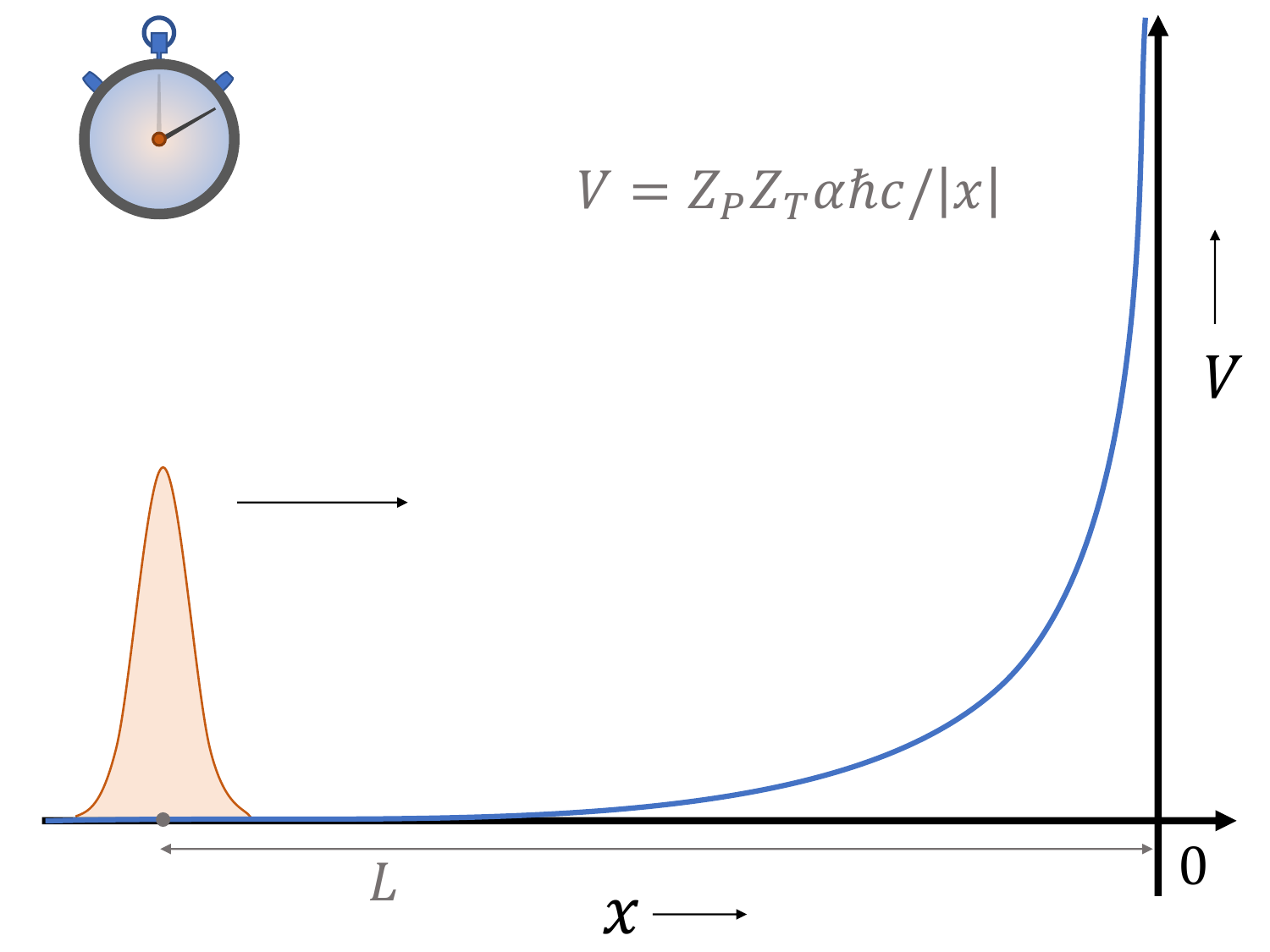}
\caption{Schematic of the experiment. The projectiles are prepared as Gaussian wave packets, which evolve in the Coulomb potential of the target nucleus fixed at the origin. The symbols have their usual meanings and are defined explicitly in Sec.~\ref{sec:formalism}.}
\label{FIG_SCHEME}
\end{figure}

The quantum setup is shown schematically in Fig.~\ref{FIG_SCHEME}.
We obtain the average quantum dynamics of the projectile with the Cayley's operator as discussed in Chapter~\ref{ch:chapter2}. Since the wave function is expanding continuously during the collision event, the simulation becomes numerically demanding as time progresses. To circumvent this we have developed a method to dynamically allocate the size of discretization of space, details of which are given in Section~\ref{appendix:DynamicAllocation}. Finally, in the classical picture the projectile particles are modelled as point charges shot towards the target from a distance $L$ with a kinetic energy $T_0$. In our quantum simulations, we model the projectiles with Gaussian wave packets centered at the same distance $L$, with a width $\sigma$ and an average momentum $p_0=\sqrt{2mT_0}$. Therefore, the wave function describing the quantum state of the projectile at $t=0$ is given by
\begin{equation}
\psi(x,0) = \qty( \frac{1}{2\pi\sigma^2} )^{1/4} \exp(-\frac{(x+L)^2}{4\sigma^2} + i\frac{p_0}{\hbar}(x+L)).
\label{eq:rutherford_initialstate}
\end{equation}
We now describe several parameters that show differences between the classical and the average quantum dynamics. Following the original Rutherford experiment, we consider the collision of alpha particles with gold nuclei, and hence we set $Z_P=2$ and $Z_T=79$ throughout this chapter. Typical classical and quantum trajectories are presented in Fig.~\ref{fig:Traj_10K5MeV}.
In both cases $L$ = 10 pm and $T_0$ = 5 MeV, i.e., the typical energy of alpha particles in the original experiment.
We now systematically discuss the following quantities: distance of closest approach, the effects of a finite spread of the projectile's initial wave function, the time of collision, and the asymptotic behavior of trajectories.

\section{The quantum distance of closest approach} \label{sec:doca}

\begin{figure}[!t]
	\centering
	
	\subfloat[Classical and quantum  trajectories. The right panel shows a magnified view of the collision event. The quantum trajectories are the results from the dynamical simulations.]{\includegraphics[width=\linewidth]{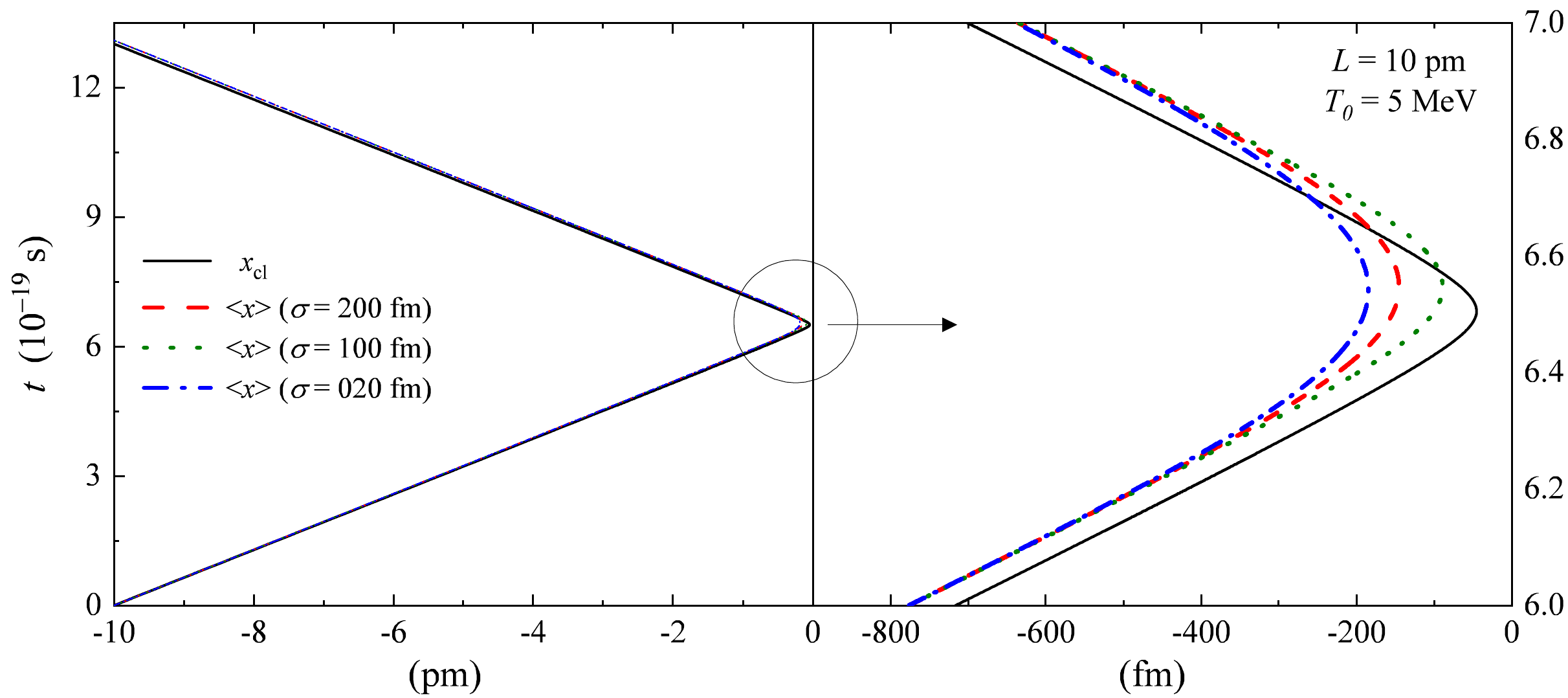}\label{fig:Traj_10K5MeV}}
	\hfill
	\subfloat[Snapshots of the wave packet at different times for the case of $\sigma = 100$ fm.]{\includegraphics[width=\linewidth]{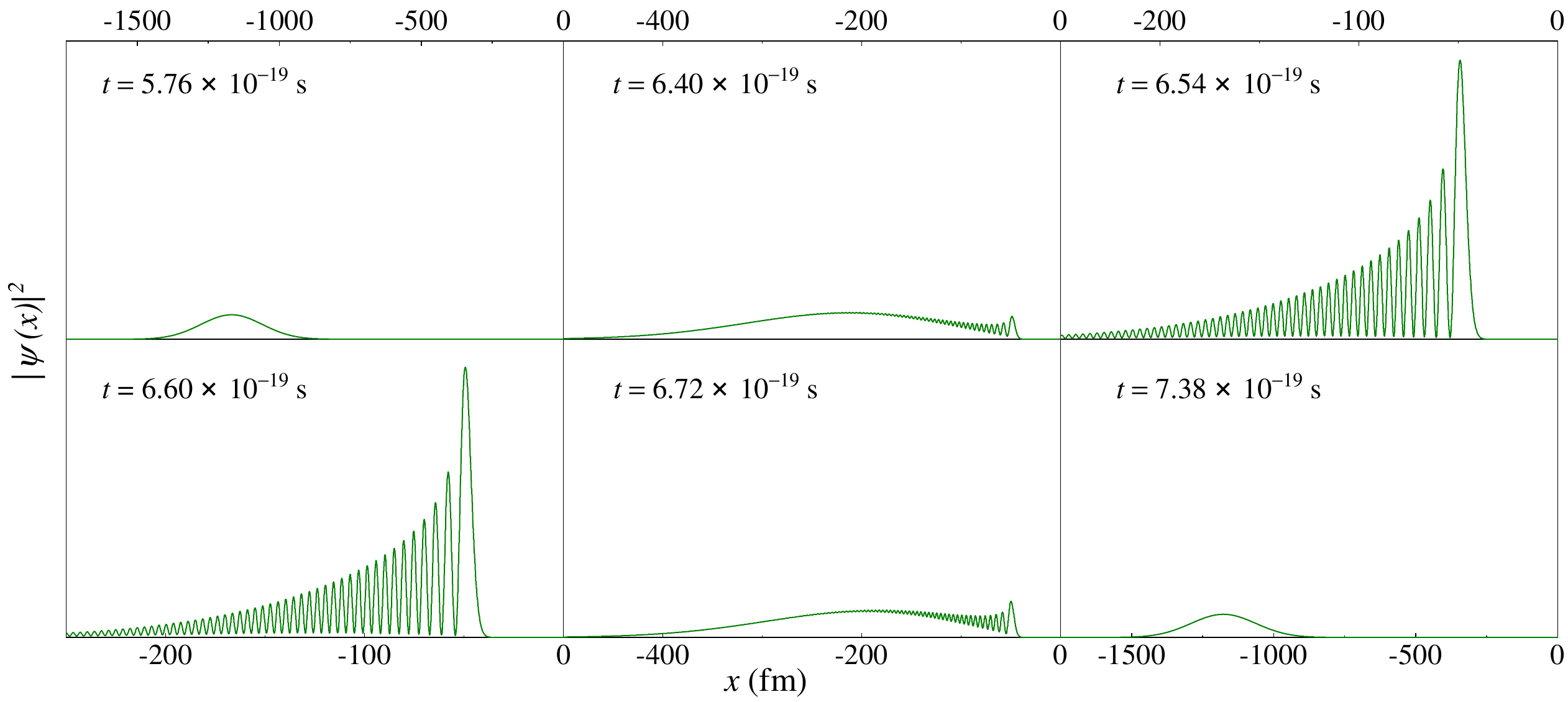}\label{fig:wavefunc_10K5MeV}}
	
\caption{Time-dependent properties of the alpha particle wave packets shot in the Coulomb potential of a gold nucleus fixed at the origin. The projectiles are shot from the distance $L = 10$ pm with the initial width of Gaussian wave packet $\sigma$ and average momentum $p_0 =\sqrt{2mT_0}$, where $T_0 = 5$ MeV, and $m$ is the mass of an alpha particle. $x_\text{cl}$ represents the classical path, and $\ev{x}$ is the expected position of the wave packet. Numerical details are given in Sec.~\ref{appendix:constants_rutherford}.}
\label{fig:10K5MeV}
\end{figure}

As seen in Fig.~\ref{fig:Traj_10K5MeV} the quantum projectiles are, on average, reflected from a bigger distance to the target. This can be intuitively understood by invoking either convexity of the Coulomb interaction or the Heisenberg uncertainty principle. In the latter case, note that a quantum particle cannot be stopped completely. There is always some momentum dispersion which leads to non-zero kinetic energy. It follows that the maximal potential energy cannot be as big as in the classical case, and hence the distance of the closest approach increases. In the former case, Jensen's inequality at time $t = 0$ allows us to write
\begin{equation}
|\ev{\hat F}| \sim \ev{\frac{1}{\hat x^2(0)}} \geq \frac{1}{\ev{\hat x(0)}^2} = \frac{1}{L^2} =  \frac{1}{x_{\mathrm{cl}}^2(0)} \sim |F_{\mathrm{cl}}|,
\label{inequality_teq0}
\end{equation}
Accordingly, the average quantum mechanical force is stronger (more repulsive) than the force experienced by the classical particle. As a result, the quantum projectile moves slower than its classical counterpart at the very beginning of its journey. We emphasize that this inequality holds at the beginning of the evolution, and at later times it may reverse (as we will show later).
Nevertheless, the quantum projectile never gets as close to the target as the classical one. In the next section we derive conditions for the quantum distance of closest approach to differ from its classical counterpart minimally.

Fig.~\ref{fig:wavefunc_10K5MeV} shows the wave function at different times for the case of $\sigma = 100$ fm. As the wave packet approaches closer to the target, the leading edge of the wave packet is reflected back. This interferes with the trailing incident part to create rapid oscillations, which eventually die out as the wave packet travels back attaining a near-Gaussian shape~\cite{EJP-20.29,AJP-66.252,AJP-35.177}. As in the case of collision with a hard wall, the qualitative similarity between $\ev{x}$ and $x_\text{cl}$ is not evident when one looks at the wave packet itself~\cite{PhyScr-71.136}.

\begin{figure}
\centering
\includegraphics[width=0.5\linewidth]{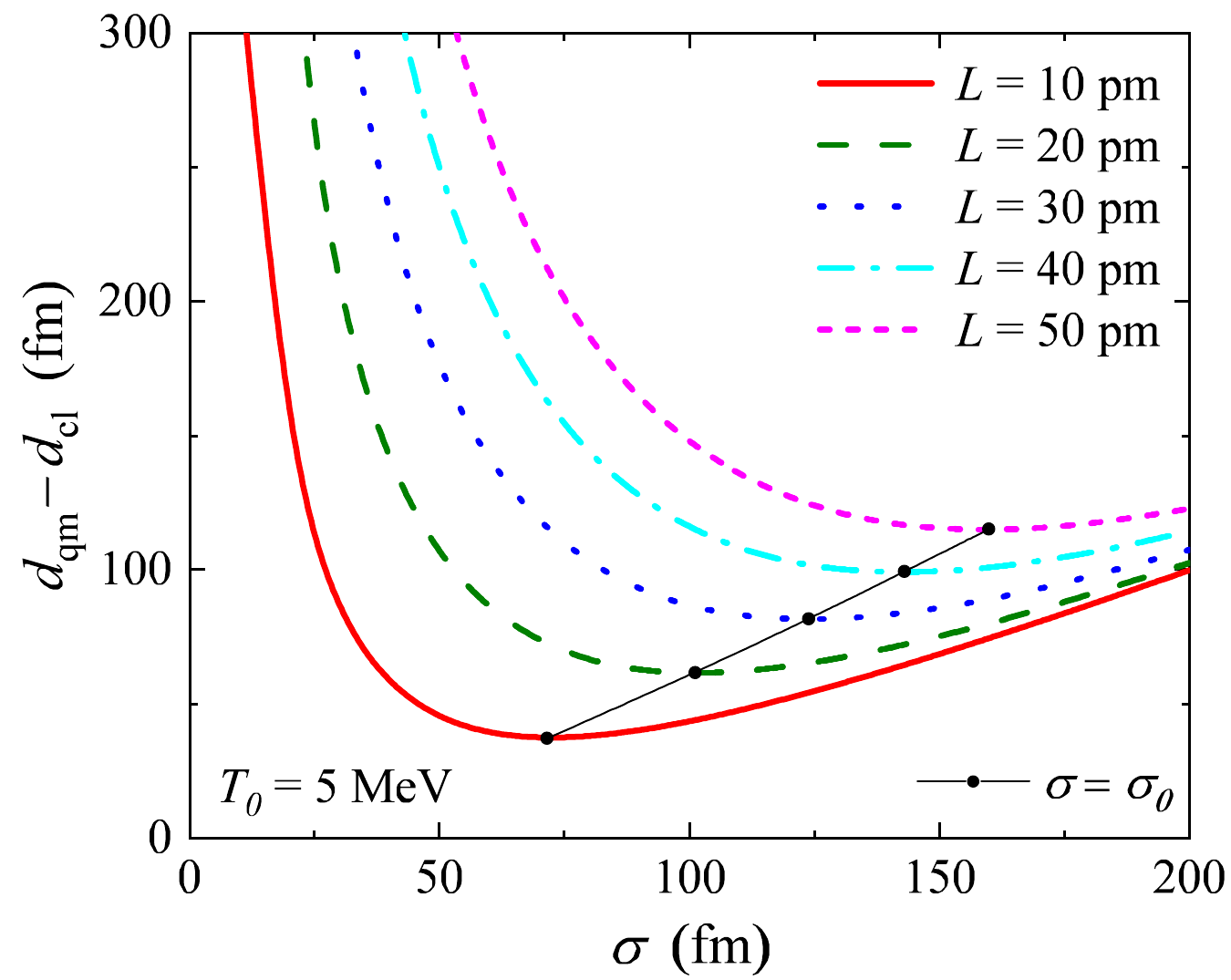}
\caption{Difference between the quantum and classical distances of closest approach. The numerically found minima of the difference are marked with dots for various launching distances. In the main text we argue that they are achieved for the initial width of the projectile given by $\sigma_0 = \sqrt{\hbar L/2p_0}$, where $p_0$ is the momentum of a classical alpha particle with kinetic energy $T_0$. Numerical details are given in Sec.~\ref{appendix:constants_rutherford}.}
\label{fig:FindingSig0_5MeV}
\end{figure}

We define the quantum distance of closest approach as the smallest average position to the target, i.e.,
\begin{equation}
d_{\mathrm{qm}} = \min(|\ev{\hat x}|).
\end{equation}
As seen in Fig.~\ref{fig:Traj_10K5MeV} this quantity depends on the initial spread of the wave function of the projectile.
We now give physical arguments which determine the optimal initial spread $\sigma_0$, for which the difference $d_{\mathrm{qm}} - d_{\mathrm{cl}}$ is the smallest. Since the projectile is launched from a large distance $L \gg  d_{\mathrm{cl}}$,
the time of collision satisfies $\tau_{\mathrm{cl}} \approx mL/p_0$.
We approximate the position spread before the collision by the value obtained for a free quantum evolution:
\begin{equation}
\sigma(\tau_{\mathrm{cl}}) = \sigma\sqrt{1+ \frac{\hbar^2 \tau_{\mathrm{cl}}^2}{4m^2\sigma^4}} = \sigma \sqrt{1+\left(\frac{\hbar L}{2 p_0}\right)^2\frac{1}{\sigma^4}}.
\label{eq:sig0}
\end{equation}
We established in Eq.~\eqref{inequality_teq0} that the quantum wave packet feels a stronger force compared to its classical counterpart. As an implication the wave packet, on average, is reflected from a farther distance, i.e., $d_\text{qm} > d_\text{cl}$. For a given potential, the quantum averages at a given time are closer to their classical values for wave functions that are more and more concentrated in space. Therefore the difference $d_\text{qm} - d_\text{cl}$ is smaller for narrow wave functions, and it is minimal when $\sigma(\tau_{\mathrm{cl}})$ is minimal. Eq.~\eqref{eq:sig0} suggests that this happens for the initial spread given by
\begin{equation}
\sigma_0 = \sqrt{\frac{\hbar L}{2p_0}},
\label{EQ_IN_SPREAD}
\end{equation}
and corresponds to the spread at the collision time $\sigma(\tau_{\mathrm{cl}}) = \sqrt{2} \, \sigma_0$.
Fig.~\ref{fig:FindingSig0_5MeV} shows that this is in an excellent agreement with the numerically obtained minimas in $d_\text{qm}$.

\begin{figure}[!t]
\centering
\includegraphics[width=\linewidth]{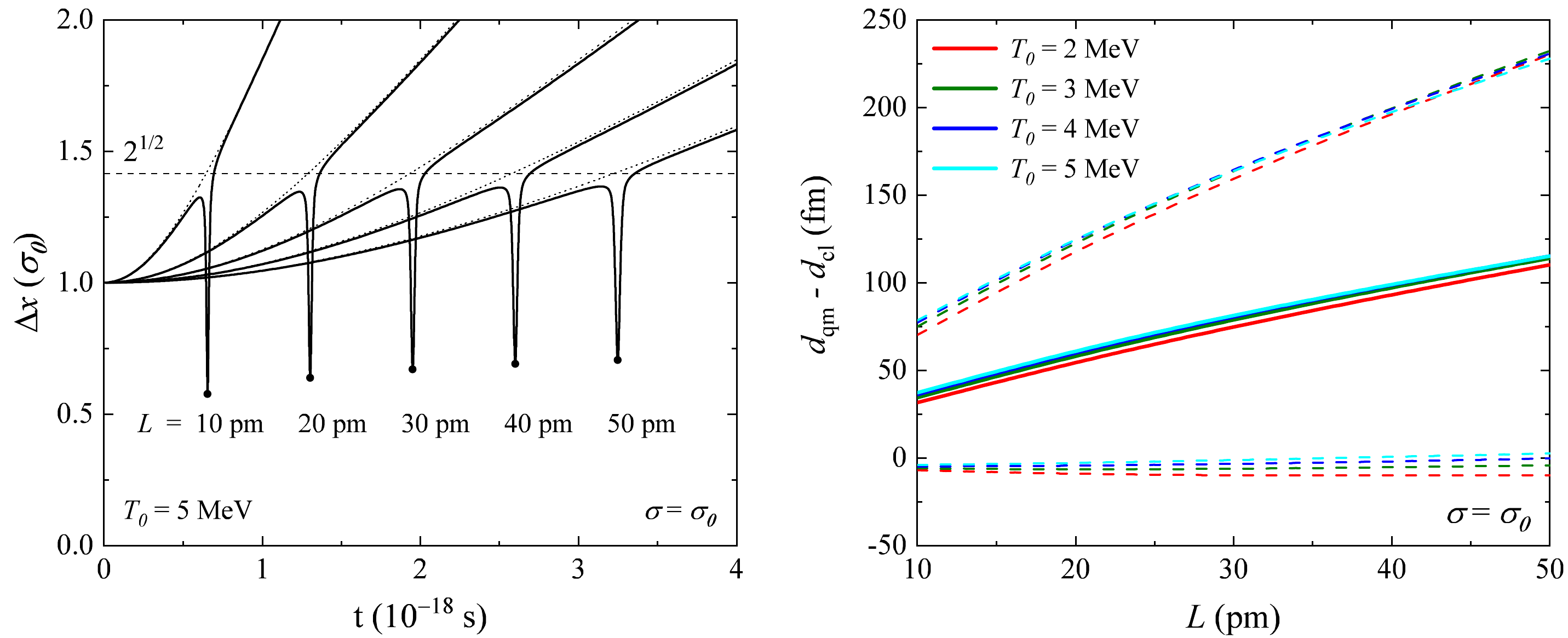}
\caption{Quantum distance of closest approach and the position spread during the collision. The initial width of wave packets is $\sigma_0 = \sqrt{\hbar L/2p_0}$, where $p_0$ is the momentum of a classical alpha particle with kinetic energy $T_0$. The left panel shows the position spreads of alpha particles with $T_0 = 5$ MeV for various launch distances $L$, with the dotted lines showing the spreads of a freely evolving particle. The black dots show the time of quantum collision, i.e., when the spread is minimum and the average position is $d_\text{qm}$. The right panel shows that the classical distance of closest approach is within a standard deviation (dashed lines) from the quantum distance of closest approach (the lower dashed line is close to zero). The energy dependence is negligible. Numerical details are given in Sec.~\ref{appendix:constants_rutherford}.}
\label{fig:qdoca}
\end{figure}

Furthermore, we establish the range accessible to the quantum distance of closest approach for the energies considered in this work. The left panel of Fig.~\ref{fig:qdoca} shows the position spread of wave packets launched from various distances.
The collisions correspond to the sudden drops of the position spread, i.e., the wave packet gets compressed when the projectile approaches the target. The distance of closest approach matches the minima in these curves, which are all smaller than the initial spread of Eq.~\eqref{EQ_IN_SPREAD}, i.e., the uncertainty in $d_\text{qm}$ is less than $\sigma_0$. Next, we show in the right panel of Fig.~\ref{fig:qdoca} that the classical distance of closest approach is within the range of one standard deviation from the quantum distance of closest approach. Finally, this translates to
\begin{equation}
d_{\mathrm{cl}} < d_{\mathrm{qm}}  < d_{\mathrm{cl}} + \sqrt{\frac{\hbar L}{2 p_0}}.
\end{equation}

\subsection{Origin of the optimal spread}

Mathematically, Eq.~\eqref{eq:EeqH} encoding the equivalence between the classical and Ehrenfest dynamics, is satisfied for any potential if the wave function is given by the Dirac delta.
The more the wave function is concentrated in space, the more similar classical and average quantum positions are.
Physically, however, due to the Heisenberg uncertainty principle, there is no evolution which preserves the Dirac delta, and we show that this leads to the optimal initial spread.
For the values presented in Fig.~\ref{fig:Traj_10K5MeV}, the optimal spread is about $50$ fm (see green dotted line).
If the initial spread is larger (compare with the red dashed line), the wave function spreads further and the deviation from the classical trajectory is higher. But for smaller initial spreads (compare with the blue dashed-dotted line), the free quantum evolution dominating initially predicts faster spreading due to large initial momentum dispersion.  Therefore, the standard deviation becomes larger by the time the wave packet reaches close to the target, and the deviation from the classical trajectory is again higher.

\subsection{Time asymmetry of quantum collision}

As already seen in Fig.~\ref{fig:Traj_10K5MeV}, the classical and quantum collisions happen at different times. The classical time $\tau_{\mathrm{cl}}$ is obtained by requiring vanishing momentum in Hamilton's solution.
The quantum collision time $\tau_{\mathrm{qm}}$ corresponds to vanishing average momentum in Ehrenfest's solution.
This time matches with the minima in the position spread shown in Fig.~\ref{fig:qdoca}.

\begin{figure}[!t]
\centering
\includegraphics[width=\linewidth]{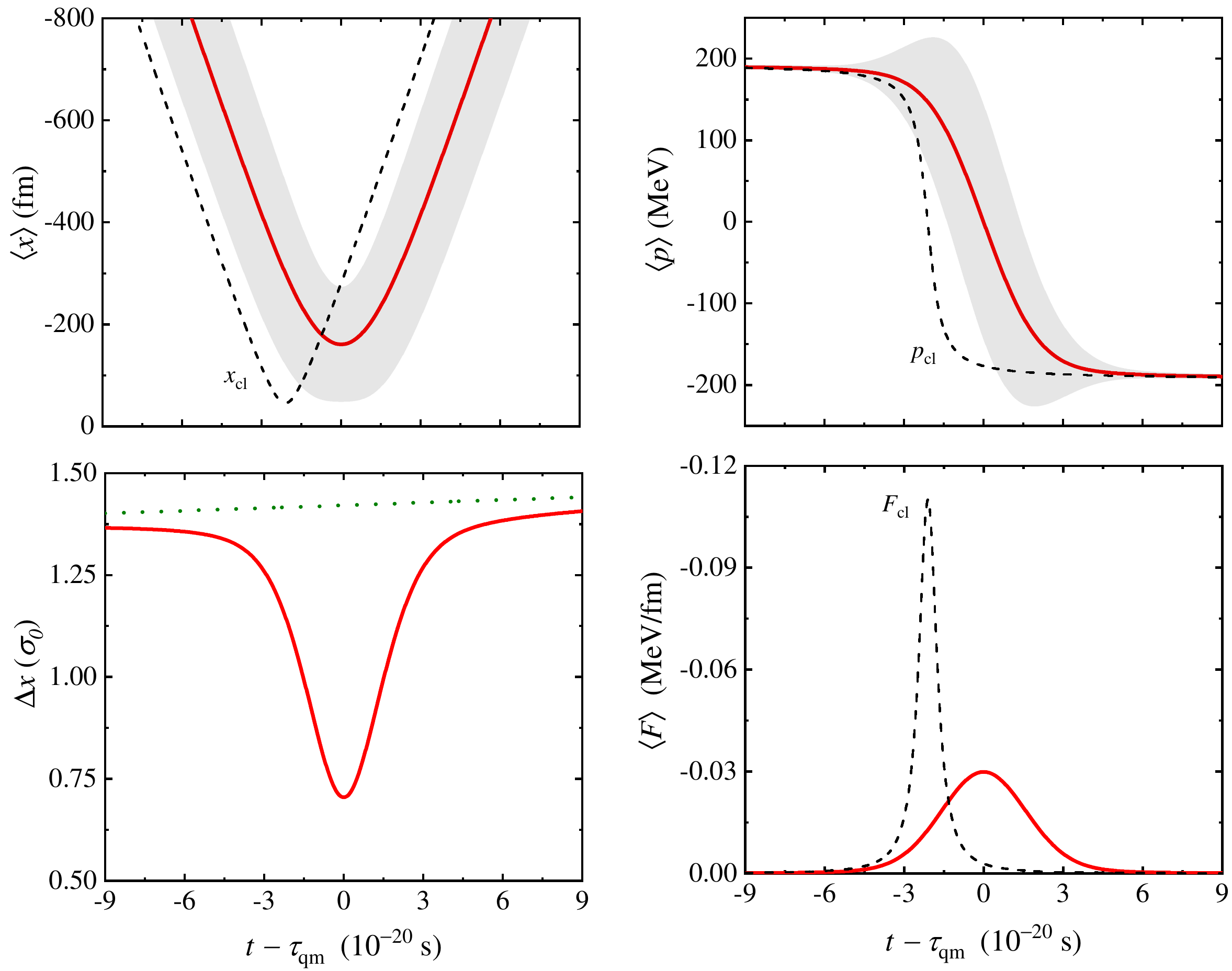}
\caption{Expected quantum properties (solid red lines) of an alpha particle about the quantum collision time. The dashed black lines present classical predictions. The dotted green line in the lower-left panel gives position spread for a freely evolving alpha particle wave packet (as if there was no Coulomb interaction). The gray shaded regions mark one standard deviation. The projectiles are launched from a distance $L = 50$ pm. The initial width of the wave packet is $\sigma_0 = \sqrt{\hbar L/2p_0}$, where $p_0$ is the momentum of a classical alpha particle with kinetic energy $T_0 = 5$ MeV. Numerical details are given in Sec.~\ref{appendix:constants_rutherford}.}
\label{fig:MicroProp_50K5MeV}
\end{figure}

A distinguishing feature of the quantum trajectory is its asymmetry with respect to the collision time.
Recall that the classical trajectory is symmetric in time, i.e., $x_\text{cl}(\tau_{\mathrm{cl}} - \bm{\Delta} t) = x_\text{cl}(\tau_{\mathrm{cl}} + \bm{\Delta} t)$, and similarly for momentum $p_\text{cl}(\tau_{\mathrm{cl}} - \bm{\Delta} t) = - p_\text{cl}(\tau_{\mathrm{cl}} + \bm{\Delta} t )$. In particular, the projectile returns back to its original launch distance exactly at time $T = 2 \, \tau_{\mathrm{cl}}$. Our calculations show that the quantum projectile makes a collision at a different time $\tau_{\mathrm{qm}} (> \tau_{\mathrm{cl}})$, and the evolution for $t > \tau_{\mathrm{qm}}$ is not a mirror image of evolution for $t < \tau_{\mathrm{qm}}$. The return journey of a quantum particle takes a longer time than the onward one, and the wave packet returns at its launch position at time $T > 2 \, \tau_{\mathrm{qm}} (> 2 \, \tau_{\mathrm{cl}})$. 
For the case in Fig.~\ref{fig:MicroProp_50K5MeV} ($\sigma = 159.85$ fm), the time taken for the return journey is longer by $\approx 1.3 \times 10^{-22}$ s. Such effects are more prominent for wave functions with a larger momentum variance, e.g., the time difference is $\approx 1.1 \times 10^{-21}$ s for the case of $\sigma = 20$ fm in Fig.~\ref{fig:10K5MeV}.

The discussed asymmetry in quantum trajectories is yet another manifestation of the convexity in Coulomb potential. Consider a point at distance $X$ from the target. Irrespective of whether it is travelling towards or away from the target, a classical particle feels the exact same force when it passes through that point, i.e., $-V'(X)$. Accordingly, the Hamilton's solution of Eq.~\eqref{eq:hamiltoneqs} has to be symmetric about $\tau_{\mathrm{cl}}$. The quantum case is much more interesting; Figs.~\ref{fig:qdoca} and~\ref{fig:MicroProp_50K5MeV} show that the position spread immediately after the collision is larger than what it was before. This means that the width of the wave packet, when it passes through point $X$, is larger during its return journey. As a consequence, the average force experienced by the quantum projectile at that point, i.e. $-\ev{V'(X)}$, is different in the onward and the return journeys. Accordingly the Ehrenfest's solution of Eq.~\eqref{eq:ehrenfesteqs} has to be asymmetric about $\tau_{\mathrm{qm}}$. Finally, the classical trajectory $x_{\mathrm{cl}}$ and $p_{\mathrm{cl}}$ is not even statistically within the quantum prediction, i.e. it is outside $\ev{\hat x} \pm \bm{\Delta} x$ and $\ev{\hat p} \pm \bm{\Delta} p$ (see Fig.~\ref{fig:MicroProp_50K5MeV}).

We started our discussion by showing that initially the quantum projectile is repelled more than the corresponding classical counterpart.
For example, alpha particle of $T_0$ = 5 MeV launched from $L$ = 50 pm feels an average force $\ev{\hat F} \approx 1.000031 F_{\mathrm{cl}}$ at $t=0$.
Since for later times $\ev{x} < x_\text{cl}$ (negative, see Fig.~\ref{fig:Traj_10K5MeV}) we obtain
\begin{equation}
|\ev{\hat F}| \sim \ev{\frac{1}{\hat x^2(0)}} \geq \frac{1}{\ev{\hat x(t)}^2} <  \frac{1}{x_{\mathrm{cl}}^2(t)} \sim |F_{\mathrm{cl}}|.
\label{inequality_tgt0}
\end{equation}
The second inequality starts dominating closer to the target, and during the collision the classical force far exceeds the quantum one as seen in Fig.~\ref{fig:MicroProp_50K5MeV}.

\section{Quantum tunneling and the WKB limit}

\begin{figure}
\centering
\includegraphics[width=0.55\linewidth]{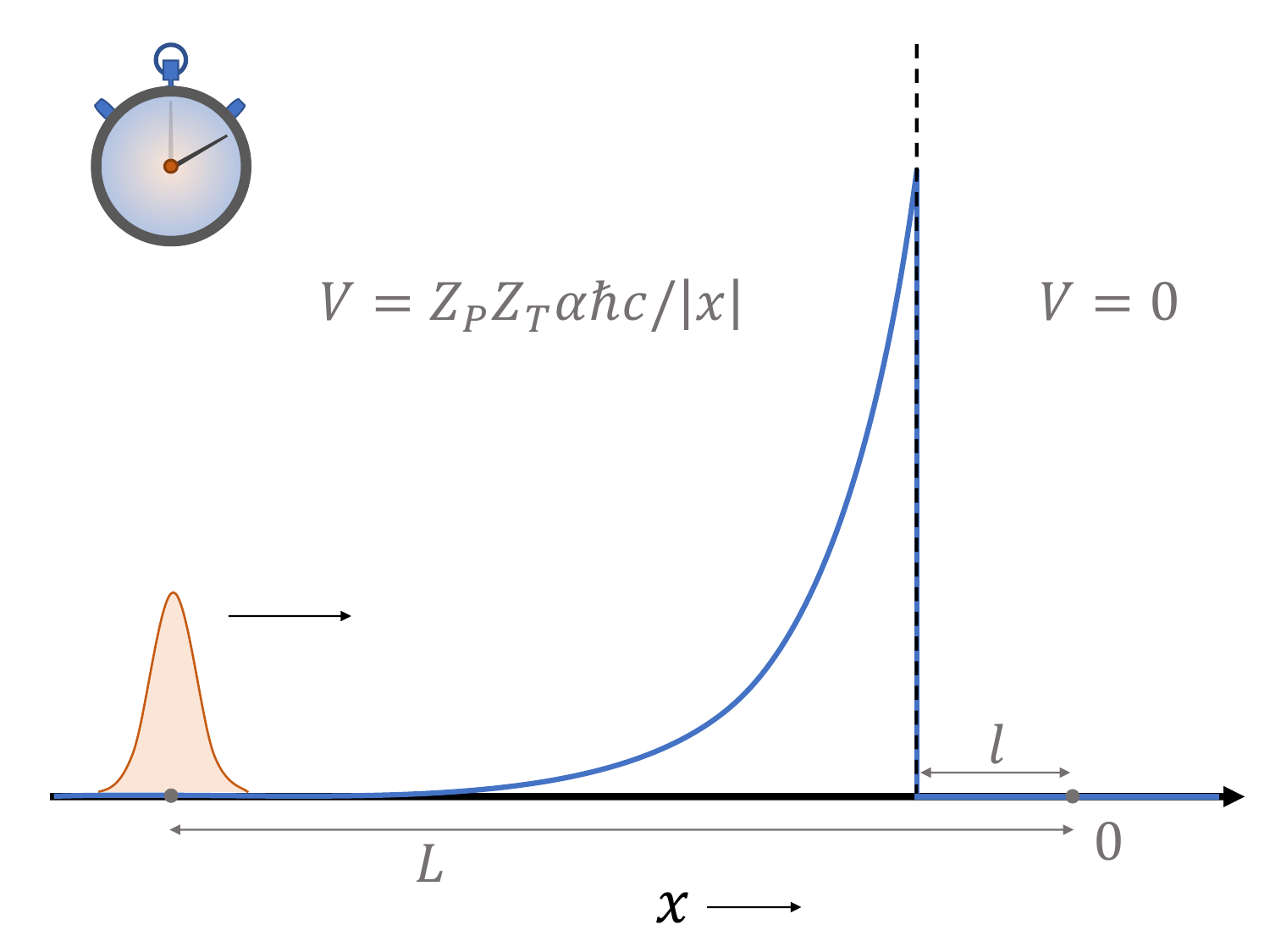}
\caption{Tunneling through the Coulomb barrier. The barrier is plotted in blue. The projectiles are prepared as Gaussian wave packets and shot from the left side. The size of the well, $l = 25$ fm, is the typical distance just outside the range of strong interaction.}
\label{FIG_BARRIER}
\end{figure}

Having established the differences between the classical and average quantum dynamics, we now move to a possibility of their experimental verification. The basic idea is to look for the onset of nuclear reactions. We model the potential in the vicinity of the nucleus by the Coulomb barrier given in Fig.~\ref{FIG_BARRIER} and defined by
\begin{equation}
V(x) =
\begin{cases}
Z_{P}Z_{T}\alpha \hbar c/|x|, \hspace{5mm} x \leq l, \\
0, \hspace{27mm} x > l.
\end{cases}
\label{barrier}
\end{equation}
 Under the classical model of the collision, if the projectile can cross the barrier it can be observed as a reaction. In such a case, the classical distance of closest approach is smaller than the size where the barrier is truncated. We take as a well-established fact that alpha particles cannot be prepared with only one momentum value and repeated momentum measurement on the projectile before the collision returns a normal distribution.
For this reason, within the classical model, we deal with an ensemble of projectiles with Gaussian momentum probability density. This is calculated by taking a Fourier transform of the initial wave function in Eq.~\eqref{eq:rutherford_initialstate}:
\begin{eqnarray}
\abs{\phi(p,0)}^2 &=& \abs{ \frac{1}{\sqrt{2\pi}} \int_{-\infty}^{+\infty} dx \ \psi(x,0)  \  \exp\qty( -i\frac{p}{\hbar}x ) }^2		\nonumber \\
 &=&  \frac{1}{(2\pi)^{3/2}\sigma} \ \abs{  \int_{-\infty}^{+\infty} dx \ \exp(-\frac{(x+L)^2}{4\sigma^2} + i\frac{p_0}{\hbar}(x+L))  \  \exp\qty( -i\frac{p}{\hbar}x ) }^2		\nonumber \\ 
&=&	 \frac{\sigma\sqrt{2}}{\sqrt{\pi}\hbar} \ \exp\qty( -2\frac{\sigma^2}{\hbar^2}(p-p_0)^2  ).
\end{eqnarray}
Only those particles will cross the barrier which are launched with an energy larger than the Coulomb barrier, i.e., $p^2/2m + V(L) > V(l)$. This gives the limiting (minimum) momentum required to cross the barrier as
\begin{equation}
p_\text{lim} = \sqrt{2m\qty[V(l)-V(L)]}  =  \sqrt{ 2 m  E_0 \qty( \frac{1}{l} - \frac{1}{L} ) d_\text{cl} } .
\end{equation}
Accordingly, the classical crossing probability $P_{\mathrm{cl}}$ is given by
\begin{eqnarray}
P_\text{cl} = \int_{p_\text{lim}}^{+\infty} dp \ \abs{\phi(p,0)}^2
&=&	 \frac{\sigma}{\hbar}\sqrt{\frac{2}{\pi}} \int_{p_\text{lim}}^{+\infty} dp \ \exp\qty( -2\frac{\sigma^2}{\hbar^2}(p-p_0)^2  )		\nonumber	\\
&=&		\begin{cases}
\frac{1}{2}\qty[ 1 - \erf\qty( \frac{\hbar}{\sigma\sqrt{2}} (p_\text{lim}-p_0) )  ]	,	\hspace{5mm} p_\text{lim} > p_0,	\\
\vspace{1mm}
\frac{1}{2}\qty[ 1 + \erf\qty( \frac{\hbar}{\sigma\sqrt{2}} (p_0-p_\text{lim}) )  ]	,	\hspace{5mm} p_\text{lim} < p_0,	
\end{cases}
\end{eqnarray}
where erf is the error function. In a concise form,
\begin{equation}
P_\text{cl} = \frac{1}{2}\qty[ 1 - \text{sign}(p_\text{lim}-p_0) \erf\qty( \frac{\hbar}{\sigma\sqrt{2}} \abs{p_\text{lim}-p_0}  )  ].
\end{equation}
This quantity is shown by the bold dashed line in the left panel of Fig.~\ref{fig:TunnProb_50K}. Conversely, from the measured probability of nuclear reaction, one can estimate the variance in position and momentum of the classical ensemble and compute the mean distance of closest approach.

Within the quantum model, the initial randomness is encoded by the Gaussian wave function. We compute its evolution in the presence of the Coulomb barrier and determine the probability that the particle tunnels through
$P_T = \lim\limits_{t \rightarrow \infty} \int_{l}^{\infty} dx \ |\psi(x,t)|^2$.
The result is plotted as a solid curve in the left panel of Fig.~\ref{fig:TunnProb_50K} and it is considerably different from the classical result for initial wave functions with position spread bigger than about $10$ fm.
For smaller position spreads the quantum and classical curves are approaching each other. Finally, for Dirac delta position distribution, in both cases we have flat momentum distribution, and hence half of the particles tunnel through.
The range of very small position spreads is not accessible in our numerical calculations because the wave function spreads very fast such that the dynamics of approaching the barrier already consumes all of the computational resources. We again emphasize that the nuclear reaction cross-section directly translates to the spread of the projectile's initial wave function. This provides an interesting way of experimentally determining this spread.

Note also that simultaneous measurement of the tunneling probability and the initial position or momentum spread is capable of disproving the classical model.
For example, at $\sigma = 10$ fm the classical probability is more than three orders of magnitude smaller than the quantum one, which is at the measurable level of $10^{-3}$ (one in a thousand).

\subsection{The Wentzel-Kramers-Brillouin approximation}
\label{sec:qtunneling}

Quantum tunneling is of course a well-studied phenomenon, with many interesting applications even in astrophysics~\cite{Nature-106.14,ZP-54.656}.
Traditionally, the projectile is assumed as an incident plane wave and the tunneling probability is approximated within the time-independent theory. One such treatment is the Wentzel-Kramers-Brillouin (WKB) approximation~\cite{ZP-39.828,ZP-38.518}:
\begin{eqnarray}
P_T^\text{WKB} &=& \frac{\hbar}{ \sqrt{2m \qty[V(l)-E_0] } } \ \exp\qty( -\frac{2}{\hbar} \int_{d_{cl} }^{l} dx \ \sqrt{ 2m \left[ V(x) - E_0 \right] } )	\nonumber \\
&=& \frac{\hbar}{ \sqrt{2m E_0 \qty(\frac{d_\text{cl}}{l}-1) } } \ \exp\qty[ -\frac{2}{\hbar} \int_{d_
\text{cl} }^{l} dx \ \sqrt{ 2m E_0 \qty(\frac{d_\text{cl}}{x}-1) } ].
\label{P_Cross_cl}
\end{eqnarray}

\begin{figure}
\centering
\includegraphics[width=\linewidth]{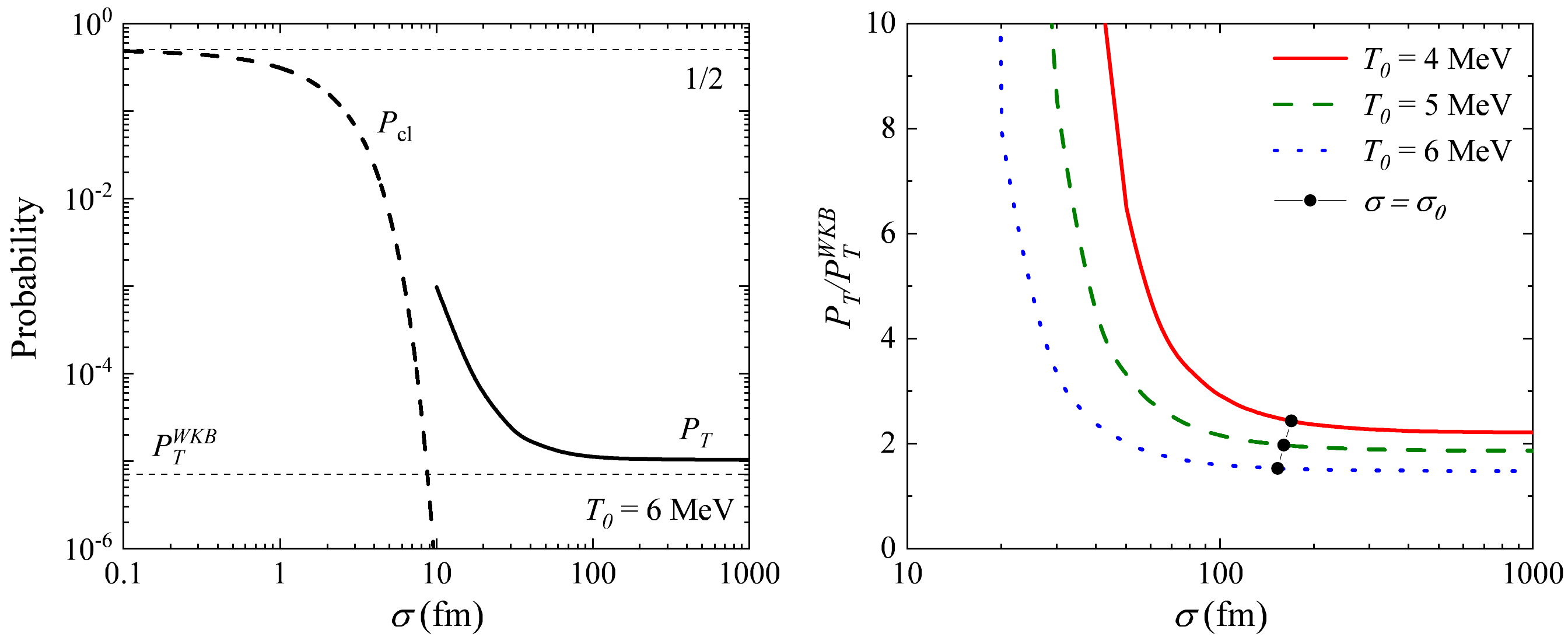}
\caption{Tunneling of alpha particle launched towards a Coulomb barrier of a gold nucleus. $\sigma$ is the spread of the initial position probability density of the projectile. $P_T$ is the tunneling probability obtained in the simulations of quantum dynamics. $P_T^\text{\emph{WKB}}$ is the WKB approximation of the same. $P_\text{cl}$ is the probability of barrier crossing obtained in the classical model [see Eq.~\eqref{P_Cross_cl}]. The initial launch distance is $L = 50$ pm. Numerical details are given in Sec.~\ref{appendix:constants_rutherford}.}
\label{fig:TunnProb_50K}
\end{figure}

We now provide conditions under which this approximation matches the time-dependent approach presented here. Some advancements have already been made in the limiting case of low-energy projectiles with a negligible momentum variance~\cite{PLA-378.1071}, and it is expected that many peculiar effects arise with a large momentum spread~\cite{PR-40.621,PLA-220.41,PLA-225.303,JPA-37.2423,PLA-378.1071}.
In the left panel of Fig.~\ref{fig:TunnProb_50K} we plot the tunneling probability in the WKB approximation by the horizontal dashed line at the bottom. The WKB result tends to the dynamical one in the limit of small initial momentum spread (large position spread).
However, it does not exactly approach the tunneling probability obtained in the dynamical simulation.
The right panel of Fig.~\ref{fig:TunnProb_50K} compares the two results for different initial energies of the projectiles. One can draw a boundary between the range where the two results are comparable and where they are very different.
It turns out that it is given by the optimal $\sigma_0$ we have derived above. In our simulations we varied the launching distance about $50$ pm,
and observe that the results are practically independent of $L$.

\section{Rutherford's experiment with photons}

An interesting regime of Rutherford experiment is when the projectiles are moving at relativistic speeds, e.g., the photons moving through a medium with refractive index $\sim 1/x$~\cite{AJP-81.405}. 
While the non-relativistic wave packets shown in Fig.~\ref{fig:FindingSig0_5MeV} of the present work do not recover the classical solution, in the Rutherford scattering of photons the classical limit is achieved~\cite{AJP-81.405}. This is a consequence of the non-dispersive behaviour of photonic systems. For non-relativistic particles the variance in momentum directly translates to a variance in the velocity, which leads to a growing position spread in time.
 Unlike alpha particles, photonic systems are non-dispersive as the variance in momentum does not correspond to a variance in the velocity. In other words the wavelength of photons does not change with time as they travel through free space. The same can be readily proved by solving the Ehrenfest's dynamics with the relativistic Hamiltonian for a photon: $\hat H=c\hat p$, where $c$ is the speed of light in vacuum. One can use the Ehrenfest's theorem to prove that the variance does not change:
\begin{eqnarray}
\dv{t} \bm{\Delta} x^2  = \dv{t} \qty( \ev{\hat x^2} - \ev{\hat x}^2 )
&=& \dv{t} \ev{\hat x^2} - 2  \ev{\hat x} \dv{t} \ev{\hat x}
\nonumber	\\
&=&	\frac{c}{i\hbar} \ev{\comm{\hat x^2}{\hat p}} -  2  \ev{\hat x} \frac{c}{i\hbar} \ev{\comm{\hat x}{\hat p}}
\nonumber	\\
&=&	\frac{c}{i\hbar} \times \ev{ 2i\hbar\hat x } -  2  \ev{\hat x} \frac{c}{i\hbar} \times i\hbar
\nonumber	\\
&=& 0.
\end{eqnarray}
Accordingly, apart from the negligible impact of refractive index, there is no change in the position spread of a photonic wave function as it evolves in time. One can therefore have an extremely narrow wave packet, and hence particle-like dynamics, all along the collision event. This explains the recovery of classical solutions in photonic Rutherford scattering~\cite{AJP-81.405}.

\section{Prospects and limitations}

The time-dependent scattering problems, like the one discussed here, involve the notions of positions and trajectories. While one can talk about the expected positions for a wave packet, no such analogous counterpart exists for individual plane waves~\cite{ActaPhyPolB-33.2059}. Hence, a consistent theory can only be built on the dynamics of localised wave functions~\cite{AJP-81.405,ActaPhyPolA-101.369}. 

The internal structures of the projectile and the target will give rise to various exchange effects~\cite{NP-78.409,NPA-148.529,PR-155.29,PRC-61.054610}, which will be more prominent at higher energies and must be accounted for by inclusion of exchange terms in the Hamiltonian~\cite{NP-47.652,PTP-122.1055,PLB-772.1}. An even fuller approach would be to replace the Coulomb potential with more realistic interactions~\cite{JPCC-122.14606,MicMach-11.319,JPCA-119.6897}. 

Finally, if we consider a two or three-dimensional setting, different parts of the wave packet will acquire different phases as time passes. This will give rise to a diffraction pattern in the angular distribution of the total amplitude~\cite{AJP-52.60}. An extension of our work in two-dimensions promises to calculate the possible dependence of this pattern on the initial spread of the projectile wave function.

\section{The case of two colliding wavepackets}

Untill now we represented the projectile as an incident Gaussian wave packet and target as a simplistic Coulomb potential. Even though this works well in the considered energy range, this is an approximation. We now consider the target also as a wave packet and study the scattering from the point of view of dynamics of two colliding wave packets~\cite{AJP-68.1113}. 

For example, consider the case of two nuclei $A$ and $B$ launched from $x = \pm L$ with an equal energy $T_0$, i.e., with a momentum $p_0 = \sqrt{2mT_0}$. If they are prepared in identical Gaussian wave packets with a position spread $\sigma$ at $t=0$, the transformations described in Chapter~\ref{ch:chapter2} imply that in the COM frame the problem decouples into two independent single particle dynamics. The COM and the reduced mass, at $t=0$, are described as localised Gaussian wave packets of position spreads $\sigma/\sqrt{2}$ and $\sqrt{2}\sigma$, respectively. Note that the reduced mass wave packet is broader as compared to the projectiles by a factor of $\sqrt{2}$.

Given the symmetry of collision, the COM (mass $2m$) has a zero momentum and sits at the origin, and the reduced mass (mass $m/2$) is launched from $x=+2L$ towards the origin with a momentum $p_0$. We can now easily prove that the relative distance is minimised when the reduced mass wave packet has a width of $\sqrt{\hbar L/p_0}$, which corresponds to the optimal width of the projectiles as $\sqrt{\hbar L/2p_0}$, but with the quantum distance of closest approach given by
\begin{equation}
d_{\mathrm{cl}} < d_{\mathrm{qm}}  < d_{\mathrm{cl}} + \sqrt{\frac{\hbar L}{ p_0}}.
\end{equation}

\section{Numerical details} \label{appendix:constants_rutherford}

Numerical calculations are performed in natural units of $c=1$. Accordingly, the conversion constant $\hbar c = 197.3269804$ MeV fm. We follow these units within this chapter. Fine-structure constant, $\alpha$ = $1/137.035999084$, and the mass of the alpha particle is $3727.3794066$ MeV.

\begin{itemize}[leftmargin=*]

\item In Fig.~\ref{fig:10K5MeV}: $L = 10$ pm and $T_0 = 5$ MeV, which implies $p_0 = 193.06$ MeV.

\item In Fig.~\ref{fig:FindingSig0_5MeV}: $T_0 = 5$ MeV implying $p_0 = 193.06$ MeV; $\sigma_0$ = $71.49$, $101.10$, $123.82$, $142.97$, and $159.85$ fm, for $L$ = $10$, $20$, $30$, $40$, and $50$ pm, respectively.

\item In Fig.~\ref{fig:qdoca}:
\begin{itemize}[leftmargin=*]
\item $T_0 = 2$ MeV implies $p_0 = 122.10$ MeV; $\sigma_0$ = $89.89$, $127.12$, $155.69$, $179.78$, and $201.00$ fm, for $L$ = $10$, $20$, $30$, $40$, and $50$ pm, respectively.
\item $T_0 = 3$ MeV implies $p_0 = 149.55$ MeV; $\sigma_0$ = $81.22$, $114.87$, $140.69$, $162.45$, and $181.62$ fm, for $L$ = $10$, $20$, $30$, $40$, and $50$ pm, respectively.
\item $T_0 = 4$ MeV implies $p_0 = 172.68$ MeV; $\sigma_0$ = $75.59$, $106.90$, $130.92$, $151.18$, and $169.02$ fm, for $L$ = $10$, $20$, $30$, $40$, and $50$ pm, respectively.
\item $T_0 = 5$ MeV implies $p_0 = 193.06$ MeV; $\sigma_0$ = $71.49$, $101.10$, $123.82$, $142.97$, and $159.85$ fm, for $L$ = $10$, $20$, $30$, $40$, and $50$ pm, respectively.
\end{itemize}

\item In Fig.~\ref{fig:MicroProp_50K5MeV}: $L = 50$ pm and $T_0 = 5$ MeV, which implies $p_0 = 193.06$ MeV and $\sigma_0 = 159.85$ fm.

\item In Fig.~\ref{fig:TunnProb_50K}: $L = 50$ pm, which implies $\sigma_0$ = $169.02$, $159.85$, and $152.73$ fm, for $T_0$ = $4$, $5$, and $6$ MeV, respectively.

\end{itemize}

\begin{figure}[!t]
	\centering
	
	\subfloat[The first test: convergence of the quantum distance of closest approach.]{\includegraphics[width=0.9\linewidth]{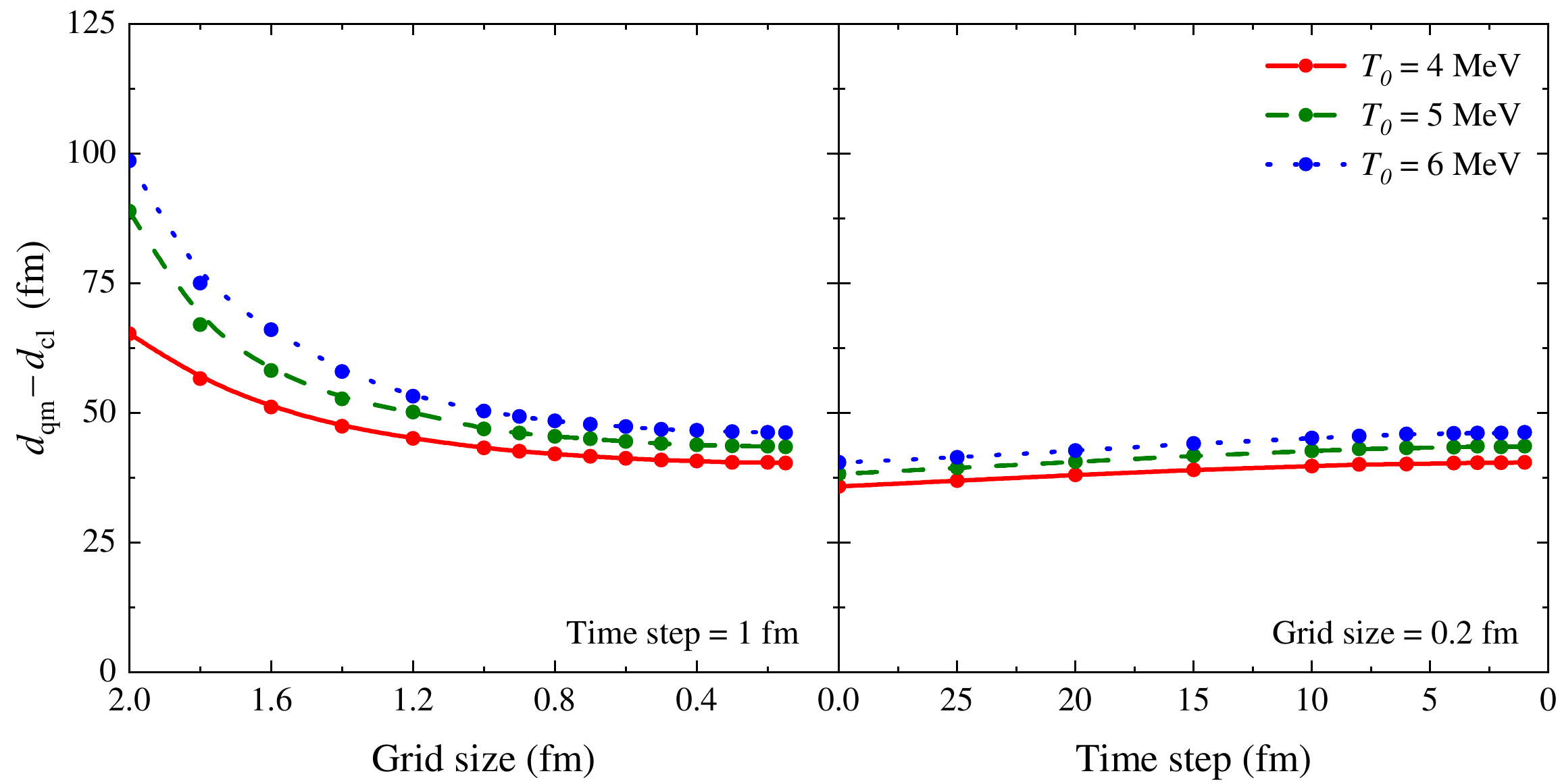}\label{fig:convg_10K5MeV_dqm}}

	\subfloat[The second test: constancy of the expected Hamiltonian during the scattering event. The relative errors in $\ev{\hat H}$ are calculated w.r.t. $\ev{\hat H(0)}$.]{\includegraphics[width=\linewidth]{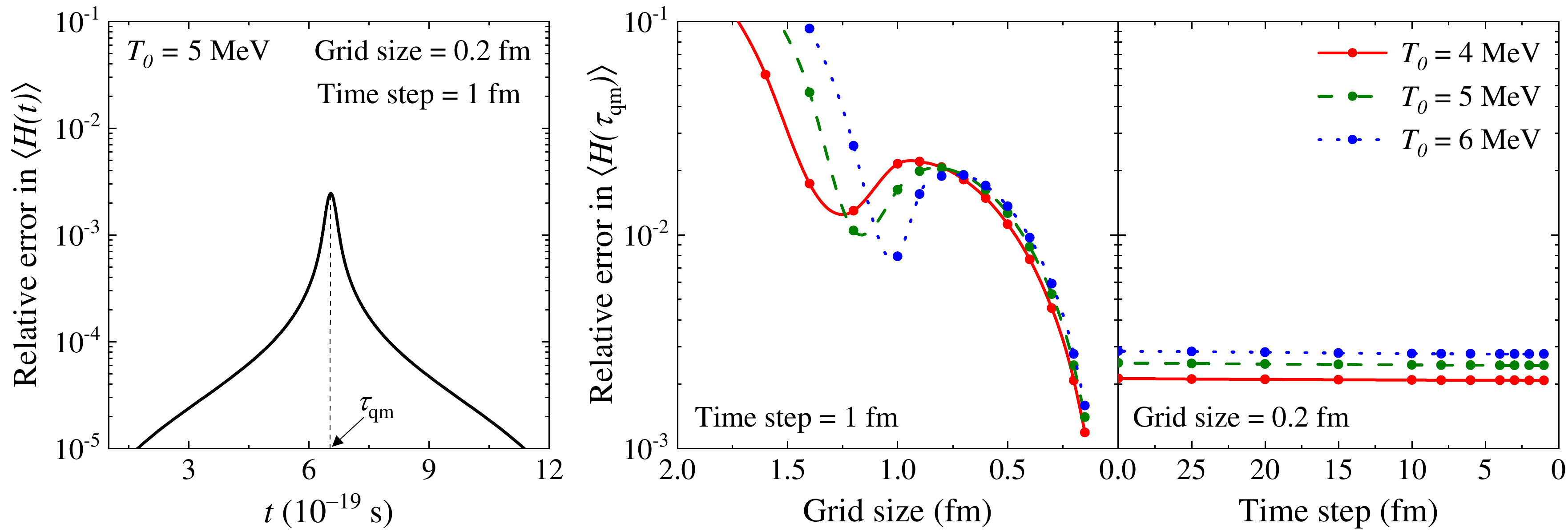}\label{fig:convg_10K5MeV_energy}}
	
\caption{Convergence of numerical results. The alpha particle Gaussian wave packet is initially centered at a separation $L=10$ pm with a width of $\sigma = 100$ fm. $T_0$ denotes the kinetic energy of an equivalent classical projectile. $d_\text{qm}$ and $d_\text{cl}$ represent the quantum and classical distances of closest approach, and $\hat H$ is the Hamiltonian.}.
\label{fig:convg-10K5MeV}
\end{figure}

\subsection{Tests of convergence}

We have employed two different convergence tests to perform the error analysis: (i) the convergence of the quantum distance of closest approach, and (ii) the constancy of the expected Hamiltonian all along the scattering event. In Fig.~\ref{fig:convg-10K5MeV} we present the error analysis assuming $L=10$ pm and $\sigma = 100$ fm (the typical width of Gaussian wave packets considered here).

Fig.~\ref{fig:convg_10K5MeV_dqm} shows that there is a very good convergence in $d_\text{qm}$ for grid size $\lesssim 0.3$ fm and time step $\lesssim 2$ fm. The first panel of Fig.~\ref{fig:convg_10K5MeV_energy} shows the variation of the relative error in expected Hamiltonian, i.e. $1-\ev{\hat H}/\ev{\hat H(0)}$, as a function of time for a fixed grid size and time step. For most times this error is negligible, except at $t=\tau_{qm}$ where it attains its peak. The second panel shows this peak error is negligible for grid size $\lesssim 0.3$ fm. The last panel shows that an increased time step has no significant impact on the error in energy. Since this was not the case for the error in $d_\text{qm}$ (see the second panel in Fig.~\ref{fig:convg_10K5MeV_dqm}), we use two tests of convergence for reliable calculations. A grid size of $0.2$ fm, with a time step of $1$ fm (equivalent to 1 fm/$c$ in SI units), ensures a good precision in all of our calculations. We have accordingly set the same throughout this chapter.

\subsection{Dynamic grid allocation}
\label{appendix:DynamicAllocation}

Dynamical simulations of long-range interactions require huge computational resources. Moreover, the intrinsic time scales (the timescale on which the wave function is changing) are much smaller than the time elapsed in the collision events. This makes it challenging to study the scattering in a fully quantum mechanical time-dependent theory~\cite{FBS-56.727,PRC-73.054608,CMAME-193.1733}. Within the scope of this chapter, the wave function remains confined near a single point; this allows us to perform a dynamic allocation of the grid and recast the problem as a sequential chain of small simulations.
\begin{itemize}
\item We start by defining a box around the initial launch distance such that the wave packet is highly concentrated at its center. The simulation is initiated, and the wave packet evolves in the chosen potential. 
We keep track of the peak and the width of the probability density function to make sure there is no reflection from the boundaries of the box.
\item  Once the tail of the probability distribution starts approaching either of the boundaries, we define a new box that re-confines the wave packet at its center. 
Note that at all times the wave packet's center is separated from numerical boundaries by a distance which is at least seven times the position spread of the wave function. 
\item The final wave function from the old grid (box) is interpolated and used as an initial condition to start the calculation onto the new grid.
The grid allocation and interpolation keep repeating until the wave packet returns back to the original launch distance.
\end{itemize} 
This chain of small simulations is verified to precisely simulate whole of the event.

\section{Summary}

Average quantum dynamics in potential with well-defined convexity properties does not approach its classical counterpart.
We have shown this in the case of Coulomb interaction between the alpha particles and the stationary gold nuclei in a head-on collision. The differences include the distance of closest approach, time of collision, and time symmetry of the dynamics.
We sketched an experiment aimed at verifying these predictions.
It could be rather challenging as we focused on head-on collisions in this work.
It would be interesting to work out possible differences in the 2D setting and compute how the differential cross-sections depend on the spread of the initial wave function of the alpha particle.
Such predictions might be easier to verify in a laboratory.
Finally, the model could be extended to study fully quantum mechanically astrophysical nuclear reactions.

%% file: science/ch4-Gravity.tex
\chapter{Gravitational Entanglement between Freely Falling Masses}
\label{ch:chapter4}

We describe a method for a precise study of gravitational interaction between two nearby quantum masses. Since the displacements of these masses are much smaller than the initial separation between their centers, the displacement-to-separation ratio is a natural parameter in which the gravitational potential can be expanded. We show that entanglement in such experiments is sensitive to the initial relative momentum only when the system evolves into non-Gaussian states, i.e., when the potential is expanded at least to the third-order (cubic) term. A pivotal role of the force gradient as the dominant contributor to position-momentum correlations is demonstrated. We establish a closed-form expression for the amount of entanglement, which shows a linear dependence on relative momentum. From a quantum information perspective, the results find applications as a momentum witness of non-Gaussian entanglement. Our methods are versatile and apply to any number of central interactions expanded to any order.

\section{Introduction}

\begin{figure}
	\centering
	\includegraphics[width=0.6\linewidth]{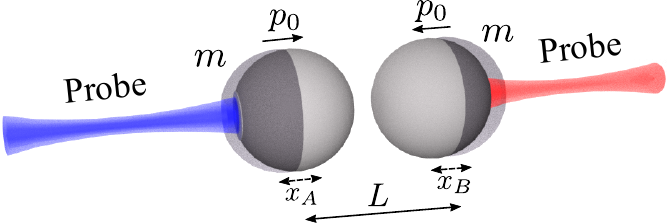}
	\caption{\label{fig:expsetup}
		Setup under consideration. Two identical spheres of mass $m$ are released from the ground state of identical harmonic traps with an equal and opposite momentum $p_0$ along the line joining their centers. The centers are initially separated by a distance $L$, and displacements from them are denoted by
		$x_A$ and $x_B$. After time $t$ entanglement is estimated with the help of the probing lasers.}
\end{figure}

The experiment we have in mind in this chapter could be realized within the field of optomechanics~\cite{RMP_86.1391}, which already succeeded in cooling individual massive particles near their motional ground state~\cite{NJP_11.073032,Science_372.6548,Nature.478.89}, and in entangling cantilevers to light and themselves~\cite{Palomaki-Science,NatPhot.15.817,Nature.556.473}.
In such a setup the particles are separated much more than their displacements.
For example, two Osmium spheres (the densest natural material) each of mass 100 $\mu$g (radius $0.1$ mm) with an initial inter-surface distance of $0.1$ mm move by less than a nanometer within $1$ second of evolution~\cite{npjQI_6.12}, vividly illustrating the weakness of gravity. Since the situation under consideration is non-relativistic, the relevant interaction is characterized by the quantum Newtonian potential.
Given that the displacements are small compared to the initial separation between two spheres, a natural parameter in which the potential can be expanded is the displacement-to-separation ratio~\cite{npjQI_6.12,JOPB_23.235501,Datta_2021,PhysRevLett.128.110401,Roccati2022}. The goal we propose here is to identify phenomena that can only occur if the potential is expanded to a particular order, thus witnessing the relevance of at least this order in experiments.

From this perspective, the gravitational entanglement proposals, in addition to providing clues about the quantum nature of gravity, also supply tests of the form of gravitational interaction.
For example, entangling two initially disentangled masses requires at least the second-order term~\cite{PhysRevLett.119.240401,PhysRevLett.119.240402,npjQI_6.12}. Here we show a method that witnesses the third-order term and has an advantage of a simple modification of the entanglement scheme with confined particles. Hence, an experiment designed to probe gravitational entanglement can also be used to witness even weaker gravitational coupling.

Our basic idea is to push the particles towards each other as it is intuitively expected that such obtained stronger gravity will lead to higher accumulated entanglement. Yet,
we demonstrate that the quantum entanglement generated by gravitational potential truncated at the second order is \emph{insensitive} to the relative motion of the two masses. This is shown explicitly with an analytical solution for the time evolution of the corresponding covariance matrix~\cite{PRA_65.032314,PRA_70.022318,PRA_72.032334}.
Our intuition is only recovered with the
potential containing at least the third-order term, i.e., when the system evolves into a non-Gaussian state. A closed-form expression for entanglement is established, which agrees with numerical simulations, showing a linear dependence on the relative momentum and the critical role played by the force gradient across the reduced mass wave packet. The introduced methods are applicable to any central interaction, even when many of them are present side by side. Moreover, the closed forms can be obtained for expansions to arbitrary order. They also show remarkable robustness, e.g., even the impact of the fourth-order term on the non-Gaussianity quantifier can be captured numerically despite an astonishingly weak gravitational interaction.

\section{Experimental setup}

\begin{figure}
	\centering
	\includegraphics[width=\linewidth]{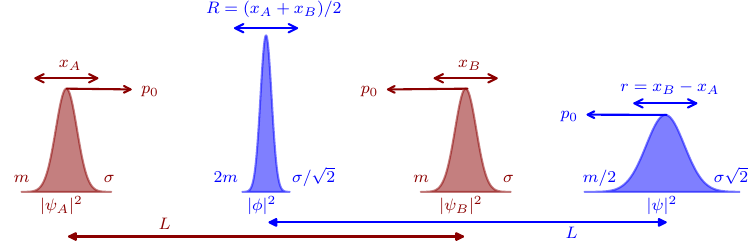}
	\caption{The same as Fig.~\ref{fig:TMGS_genreal}, but for two identical masses $m$ prepared in Gaussian states having the same width $\sigma$ and equal and opposite momentum $p_0$.}
	\label{fig:TMGS_special}
\end{figure}

Consider the setup schematically represented in Fig.~\ref{fig:expsetup}, where we also introduce our notation.
The initial wave function is assumed to describe two independent masses, each in a natural Gaussian state with position spread $\sigma$: $\Psi(x_A,x_B,t=0) = \psi_A(x_A) \ \psi_B(x_B)$, where
\begin{eqnarray}
	\psi_A(x_A) &=& \qty( \frac{1}{2\pi\sigma^2} )^{1/4} \exp(-\frac{x_A^2}{4\sigma^2} + i\frac{p_{0} }{\hbar}x_A ),		 \\
	\psi_B(x_B) &=& \qty( \frac{1}{2\pi\sigma^2} )^{1/4} \exp(-\frac{x_B^2}{4\sigma^2} - i\frac{p_{0} }{\hbar}x_B ).   
\end{eqnarray}
Note that without loss of generality we chose the momenta to be opposite and equal. The Hamiltonian in the non-relativistic regime is given by
\begin{equation}
	\hat{H} = \frac{ \hat{p}_A^2}{2m}+\frac{ \hat{p}_B^2}{2m} - \frac{Gm^2}{L+( \hat{x}_B- \hat{x}_A)}.
\end{equation}
Since this is a two-body problem, 
we showed in Sec.~\ref{appendix:TDSE-Transf2COM} that
the initial wave function separates as $\Psi(x_A,x_B,t=0) = \phi(R,t=0) \ \psi(r,t=0)$,
where
\begin{eqnarray}
	\phi(R,t=0) &=& \qty( \frac{1}{\pi\sigma^2} )^{1/4} \exp(-\frac{R^2}{2\sigma^2}), 	\label{eq:COMwavefunction_t0}
\\
	\psi(r,t=0) &=& \qty( \frac{1}{4\pi\sigma^2} )^{1/4} \exp(-\frac{r^2}{8\sigma^2} - i\frac{p_{0} }{\hbar}r ).	\label{eq:RMwavefunction_t0}
\end{eqnarray}
The wave functions $\phi$ and $\psi$ describe the motion of the COM and the reduced mass, respectively. Recall that the COM wave packet admits a smaller width of $\sigma/\sqrt{2}$, and the reduced mass wave packet has a larger width of $\sigma\sqrt{2}$. The corresponding relations are illustrated in Fig.~\ref{fig:TMGS_special}.
In this frame the Hamiltonian decouples as
\begin{equation}
	\hat{H} =  \hat{H}_R +  \hat{H}_r = \qty(  \frac{ \hat{P}^2}{4m} ) + \qty(  \frac{ \hat{p}^2}{m} - \frac{Gm^2}{L+ \hat{r}} ),
	\label{eq:hamseparable}
\end{equation}
where $ \hat{P} = -i\hbar\partial/\partial R$ and $ \hat{p} = -i\hbar\partial/\partial r$ are the momentum operators for the COM and the reduced mass, respectively. A separable Hamiltonian implies that the two-body wave function retains its product form at all times, i.e., $\Psi(x_A,x_B,t) = \phi(R,t) \ \psi(r,t)$. 
Furthermore, the COM wave packet evolves like a free particle, i.e., its Gaussianity is preserved~\cite{JOPB_33.4447,RevModPhys.84.621,book_decoherence_Maximilian}, and the first two statistical moments characterize the quantum state fully. They are given by [see Appendix~\ref{appendix:Ehrenfest_COM_redmass} for details]
\begin{eqnarray}
\bm{\Delta} R^2 &=& \ev{\hat R^2} - \ev{\hat R}^2  = \frac{1}{2}\sigma^2 (1+\omega_0^2t^2),
	\\
\bm{\Delta} P^2 &=& \ev{\hat P^2} - \ev{\hat P}^2 	= \frac{\hbar^2}{2\sigma^2},
	\\
\textbf{Cov}( {R}, {P}) &=& \frac{1}{2}\ev{\hat R \hat P + \hat P \hat R} - \ev{\hat R}\ev{\hat P} = \frac{1}{2}\hbar\omega_0 t.
\end{eqnarray}
The state $\psi$ evolves in the gravitational potential, which we now expand in the powers of the displacement-to-separation ratio $r/L$:
\begin{equation}
	\hat H_r = \frac{p^2}{m} - \frac{Gm^2}{L+ \hat{r}}
 \approx \frac{p^2}{m} - \frac{1}{4} m \omega^2 \sum_{n=0}^{N} \frac{(-1)^n}{L^{n-2}} \hat r^n,
	\label{eq:Ham_redmass}
\end{equation}
where $N$ is the order of approximation, and we defined $\omega^2 = 4Gm/L^3$ for later convenience. 
We derive exact analytical expressions for the statistical moments of $\psi$ by solving the related Ehrenfest equations in the case of $N=2$ [see Appendix~\ref{appendix:Ehrenfest_COM_redmass} for details]:
\begin{eqnarray}
\bm{\Delta} r^2 &=& \ev{\hat r^2} - \ev{\hat r}^2 	\nonumber	\\
	&&	= \Bigg[  2\sigma^2 \Big( 1+\sinh[2](\omega t) \Big)		 + \frac{1}{8}L^2 \Big( 3 + \cosh(2\omega t)-4\cosh(\omega t) \Big) 	\nonumber \\
	&&	\ \ + \frac{Lp_0}{m\omega} \Big( \sinh(2\omega t) - 2\sinh(\omega t) \Big) 	 + \frac{4}{m^2\omega^2} \qty( p_0^2+\frac{\hbar^2}{8\sigma^2} ) \sinh[2](\omega t) \Bigg]	\nonumber 	\\
	&& \ \	- \Bigg[ \frac{1}{2}L \Big( 1 -\cosh(\omega t) \Big) -  \frac{2p_0}{m\omega} \sinh(\omega t) \Bigg]^2	\nonumber	\\
	&&	= 2\sigma^2 \qty( \cosh[2](\omega t) + \frac{\omega_0^2}{\omega^2} \sinh[2](\omega t) ),  
\label{eq:var_r2-freefall}  
\\	\nonumber	\\
\bm{\Delta} p^2  &=& \ev{\hat p^2} - \ev{\hat p}^2 	\nonumber	\\
&&	= \Bigg[ \qty(p_0^2+\frac{\hbar^2}{8\sigma^2} ) \Big( 1+\sinh[2](\omega t) \Big)    + \frac{1}{4}m\omega Lp_0\sinh(2\omega t)	\nonumber	\\
&& \ \ + \frac{1}{4}m^2\omega^2 \qty( 2\sigma^2+\frac{1}{4}L^2 ) \sinh[2](\omega t)	\Bigg]  - \Bigg[ - p_0\cosh(\omega t) - \frac{1}{4} m\omega L \sinh(\omega t) \Bigg]^2	\nonumber \\
&&	= \frac{\hbar^2}{8\sigma^2} \qty( \cosh[2](\omega t) +  \frac{\omega^2}{\omega_0^2} \sinh[2](\omega t) ), 
\label{eq:var_p2-freefall}  
\\	\nonumber	\\
\textbf{Cov}( {r}, {p}) &=& \frac{1}{2}\ev{\hat r \hat p + \hat p \hat r} - \ev{\hat r}\ev{\hat p} 	\nonumber	\\
&& = \frac{1}{2} \Bigg[		 Lp_0 \Big( \cosh(2\omega t) - \cosh(\omega t) \Big)	+ \frac{1}{8}m\omega L^2 \Big( \sinh(2\omega t) - 2\sinh(\omega t) \Big) 		\nonumber \\
&&	\	\  + \frac{2}{m\omega} \qty( p_0^2+\frac{\hbar^2}{8\sigma^2}+ \frac{1}{2}m^2\omega^2\sigma^2 ) \sinh(2\omega t)	\Bigg] 	\nonumber	\\
&&	\	\ - \Bigg[ \frac{1}{2}L \Big( 1 -\cosh(\omega t) \Big) -  \frac{2p_0}{m\omega} \sinh(\omega t) \Bigg] 	\times 	\Bigg[ - p_0\cosh(\omega t) - \frac{1}{4} m\omega L \sinh(\omega t) \Bigg]	\nonumber	\\
&&	= \frac{\hbar}{4} \qty( \frac{\omega_0}{\omega}+\frac{\omega}{\omega_0}  ) \sinh(2\omega t).
\label{eq:cov_rp-freefall}  
\end{eqnarray}
Together with the statistical moments for the COM, these determine the bipartite covariance matrix, $\bm\sigma$, in an exact closed form [see Appendix~\ref{appendix:EntangFormalism} for details]:
\begin{gather}
	\bm{\sigma}_{00}(\bm{\sigma}_{02})   = \bm{\Delta} R^2 +\!(\!-\!) \ \frac{1}{4} \bm{\Delta} r^2,   
\label{eq:covmatak-varpos}
\\
	\bm{\sigma}_{11}(\bm{\sigma}_{13})  =  \frac{1}{4} \bm{\Delta} P^2 +\!(\!-\!) \ \bm{\Delta} p^2,
\label{eq:covmatak-varmom}
\\
	\bm{\sigma}_{01}(\bm{\sigma}_{03})  =  \frac{1}{2} \textbf{Cov}( {R}, {P})  +\!(\!-\!) \ \frac{1}{2} \textbf{Cov}( {r}, {p}) .
\label{eq:covmatak-covposmom}
\end{gather} 

With the inclusion of higher-order terms in the potential, i.e., $N>2$, the corresponding Ehrenfest's equations cannot be solved analytically due to the emergence of an infinite set of coupled differential equations involving ever-increasing statistical moments. 
We therefore resort to numerical methods and calculate the time evolution of $\psi$ by implementing Cayley's form of evolution operator~\cite{book_NumericalRecipies}. The numerical evaluations for $\psi$ are combined with analytical solutions for the COM to construct the covariance matrix and the two-body wave function. In order to deal with weak gravitational interaction, we used the improved Cayley's method with the five-point stencil and discretised onto a pentadiagonal Crank-Nicolson scheme, which is further solved by implementing the LU factorization techniques. 
The details have been described in Sec.\ref{sec:pentadiagonaldiscretisation}.

The methodology just described returns an analytical form of the covariance matrix at time $t$ for potentials truncated at $N = 2$ and a numerical form of the two-body wave function for all $N$.
These are thereafter used for computing the entanglement between two masses [see Appendix~\ref{appendix:EntangFormalism} for the methodologies for estimation of quantifiers].
In particular, we use logarithmic negativity and entropy of entanglement as entanglement quantifiers.
While logarithmic negativity is known to be a faithful entanglement quantifier for Gaussian states~\cite{PRA_65.032314,PRA_70.022318,PRA_72.032334}. we will also discuss non-Gaussian pure states and hence the inclusion of the entropy of entanglement.
We first give the results for $N=2$, emphasizing the independence of relative momentum and its origin.
Then we move to $N=3$ and demonstrate that entanglement does depend on the initial momentum, and in the relevant regime, it is linear in the relative momentum.
We also analyze an indicator of non-Gaussianity (skewness) and demonstrate the precision of our methods by calculating the marginal impacts of the fourth-order term in the potential expansion. A methodology to obtain closed-form expressions for the entanglement gain with time is presented.

\section{Entanglement gain with quadratic interactions}
\label{sec:covmat_freefall}

Consider first the gravitational potential truncated at the second order. 
We obtained exact analytical forms for the independent elements of the covariance matrix. The solutions simplify if they are written in terms of already introduced $\omega$ and in terms of $\omega_0 = \hbar/2m\sigma^2$, which is the frequency of harmonic trap for which the initial state is the ground state. They are given by
\begin{eqnarray}
	\bm{\sigma}_{00}   &=& \bm{\Delta} R^2 + \ \frac{1}{4} \bm{\Delta} r^2
\nonumber 	\\	
	&&=  \qty[ \frac{1}{2}\sigma^2 (1+\omega_0^2t^2) ] + \frac{1}{4} \qty[ 2\sigma^2 \qty( \cosh[2](\omega t) + \frac{\omega_0^2}{\omega^2} \sinh[2](\omega t) ) ]		\nonumber \\
	&&= \frac{\hbar}{4m\omega_0} \qty[ 2+\omega_0^2t^2+\qty( 1+\frac{\omega_0^2}{\omega^2} )\sinh^2(\omega t) ],
\label{eq:SolCovMat_FreeFall_Start}
\\	\nonumber	\\
	\bm{\sigma}_{02}   &=& \bm{\Delta} R^2 - \ \frac{1}{4} \bm{\Delta} r^2	
\nonumber 	\\	
	&&=  \qty[ \frac{1}{2}\sigma^2 (1+\omega_0^2t^2) ] - \frac{1}{4} \qty[ 2\sigma^2 \qty( \cosh[2](\omega t) + \frac{\omega_0^2}{\omega^2} \sinh[2](\omega t) ) ]		\nonumber \\
	&&=  \frac{\hbar}{4m\omega_0} \qty[ \omega_0^2t^2-\qty(1+\frac{\omega_0^2}{\omega^2})\sinh^2(\omega t) ],
\\	\nonumber	\\
\bm{\sigma}_{11}  &=&	\frac{1}{4} \bm{\Delta} P^2 +  \bm{\Delta} p^2
\nonumber 	\\	
	&&= \frac{1}{4} \qty( \frac{\hbar^2}{2\sigma^2} ) + \frac{\hbar^2}{8\sigma^2} \qty( \cosh[2](\omega t) +  \frac{\omega^2}{\omega_0^2} \sinh[2](\omega t) )	\nonumber	\\
	&&=  \frac{m\hbar\omega_0}{4} \qty[ 2+\qty( 1+\frac{\omega^2}{\omega_0^2} )\sinh^2(\omega t) ],
\\	\nonumber	\\
	\bm{\sigma}_{13} 	&=& 	\frac{1}{4} \bm{\Delta} P^2 -  \bm{\Delta} p^2		
\nonumber 	\\	
	&& \frac{1}{4} \qty( \frac{\hbar^2}{2\sigma^2} ) - \frac{\hbar^2}{8\sigma^2} \qty( \cosh[2](\omega t) +  \frac{\omega^2}{\omega_0^2} \sinh[2](\omega t) )	\nonumber	\\
	&&= - \frac{m\hbar\omega_0}{4} \qty( 1+\frac{\omega^2}{\omega_0^2} ) \sinh^2(\omega t),
\\	\nonumber	\\
	\bm{\sigma}_{01}  &=&	\frac{1}{2} \textbf{Cov}( {R}, {P})  + \ \frac{1}{2} \textbf{Cov}( {r}, {p})	
\nonumber 	\\	
	&&=	\frac{1}{2} \qty(\frac{1}{2}\hbar\omega_0 t) + \frac{1}{2} \qty[\frac{\hbar}{4} \qty( \frac{\omega_0}{\omega}+\frac{\omega}{\omega_0}  ) \sinh(2\omega t)]	\nonumber	\\
	&&=	 \frac{\hbar}{8} \qty[  2\omega_0 t + \qty( \frac{\omega_0}{\omega}+\frac{\omega}{\omega_0} )\sinh(2\omega t) ],
\\	\nonumber	\\
\bm{\sigma}_{03} &=&	\frac{1}{2} \textbf{Cov}( {R}, {P}) - \ \frac{1}{2} \textbf{Cov}( {r}, {p})
\nonumber 	\\	
	&&= \frac{1}{2} \qty(\frac{1}{2}\hbar\omega_0 t) - \frac{1}{2} \qty[\frac{\hbar}{4} \qty( \frac{\omega_0}{\omega}+\frac{\omega}{\omega_0}  ) \sinh(2\omega t)]	\nonumber	\\
	&&= \frac{\hbar}{8} \qty[  2\omega_0 t - \qty( \frac{\omega_0}{\omega}+\frac{\omega}{\omega_0} )\sinh(2\omega t) ].
\label{eq:SolCovMat_FreeFall_Finish}
\end{eqnarray}
The logarithmic negativity for $p_0 = 0$, in the regime $\omega \ll \omega_0$ and $\omega t \ll 1$, was already approximated to~\cite{npjQI_6.12}
\begin{equation}
	E(\bm{\sigma}) \approx - \log_2 \sqrt{ 1+2\Omega^6t^6 -2\Omega^3t^3\sqrt{ 1 + \Omega^6t^6} },
	\label{eq:lneg_tk}
\end{equation}
where $\Omega^3 = \omega_0\omega^2 /6$. 
We verified that this formula indeed matches our results, and emphasize that the solutions obtained here are \emph{exact}, and hence they can be used to quantify entanglement outside of the constraints that led to Eq.~\!\eqref{eq:lneg_tk}. An example is given below.

\begin{figure}
	\centering
	\includegraphics[width=\linewidth]{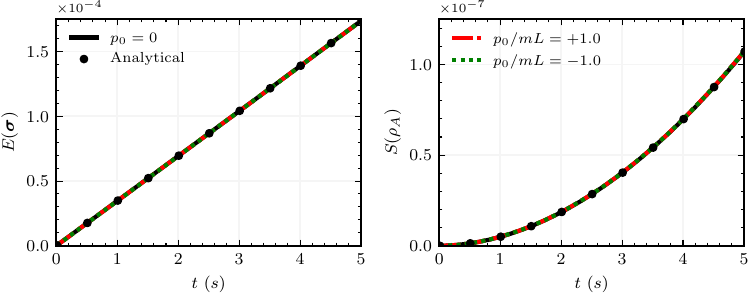}
	\caption{Accumulation of entanglement with gravitational potential truncated at the quadratic term ($N=2$). The configuration consists of identical Osmium spheres with $m = 0.25$ pg, $L=2.5$ times their radius, and $\sigma = 2.5$ nm. $p_0$ is the initial momentum. 
		Analytical results are calculated with the closed form of the covariance matrix.
		$E(\bm{\sigma})$ denotes the logarithmic negativity of covariance matrix, and $S(\rho_A)$ is the entanglement entropy. The values of $p_0/mL$ in the legends are in multiples of $6.18082292 \times 10^{-3}$ s$^{-1}$.}
	\label{fig:entV2_manymom}
\end{figure}

Perhaps the most striking feature of the covariance matrix is its insensitivity to the initial momentum $p_0$. Accordingly, all quantities derived from the covariance matrix, say entanglement or squeezing~\cite{Datta_2021,PhysRevA.101.063804}, are independent of the initial momentum.
Evidently, in this approximation, the two initially moving masses accumulate the same amount of entanglement as when they start from rest.
Furthermore, the amount of entanglement is the same, independent of whether the masses are moving toward each other or away from each other.
This is confirmed by the simulations presented in Fig.~\ref{fig:entV2_manymom}.
Not only there is no momentum dependence in the dynamics of logarithmic negativity and entropy of entanglement, they also perfectly overlap with analytical results, rendering our methods reliable and consistent. We emphasize that the configurations considered here are non-relativistic. Field theory calculations imply momentum-dependent relativistic corrections to the Newtonian potential~\cite{PhysRevD.105.106028,PhysRevA.101.052110}, and accordingly, we verify that second-order quantum entanglement generated by relativistic particles is, in principle, momentum dependent. However, for the parameters in Fig.~\ref{fig:entV2_manymom}, such corrections to the Newtonian potential energy are sixteen orders of magnitude smaller and hence we do not discuss them. We also note that Eq.~\!\eqref{eq:lneg_tk} is not applicable to the configuration considered in Fig.~\ref{fig:entV2_manymom}, because $\omega_0 \approx 25 \omega$.

\section{Force gradient as the driver of quantum correlations}

We now move to explanations of the observed results.
Mathematically, it is clear that a non-zero force gradient across the size of the wave packet is a necessary condition for entanglement.
Without it the potential is effectively truncated at $N=1$, and the total Hamiltonian is the sum of local terms.
Physically, entanglement is caused by correlations in complementary variables, here between positions and momenta. 
Due to a force gradient, the parts of the wave packets that are closer are gravitationally attracted more than the parts which are further apart. 
Hence a moment later, different momentum is developed across different positions within the wave packets, leading to quantum entanglement.

Furthermore, assuming that the force gradient is the main contributor to entanglement gain explains the independence of initial momentum.
Since the potential is truncated at $N=2$, the force gradient is constant in space.
Therefore, it is irrelevant if the particle moves to a different location in the meantime, and hence the initial momentum does not play a role in entanglement dynamics. Quantitatively, the force gradient is $F_2' = m \omega^2 / 2$, and therefore it fully describes entanglement in Eq.~\!\eqref{eq:lneg_tk} since now $\Omega^3 = \omega_0\omega^2/6 \equiv (\omega_0/ 3m) F_2'$.
In the following section we provide further evidence for the pivotal role of force gradient in entanglement dynamics due to higher-order interactions.

\begin{figure}
\centering
\includegraphics[width=\linewidth]{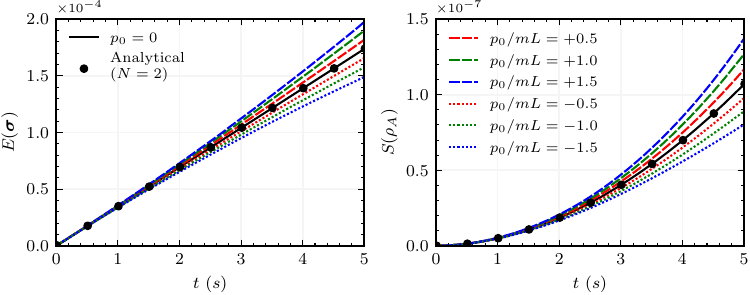}
\caption{Momentum dependence of entanglement accumulated with gravitational potential truncated at the cubic term ($N=3$). The same physical configuration as in Fig.~\ref{fig:entV2_manymom}. Analytical results are calculated from the closed-form of the covariance matrix for $N=2$, and coincide with the numerical results for $N = 3$ and $p_0 = 0$. The values of $p_0/mL$ in the legends are in multiples of $6.18082292 \times 10^{-3}$ s$^{-1}$.}
\label{fig:entV3_manymom}
\end{figure}

\section{Entanglement gain with cubic interactions}

Let us continue with the working hypothesis that the force gradient is the dominant factor in entanglement dynamics.
For the cubic potential, $N=3$, the gradient is given by $F_3'( {\hat r}) = (1-3 \hat{r}/L)m\omega^2/2$ and importantly it admits a position dependence. Accordingly, entanglement should be sensitive to the initial momentum as the gradients are different at different locations.
This is indeed observed in Fig.~\ref{fig:entV3_manymom} 
for gravitational potential truncated at the cubic term. 
Furthermore, when the two masses are moving towards each other, $p_0 > 0$ and $\ev{\hat r} < 0$, and consequently, the gradient increases, matching the growing entanglement. 
Conversely, when the masses are moving away, $p_0<0$ and $\ev{\hat r}>0$, the force gradient decreases, matching the slower entanglement gain. Quantitative statements can also be achieved.

\begin{figure}[!t]
\centering
\includegraphics[width=\linewidth]{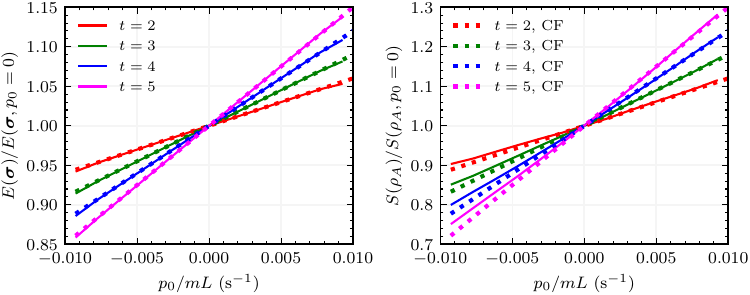}
	\caption{Comparison of the entanglement accumulated with non-zero momentum to entanglement gained from rest, with gravitational potential truncated at the cubic term ($N=3$). The ratios show a linear dependence on the momentum and are very well approximated in the regime of positive $p_0$ (masses moving towards each other) with Eqs.~\!\eqref{eq:ent_gradestimate_S} and~\!\eqref{eq:ent_gradestimate_E}.}
\label{fig:entV3_manytime}
\end{figure}

Fig.~\ref{fig:entV3_manytime} shows experimentally friendly plots of the ratio of entanglement accumulated within time $t$ with non-zero initial momentum to entanglement gained from rest.
The numerically calculated linear dependence (solid lines) can be explained with closed expressions (dotted lines) that we now explain. The force gradients for the quadratic and the cubic interactions are related by the following conversion factor: $F_3'( \hat{r}) = (1-3 \hat{r}/L) F_2'$.
The average factor therefore reads
\begin{equation}
	\frac{\ev{F_3'}}{F_2'} =	1-\frac{3}{L}\ev{ \hat{r}}.
\end{equation}
The initial  momentum $p_0$ is much larger than the momenta generated by gravity and the wave packet, on average, practically follows a free particle trajectory: $\ev{\hat r} \approx r_\text{cl} = -2p_0t/m$. Hence,
\begin{equation}
	\frac{\ev{F_3'}}{F_2'} = 1+ \frac{6p_0}{mL}t \equiv	1 + \epsilon_3(t).
	\label{eq:ampforcegrad3}
\end{equation}

Fig.~\ref{fig:entV3_manymom} shows that for vanishing initial momentum, $p_0 = 0$, the entanglement obtained with cubic and quadratic potentials is practically the same.
We therefore extrapolate that entanglement for non-zero initial momentum is related to entanglement from rest
by a simple function of the conversion factor.
The plots of Fig.~\ref{fig:entV3_manytime} are fitted with
\begin{eqnarray}
	S(\rho_A) &=& \Big[ 1 +  \epsilon_3(t)  \Big] \ S(\rho_A,p_0=0),
	\label{eq:ent_gradestimate_S}
\\	
	E(\bm{\sigma}) &=& \Bigg[ 1 + \frac{1}{2} \epsilon_3(t) \Bigg] \ E(\bm{\sigma},p_0=0),
	\label{eq:ent_gradestimate_E}
\end{eqnarray}
where $S(\rho_A,p_0=0)$ and $E(\bm{\sigma},p_0=0)$ are to be calculated based on the Gaussian covariance matrix due to the second-order potential. Note that the factor of $1/2$ next to $\epsilon_3$ in the logarithmic negativity is causing a departure from the exact conversion factor between the force gradients.
These formulae are in remarkable agreement with the computational results in the regime of positive initial momentum (masses moving towards each other, the regime of experimental interest) and also work quite well for negative initial momenta.
This again affirms that the force gradient is the primary driver of gravitational entanglement.
Furthermore, these closed forms can now be used in a plethora of configurations to estimate the amplification of entanglement for a non-zero initial momentum given entanglement from rest.

\section{Non-Gaussianity is necessary but insufficient}

\begin{figure}[!t]
	\centering
	\includegraphics[width=\linewidth]{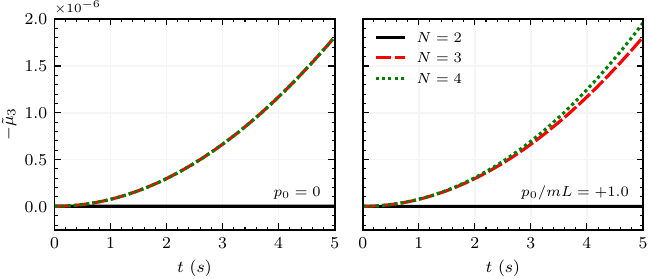}
	\caption{Non-Gaussian reduced mass dynamics.
		The physical situation is as in Fig.~\ref{fig:entV2_manymom}.
		Skewness ($\tilde{\mu}_3$) is computed for the position space distribution. 
		$N$ denotes the order of approximation, see Eq.~\!\eqref{eq:Ham_redmass}.
		The left panel is for masses initially at rest, and the right panel is for masses moving toward each other. The values of $p_0/mL$ in the legends are in multiples of $6.18082292 \times 10^{-3}$ s$^{-1}$.}
	\label{fig:skewness}
\end{figure}

The results presented so far could also be seen as a simple momentum-based witness of non-Gaussianity in a quantum state. Indeed the cubic term is responsible for non-Gaussian evolution that we now quantify in more detail.

Fig.~\ref{fig:skewness} presents the skewness $\tilde{\mu}_3$ in the evolution of the reduced mass wave function $\psi$. While $\tilde{\mu}_3$ vanishes for $N = 2$, as it should, it rises steeply for $N = 3$.
The physical reason is clear from Fig.~\ref{fig:TMGS_special} describing the change of variables between LAB and COM frames. The left end of wave function $\psi$ is attracted towards the center of mass more than the right end. Over time this makes the probability density function negatively skewed, which is indicated by $\tilde{\mu}_3 < 0$.
Fig.~\ref{fig:skewness} also demonstrates the precision of our numerical methods, which even capture marginal contributions 
of the fourth-order term to the skewness.
Note that while in optomechanical systems higher-order moments (skewness and beyond) amplify the entanglement gain~\cite{PhysRevD.105.026011}, in gravitationally coupled systems the skewness has to be accompanied by a non-zero initial momentum: see in Fig.~\ref{fig:entV3_manymom} that for $N=3$ the entanglement gained from rest is the same as that for $N=2$, and only positive initial momentum leads to a higher gain.

While non-Gaussianity is necessary for  a momentum dependence of gravitational entanglement, it does not dominate the entanglement dynamics. 
Fig.~\ref{fig:skewness} shows that skewness is practically the same for the two considered relative momenta whereas Fig.~\ref{fig:entV3_manytime} demonstrates that entanglement entropy accumulated after 5 seconds is different by 30$\%$.
Similarly, skewness is non-zero for initially stationary particles but entanglement dynamics with and without non-Gaussianity look practically the same.
To give quantitative values, we consider the stationary configuration of two Osmium spheres with $m=1$ pg separated by a distance of $L = 2.1$ times their radius.
After an evolution for 5 seconds, the entanglement gain with cubic potential is larger than the entanglement accumulated with quadratic potential by only 
$\approx 0.001, \ 0.002$, and 
$0.003\%$, for an initial spread of $\sigma = 5.00, \ 0.50$, and $0.05$ nm, respectively.
We emphasize that it is the force gradient that plays a pivotal role in entanglement dynamics. Despite practically the same levels of skewness, the force gradient, and hence the entanglement, is significantly higher when the reduced mass wave packet drifts closer to the COM.

We would also like to address the question of whether a simpler method for detecting the third-order coupling exists than based on measurements of entanglement. Indeed, note that solely the mean relative momentum signal could be used for such purposes. For gravity-like interactions, as in Eq.~\eqref{eq:Ham_redmass}, one can use the Ehrenfest's theorem to get the expected relative momentum as
\begin{eqnarray}
\dv{t} \ev{\hat p} 
&=&
\frac{1}{mi\hbar} \ev{ \comm{\hat p}{\hat p^2} }	-  \frac{m\omega^2}{4i\hbar} \sum_{n=0}^{N} \frac{(-1)^n}{L^{n-2}} \ev{ \comm{\hat p}{\hat r^n} }
\nonumber	\\
&=& \frac{m\omega^2}{4} \sum_{n=0}^{N} \frac{(-1)^n}{L^{n-2}} n  \ev{\hat r^{n-1}} ,
\end{eqnarray}
and its second derivative as
\begin{eqnarray}
\dv[2]{t} \ev{\hat p} &=& \frac{m\omega^2}{4} \sum_{n=0}^{N} \frac{(-1)^n}{L^{n-2}} n 
\ \dv{t} \ev{\hat r^{n-1}}	\nonumber	\\
&=&		\frac{m\omega^2}{4} \sum_{n=1}^{N} \frac{(-1)^n}{L^{n-2}} n \ \qty[  \frac{1}{mi\hbar} \ev{ \comm{\hat r^{n-1}}{\hat p^2} }	-  \frac{m\omega^2}{4i\hbar} \sum_{l=0}^{N} \frac{(-1)^{l}}{L^{l-2}} \ev{ \comm{\hat r^{n-1}}{\hat r^{l}}	}  ] 	\nonumber	\\
&=& \frac{\omega^2}{4} \sum_{n=2}^{N} \frac{(-1)^n}{L^{n-2}} n(n-1) \ev{\hat r^{n-2}\hat p + \hat p\hat r^{n-2}}		\nonumber	\\
&=& \omega^2 \qty[ \ev{\hat p} + \frac{1}{4} \sum_{n=2}^{N} \frac{(-1)^n}{L^{n-2}} n(n-1) \ev{\hat r^{n-2}\hat p + \hat p\hat r^{n-2}} ] .
\end{eqnarray}
Hence we derive
\begin{equation}
\frac{1}{\ev{\hat p}} \dv[2]{t} \ev{\hat p} = \omega^2 \qty[ 1 + \frac{1}{4}
    \sum_{n=3}^{N} (-1)^n n(n-1) \frac{\ev{\hat r^{n-2}\hat p + \hat p\hat r^{n-2}}}{L^{n-2}\ev{\hat p}} ].
\end{equation}
Given a non-zero relative momentum $\ev{\hat p} \neq 0$, and a quadratic potential $N=2$, the ratio is equal to a constant: $\frac{1}{\ev{\hat p}} \dv[2]{t} \ev{\hat p}  = \omega^2$.
Any time dependence of this ratio reveals higher-order coupling, thereby indicating an evolution into non-Gaussian states.
In cases where the center of mass is stationary, instead of the relative momentum, the local momentum of any particle could be used:
\begin{equation}
\frac{1}{\ev{\hat p_A}} \dv[2]{t} \ev{\hat p_A} 
= \frac{1}{\ev{\hat p_B}} \dv[2]{t} \ev{\hat p_B}
= \frac{1}{\ev{\hat p}} \dv[2]{t} \ev{\hat p}
= \omega^2 \ev{p},
 \hspace{5mm} 
\forall \ev{\hat p_A + \hat p_B} = 0 .
\end{equation}

Finally, a word on decoherence effects is in place. The common decoherence mechanisms, due to thermal photons and air molecules, have already been studied in the considered setup~\cite{PhysRevLett.119.240401,PhysRevLett.119.240402,npjQI_6.12,Rijavec_2021,Datta_2021}. It was found that the experiment is challenging, but the required coherence times are in principle realizable, e.g., for freely-falling particles in a high vacuum.
The calculations presented here only relax these requirements as the entanglement is improved when the two masses are pushed toward each other. For example, in the configuration considered in this work, the entanglement gain of $E \approx 1.75 \times 10^{-4}$ is relaxed from 5 seconds to 4 seconds with an initial momentum of $p_0/mL \approx +0.022$ s$^{-1}$. Note that an entanglement detection scheme, inspired by quantum neural networks, achieving a precision of $E \sim 10^{-4}$ has recently been put forward in Ref.~\cite{Krisnanda2022_QNN}.

\section{Contributions of higher-order terms}

\begin{figure}[!t]
	\centering
	\includegraphics[width=0.6\linewidth]{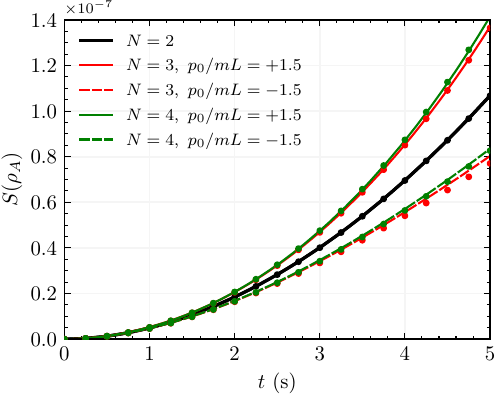}
	\caption{Comparison of entanglement accumulated with the gravitational potential expanded up to quadratic, cubic, and quartic ($N = 2, 3$, and $4$) terms, respectively.
	Solid lines show the results for positive momentum (masses moving toward each other), and dashed lines are for negative momentum.	
	The dots represent the entanglement entropy ($S$) computed with the closed formulae derived in this work.
	Compared to the quadratic case, the cubic term lowers entanglement between particles that move away from each other. 
	Compared to the cubic case, the quartic term adds a positive correction irrespective of the particles moving towards or away from each other.
	The values of $p_0/mL$ in the legends are in multiples of $6.18082292 \times 10^{-3}$ s$^{-1}$.}
	\label{fig:entV4}
\end{figure}

Given that the potential is expanded up to even higher-order terms, their contribution can also be incorporated with an appropriate conversion factor between the force gradients. 
Note that the entanglement entropy in Eq.~\!\eqref{eq:ent_gradestimate_S} is amplified in the same way as the force gradient in Eq.~\!\eqref{eq:ampforcegrad3}. Assuming that this holds for higher-order terms, the entropy amplification factor can be written as
\begin{equation}
\frac{ S(\rho_A) }{ S(\rho_A,p_0=0) } 
= 
\frac{\ev{F_N'}}{F_2'} = 1 +  \sum_{n=3}^{N} \epsilon_n(t) , 
\end{equation}
where the corrections for gravity-like interactions (inverse-square forces) arising due to the $n^\text{th}$ term in the potential expansion is given by
\begin{equation}
	\epsilon_n(t) = \frac{(-1)^n}{2L^{n-2}} \ n(n-1)   \ev{\hat r^{n-2}}.
\end{equation}
Since the gravitational force between two quantum masses is weak, for the estimation of $\ev{\hat r^n}$ we 
approximate the reduced mass wave packet to be a Gaussian, with the average position following classical trajectory and the width following the free evolution:
\begin{equation}
	\abs{\psi_0(r,t)}^2 \approx \frac{1}{\bm{\Delta} r_0 \sqrt{2\pi}} \exp\qty( -\frac{\qty(r-r_\text{cl})^2}{2\bm{\Delta} r_0^2} ),
\end{equation}
where $ \bm{\Delta} r_0^2 = 2 \sigma^2 \qty(1+\omega_0^2t^2)$.
With this approximation one obtains
\begin{eqnarray}
	\ev{\hat r^n}	&=&	\int_{-\infty}^{+\infty} dr \ r^n \abs{\psi(r,t)}^2	\nonumber \\
	&\approx& \frac{1}{\bm{\Delta} r_0 \sqrt{2\pi}}	\int_{-\infty}^{+\infty} dr \ r^n  \exp\qty( -\frac{\qty(r-r_\text{cl})^2}{2\bm{\Delta} r_0^2} ) 		\nonumber	\\
	&=&  \frac{1}{\sqrt{\pi}} \int_{-\infty}^{+\infty} dy \ \qty(r_\text{cl}+\bm{\Delta}r_0\sqrt{2}y)^n	e^{-y^2}	,	\hspace{15mm} :\qty{ y = (r-r_\text{cl})/\bm{\Delta}r_0\sqrt{2} } , \nonumber \\
	&=&	\frac{1}{\sqrt{\pi}} \int_{-\infty}^{+\infty} dy \ \sum_{m=0,1,2,\dots}^{n} {n \choose m} r_\text{cl}^{n-m} \qty(\sqrt{2}\bm{\Delta} r_0 y)^m 	e^{-y^2} \nonumber \\
	&=&	\frac{2}{\sqrt{\pi}} \sum_{m=0,1,2,\dots}^{n} {n \choose m} r_\text{cl}^{n-m} \qty(\sqrt{2}\bm{\Delta} r_0)^m \int_{0}^{\infty} dy \ y^m 	e^{-y^2} \nonumber \\
	&=&  \frac{2}{\sqrt{\pi}} \sum_{m=0,2,4,\dots}^{n} {n \choose m} r_\text{cl}^{n-m} \qty(\sqrt{2}\bm{\Delta} r_0)^m \frac{1}{2} \int_{0}^{\infty} dz \ z^{(m-1)/2} 	e^{-z}	,	\hspace{5mm} :\qty{ z=y^2 }	,\nonumber \\
	&=& \frac{1}{\sqrt{\pi}} \sum_{m=0,2,4,\dots}^{n} {n \choose m} \ r_\text{cl}^{n-m} \ \qty( \sqrt{2}\bm{\Delta} r_0 )^m \ \Gamma \qty( \frac{m+1}{2} ),
\end{eqnarray}
where $\Gamma$ is the gamma function. Hence, the correction terms for $n \ge 3$ are given by
\begin{equation}
	\epsilon_n(t) = \frac{(-1)^n}{2\sqrt{\pi}L^{n-2}} \ n(n-1) \sum_{m=0,2,}^{n-2} {n-2 \choose m} \ r_\text{cl}^{n-m-2} \ \qty( \sqrt{2}\bm{\Delta} r_0 )^m \ \Gamma \qty( \frac{m+1}{2} ).
\end{equation}

Note that the gravitational interaction is already included in $F_2'$, and the present estimation is for the factor between the force gradients of different orders only, $\ev{F_N'}/ F_2'$. Since $\epsilon_n \propto 1/L^{n-2}$, each consecutive term is diminished by a factor of $L$. Hence, a cubic order correction should be sufficient for practical applications in the near future. 
Nevertheless, one can see that the fourth-order correction to entanglement entropy is given by
\begin{equation}
	\epsilon_4(t) =	24 \frac{p_0^2 t^2}{m^2 L^2}    + 12\frac{\sigma^2}{L^2} \qty(1+\omega_0^2t^2).
\end{equation}
Unlike the third-order term, which was sensitive to the direction of momentum, the fourth-order one depends on the momentum squared, leading to a positive correction in both the scenarios of masses moving towards and away from each other.
This prediction is confirmed in Fig.~\ref{fig:entV4}, where we show the entanglement accumulated with the gravitational potential expanded up to the fourth order.
The derived formulae exactly recover the entanglement gain in the regime of positive momentum, and they work quite well in the case of negative momentum.
Note that $\epsilon_4$ also depends on the position spread, hence it might be important even for stationary configurations where the wave packet undergoes a fast expansion.

\section{The case of optomechanically levitated masses}
\label{sec:covmat_levitated}

Until now we have discussed the covariance matrix of two freely falling masses. In the following sections we also discuss the situation when the traps are not opened, i.e., when the masses are held in harmonic potentials [as depicted in Fig.~\ref{fig:system-optomech}]. In such a case the Hamiltonian is
\begin{equation}
\hat H = \frac{ \hat p_A^2 }{ 2m } + \frac{1}{2}m\omega_0^2\hat x_A^2 + \frac{ \hat p_B^2 }{ 2m } +  \frac{1}{2}m\omega_0^2\hat x_B^2 - \frac{Gm^2}{L+\hat x_B - \hat x_A},
\end{equation}
which transforms in the COM frame as [see Chapter~\ref{ch:chapter2} for explicit derivation]
\begin{equation}
\hat H = \hat H_R + \hat H_r = \qty( \frac{ \hat P^2 }{ 4m } + m\omega_0^2\hat R^2 ) +  \qty(  \frac{ \hat p^2 }{ m } +  \frac{1}{4}m\omega_0^2\hat r^2 - \frac{Gm^2}{L+\hat r} ) .
\end{equation}
It can be readily checked that $\phi(R,t=0)$ of Eq.~\eqref{eq:COMwavefunction_t0} is the ground state of $\hat H_R$. Since eigenstates are solutions for the time-independent Schr\"odinger equation, the COM does not evolve in time, and hence the statistical moments are frozen at their initial values: $\bm{\Delta} R^2 = \sigma^2/2$, $\bm{\Delta} P^2 =\hbar^2/2\sigma^2$, $\textbf{Cov}( {R}, {P}) = 0$. When the gravitational interaction is truncated up to the second order in binomial expansion,
\begin{eqnarray}
\hat H_r &=& \frac{ \hat p^2 }{ m } +  \frac{1}{4}m\omega_0^2\hat r^2 - \frac{Gm^2}{L+\hat r}
\nonumber	\\
&=& \frac{ \hat p^2 }{ m } +  \frac{1}{4}m\omega_0^2\hat r^2 - \frac{1}{4}m\omega^2( L^2 - L\hat r + \hat r^2 ) ,	\hspace{1cm} : \qty{ \omega^2 = 4Gm/L^3 },
\nonumber	\\
&\equiv& \frac{ \hat p^2 }{ m } +  \frac{1}{4}m\omega_0^2\hat r^2 - \frac{1}{4}m\omega^2L^2 +  \frac{1}{4}m\omega^2L\hat r +  \frac{1}{4}m\qty(\omega_0^2-\omega^2)\hat r^2,
\end{eqnarray}
the solutions of Ehrenfest's differential equations imply
\begin{eqnarray}
\bm{\Delta} r^2 &=& \frac{2\sigma^2}{\omega_0^2-\omega^2} \qty[ \omega_0^2 -  \omega^2 \cos[2](\sqrt{\omega_0^2-\omega^2} t) ],    
	\\
\bm{\Delta} p^2  &=&   \frac{\hbar^2}{8\sigma^2\omega_0^2} \qty[ \omega_0^2 - \omega^2  \sin[2](\sqrt{\omega_0^2-\omega^2} t) ], 
	\\
\textbf{Cov}( {r}, {p}) &=& \frac{\hbar\omega^2}{4\omega_0\sqrt{\omega_0^2-\omega^2}} \sin(2\sqrt{\omega_0^2-\omega^2} t).
\end{eqnarray}
Note that we get the same results after replacing $\omega$ by $i\sqrt{\omega_0^2-\omega^2}$ in the solutions for the freely falling case in Eqs.~\eqref{eq:var_r2-freefall}, \eqref{eq:var_p2-freefall}, and~\eqref{eq:cov_rp-freefall}. The resultant covariance matrix is
\begin{eqnarray}
	\bm{\sigma}_{00}  &=&  \frac{\hbar}{4m\omega_0} \qty[ 2 + \frac{\omega^2}{\omega_0^2-\omega^2} \sin[2](\sqrt{\omega_0^2-\omega^2} t) ],
	\\
	\bm{\sigma}_{02}   &=& - \frac{\hbar\omega^2}{4m\omega_0(\omega_0^2-\omega^2)}   \sin[2](\sqrt{\omega_0^2-\omega^2} t),
	\\
\bm{\sigma}_{11} &=& \frac{m\hbar\omega_0}{4} \qty[ 2 - \frac{\omega^2}{\omega_0^2} \sin[2](\sqrt{\omega_0^2-\omega^2} t) ],
	\\
	\bm{\sigma}_{13} &=&  \frac{m\hbar\omega^2}{4\omega_0}   \sin[2](\sqrt{\omega_0^2-\omega^2} t),
	\\
	\bm{\sigma}_{01} &=&	\frac{\hbar\omega^2}{8\omega_0\sqrt{\omega_0^2-\omega^2}} \sin(2\sqrt{\omega_0^2-\omega^2} t),
	\\
\bm{\sigma}_{03} &=& - \frac{\hbar\omega^2}{8\omega_0\sqrt{\omega_0^2-\omega^2}} \sin(2\sqrt{\omega_0^2-\omega^2} t).
\end{eqnarray}
Each element contains a sinusoidal term: the covariance matrix, and hence the entanglement, oscillates periodically with a time period 
$\tau = \pi/2\sqrt{\omega_0^2-\omega^2}$.
Note that the weakness of gravity, $\omega \ll \omega_0$, implies $\tau \approx \pi/2\omega_0$, which recovers the approximate results of Refs.~\cite{npjQI_6.12,phdthesis-Tanjung,bscthesis-GuoYao}. The logarithmic negativity and the entanglement entropy can be derived in an exact closed form:
\begin{eqnarray}
E(\bm{\sigma}) &=& - \frac{1}{2} \log_2\qty( 1+2\mathcal{E}-2\sqrt{\mathcal{E}^2+\mathcal{E}} ),
\\
S(\rho_A) &=&	 \qty(\mathcal{S}+\frac{1}{2}) \log_2\qty(\mathcal{S}+\frac{1}{2})	-	\qty(\mathcal{S}-\frac{1}{2}) \log_2\qty(\mathcal{S}-\frac{1}{2}),
\end{eqnarray}
with
\begin{eqnarray}
\mathcal{E} &=& \frac{ \omega^4 }{ 4\omega_0^2(\omega_0^2-\omega^2) }  \sin^2 \qty( \sqrt{\omega_0^2-\omega^2} t  ) ,
\\
\mathcal{S}  &=& \frac{1}{2} \sqrt{ 1 + \frac{\omega^4}{4\omega_0^2\qty(\omega_0^2-\omega^2)} } \sin^2 \qty( \sqrt{\omega_0^2-\omega^2} t  ) .
\end{eqnarray}
Under the approximation $\omega \ll \omega_0$, they reduce to
\begin{eqnarray}
E(\bm{\sigma}) &\approx& \frac{\omega^2}{2\ln(2) \omega_0^2} \sin(\omega_0t) ,
\label{eq:NegHarmTrap_T0}
\\
S(\rho_A) &\approx&  \frac{\omega^4}{16 \omega_0^4}\sin^2(\omega_0t) \qty[ 1 - 4\log_2\qty( \frac{\omega}{2 \omega_0} \sqrt{\sin(\omega_0t)} )  ] ,
\end{eqnarray}
which exactly matches with previous finding for the entanglement negativity~\cite{npjQI_6.12}.

\section{Compatibility with arbitrary central interactions}

While our discussions up to now were mainly focused on the gravity-induced entanglement, the methods we have presented are applicable more generally. First of all, they hold not just for gravity-like, but for arbitrary central interactions.
We just need to expand the potential in a binomial series similar to Eq.~\eqref{eq:Ham_redmass}. Note that the entanglement is characterized by the parameters $\omega$ and $\epsilon_3$. 
As we derived, for identical masses coupled via Newtonian gravity:
\begin{equation}
\omega^2 = \frac{4Gm}{L^3},
\hspace{1cm}
\epsilon_3(t) = \frac{6p_0}{mL} t.
\end{equation}
The general rule is quite simple. Once the potential is expanded in a binomial series of the relative displacement, the coefficient of $r^2$ is to be compared with $-m\omega^2/4$, and $\epsilon_3$ is to be calculated by comparing the force gradients as $\ev{F_3'}/F_2' = 1+\epsilon_3$. One can then verify that for the Coulomb potential between charges $q_1$ and $q_2$ embedded into the masses we obtain
\begin{equation}
\omega^2 = \frac{4q_1q_2\alpha \hbar c}{e^2mL^3},
\hspace{1cm}
\epsilon_3(t) = \frac{6p_0}{mL} t.
\end{equation}
where $\alpha$ is the fine structure constant and $e$ is the electronic charge. As we shall see towards the end of this thesis, this can be used to quantify the entanglement gain in the Rutherford scattering experiment. For an arbitrary central interaction with a potential
\begin{equation}
	V(x_A,x_B) = -\frac{C}{(X+x_B-x_A)^{j}},
\end{equation}
we obtain
\begin{equation}
	\omega^2 = \frac{2j(j+1)C}{mX^{j+2}}, 
\hspace{1cm}
\epsilon_3(t) = \frac{2(j+2)p_0}{mX} t.
\end{equation}

There are situations where the force is known, but solving for the potential is quite difficult, and sometimes uncertain due to non-unique boundary conditions. In that case the general rule would be to expand the force in a binomial series and compare the coefficient of $r$ with $m\omega^2/2$. For example, if the force is of the form
\begin{equation}
	F(x_A,x_B) = -\frac{C}{(X+x_B-x_A)^{j}},
\label{eq:CentralForce-General}
\end{equation}
we obtain
\begin{equation}
	\omega^2 = \frac{2jC}{mX^{j+1}},
\hspace{1cm}
\epsilon_3(t) = \frac{2(j+1)p_0}{mX}t.
\end{equation}
Note that the functional forms of $\epsilon_3$ written in this section are valid only for weak interactions, because the expectation value of relative displacement is approximated with free evolution. For stronger potentials one has to take a step back and use $\epsilon_3 = -3\ev{\hat r}/L$, where $\ev{\hat r}$ has to be approximated either analytically or numerically.

\section{Gravity and Casimir acting side by side}

The methods we have established so far also work for multiple central forces acting simultaneously. If we write the interaction as a sum
\begin{equation}
V(x_A,x_B) = \sum_k  V_k(x_A,x_B),	\hspace{5mm} : \qty{
F(x_A,x_B) = \sum_k  F_k(x_A,x_B) },
\end{equation}
the equivalent $\omega$ characterizing the Gaussian covariance matrix is simply given by a Pythagoras-like theorem, and the equivalent $\epsilon_3$ governing the entanglement amplification due to the cubic-order term is calculated as a weighted sum:
\begin{equation}
	\omega^2 = \sum_k \omega^2_k, 
\hspace{1cm}
\epsilon_3(t) = \frac{1}{\omega^2} \sum_k \omega_k^2 \epsilon_3(t)_k,
\end{equation}
where $\omega_k$ and $\epsilon_3(t)_k$ characterise the individual interactions.

\begin{figure}
\centering
\includegraphics[width=\linewidth]{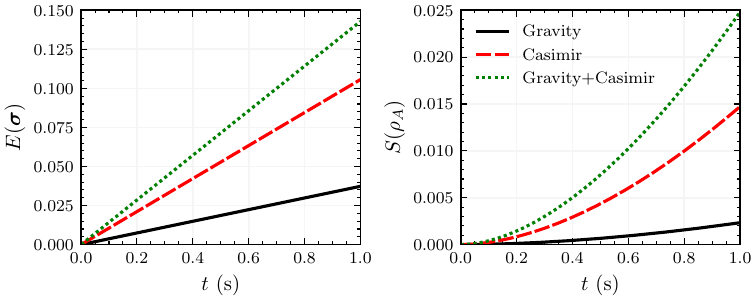}
\caption{Gain of Gaussian entanglement between two Osmium spheres due to gravity, Casimir, and due to both of them acting side by side. The mass of each sphere is $1$ mg, and  they are prepared  in Gaussian states with a spread of $1$ pm with their centers separated by a distance which is $2.1$ times their radius. The entanglement quantifiers $E$ and $S$ are calculated based on the Gaussian covariance matrix derived in this work.}
\label{fig:Gravity+Casimir_m1mg_s1pm_L2.1R}
\end{figure}
This is particularly useful from an experimental point of view as, in practice, it might be difficult to screen all interactions except gravity. For example, the gravitational and the Casimir interaction are likely to be present side by side. The Casimir energy due to interaction between the surfaces of two spheres, under proximity force approximation ($L \gtrsim 2R_0$) is~\cite{PhysRevLett.99.170403}
\begin{equation}
V(x_A,x_B) = -\frac{\pi^3 \hbar c R_0}{1440(L-2R_0 + x_B-x_A)^2},
\end{equation}
where $R_0$ is the radius of each sphere. This implies
\begin{equation}
	\omega^2  =  \frac{\pi^3\hbar cR_0}{120m(L-2R_0)^4},
\hspace{1cm}
\epsilon_3(t)  =  \frac{8p_0}{m(L-2R_0)} t.
\end{equation}
For example, consider the case of two Osmium spheres of mass $1$ mg prepared in Gaussian states with $\sigma = 1$ pm at a separation of $L=2.1R_0$. The gain of Gaussian entanglement due to gravity, Casimir, and due to both of them acting simultaneously, is shown in Fig.~\ref{fig:Gravity+Casimir_m1mg_s1pm_L2.1R}. 
Note that the entanglement due to gravity and Casimir acting together is not equal to the sum of the entanglement they would generate separately.

\begin{figure}
\centering
\includegraphics[width=\linewidth]{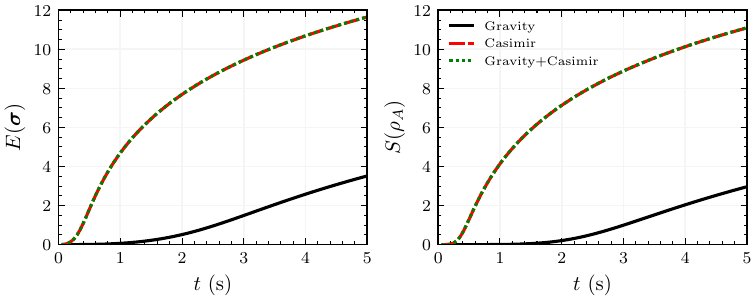}
\caption{Same as Fig.~\ref{fig:Gravity+Casimir_m1mg_s1pm_L2.1R}, but for $m = 100 \ \mu$g and $\omega_0 = 100$ kHz~\cite{npjQI_6.12}, and $L=2.1R_0$.}
\label{fig:Gravity+Casimir_m100ug_w0100kHz_L2.1R}
\end{figure}
Considering a configuration of $m = 100 \ \mu$g, $\omega_0 = 100$ kHz, and $L=2.1R_0$, studied in~\cite{npjQI_6.12}, in Fig.~\ref{fig:Gravity+Casimir_m100ug_w0100kHz_L2.1R} we show the entanglement gain with gravity, Casimir, and with both of them combined. Here the Casimir entanglement practically dominates everything, thereby signifying the need for its consideration, and hence the utility of our framework, in experiments aimed at observing the gravitational entanglement between massive objects.

\section{Galilean relativity and a drifting COM}
\label{sec:GalileanTransformation}

We made a change of reference frames so as to dissect the bipartite evolution into two independent single-particle dynamics. The first one is the free evolution of the COM, and the second one is the evolution of reduced mass in gravitational potential. Under the assumption that the two spheres are imparted with equal and opposite momentum, the COM is stationary on average. While this simplifies our theoretical framework substantially, such a configuration may be difficult to achieve in an actual experiment. It is much easier to push one of the masses while the other one is kept at rest. In such a case the COM moves rectilinearly with a constant velocity. The Galilean principle of relativity demands that the laws of non-relativistic physics must be invariant in all inertial frames of reference. Consequently, the centered moments of the moving COM should evolve in the same way as for the stationary COM~\cite{GaussEvolFreeSpace_SMBlinder}. This is readily cross-checked as we get the exact same correlation and variances after incorporating a non-zero momentum in the initial conditions for solving COM Ehrenfest's equations. In conclusion, a uniformly moving COM has no role in generating quantum (or, for that matter, classical) correlations. Only the relative momentum matters, and as long as it remains the same, the individual momenta can be tweaked as per convenience.

\section{Numerical details}

Numerical calculations are performed in natural units of $c=1$, and hence the conversion constant $\hbar c = 197.3269804$ keV pm. 
The density of Osmium and Silica is $22.5872$ and $2.65$ g/cm$^3$, respectively. 
An error analysis implies that, in the numerical time evolution of the reduced mass wave function, a grid size of $\lesssim 0.2$ pm with a time step of $\lesssim 10 \ \mu$s is required to maintain accuracy in the extreme cases of the largest momentum considered in this work. Accordingly, we set a grid size of $0.1$ pm and a time step of $5 \ \mu$s throughout this work.
Note that the first term in the gravitational potential of Eq.~\!\eqref{eq:Ham_redmass} is just a constant energy offset, which only contributes to an irrelevant global phase in the quantum dynamics. Moreover, this term is the most prominent of all magnitude-wise. We have therefore ignored it in numerical simulations so as to maximally utilize the precision for the (relevant) higher-order terms.

\section{Summary}

We have shown that experiments aimed at the observation of gravitational entanglement can also be used as precision tests of gravitational coupling.
In particular, entanglement dependence on the relative momentum of interacting particles indicates third-order coupling. 
Furthermore, the amount of entanglement accumulated in a fixed time interval grows linearly with the relative momentum when the particles are pushed toward each other.
We presented a closed expression for the amount of entanglement as a function of relative momentum based on the derived exact covariance matrix for Gaussian dynamics extended to the third-order and higher-order couplings. The methods introduced apply to arbitrary central interaction, as well as to arbitrary number of central interactions present side by side.

%% file: science/ch5-MOND.tex
\chapter{Probing Galactic Rotation with Entanglement of Microspheres}
\label{ch:chapter5}

While a wide variety of astrophysical and cosmological phenomena suggest the presence of Dark Matter, all evidence remains via its gravitational effect on the known matter. As such, it is conceivable that this evidence could be explained by a modification to gravitation or concepts of inertia. Various formulations of modified gravity exist, each giving rise to several non-canonical outcomes. This motivates us to propose experiments searching for departures from (quantum) Newtonian predictions in a bipartite setting with gravitational accelerations $\lesssim 10^{-10}$ m/s$^2$, i.e., where the effective force needs to be stronger than Newtonian to account for the Dark Matter effects. Since quantum particles naturally source weak gravitation, their non-relativistic dynamics offers opportunities to test this small acceleration regime. We show that two nearby quantum particles accumulate significantly larger entanglement in modified gravity models, such as the Modified Newtonian Dynamics (MOND). We demonstrate how the temperature can be fine-tuned such that these effects are certified simply by witnessing the entanglement generated from uncorrelated thermal states, eliminating the need for precise noise characterization.

\section{Introduction}

\begin{figure}
\centering
\includegraphics[width=0.5\linewidth]{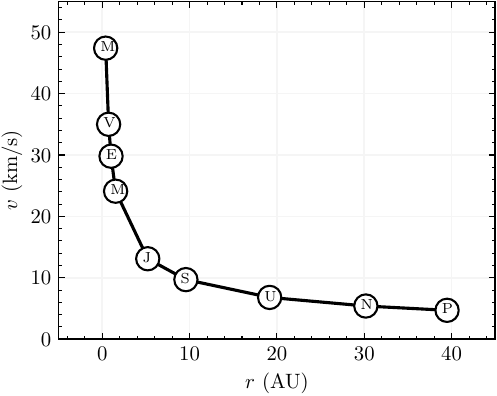}
\caption{Planetary rotation curves in the solar system. $v$ denotes the orbital speed of the planet with an average distance $r$ from the Sun. Observational data taken from the ``NASA Planetary Fact Sheet''~\cite{NASA-PlanetSheet}.}
\label{fig:OrbSpeeds_Planets}
\end{figure}

Consider the motion of planets in our solar system, where most of the mass is concentrated at the center, i.e., in the Sun.
For an arbitrary planet located at an average distance $r$, a balancing of centrifugal and gravitational forces implies that orbital velocity should fall as the square root of the distance: $v \propto 1/\sqrt{r}$. Famously known as the Keplerian decline, this has been observed to hold true in our solar system [see Fig.~\ref{fig:OrbSpeeds_Planets} for actual data].

Spiral galaxies have a lot in common with the solar system. Most of their mass is also concentrated towards the center, but the stars do not show any asymptotic Keplerian decline~\cite{sanders_2010}. 
The orbital speeds never fall, and the rotation curves saturate beyond the region where most of the mass is concentrated~\cite{Babock_AndromedaGalaxy}.
Consequently, the stars in the outer regions appear to be orbiting so fast that, with such small gravity of the visible matter, they should fly away.
Since we do not see it happening, there is much more gravity in the system than we expect based on the observed mass.
Where is that extra gravity coming from? 

One of the dominant proposals along this line is the existence of an invisible matter distributed throughout the galaxies~\cite{Zwicky2009}. This invisible matter, called Dark Matter (DM), is postulated to generate the missing gravity.
Despite being the most widely accepted explanation, the DM has not been directly detected or confirmed by any experiment~\cite{DarkMatter-RMP}.
Thinking on a different trajectory, an Israeli physicist Mordehai Milgrom postulated that there is no DM, and our understanding of gravity is incomplete.
He noticed that the galactic rotation curves saturate beyond a point where the acceleration due to gravity falls below $a_0 \approx 1.2 \times 10^{-10}$ m/s$^2$~\cite{Milogram_MOND,Famaey_2005_MOND_in_Milky_Way,Gentile_2011_Things_about_MOND}, and proposed proposed to modify Newton's law of gravity in this regime such that it explains the observations based on just the known mass.
Based on empirical assumptions, a new law of gravitation was formulated where the force falls as distance squared, until for tiny accelerations, $\lesssim a_0$, where it plateaus to fall as the distance. This idea is famously known as the Modified Newtonian Dynamics (MOND)~\cite{Milogram_MOND}.
While the MOND conflicts with a lot of fundamental physics, no alternative matches its predictive power.
Accordingly, there has been an effort to formalize MOND by its
derivation from underlying fundamental principles~\cite{Bekenstein_Lag4MOND,Bekenstein-TeVes,Zlosnik-TeVes}.

Why are we discussing this topic because of the recent development in optomechanics where physicists have successfully cooled Silica spheres of radius $\approx 75$ nm down to a temperature of $\sim 10 \ \mu$K~\cite{Sciene-Aspelmeyer}. It can be easily seen in Fig.~\ref{fig:accmond-aspelmeyer} that two such masses placed nearby generate an internal acceleration deep into the limit where MOND is supposed to be relevant.

A modified gravitational coupling between the two masses will have a noticeable impact on their gravitational entanglement.
Fortunately, the MOND potential is of central nature, and hence compatible with the entanglement machinery developed in the previous chapter. The MONDian correlations are significantly larger than their Newtonian counterpart, which relaxes the difficulties in their experimental observation. 
Going one step further, we develop a strategy where the temperature can be fine-tuned such that departures from Newtonian gravity are certified simply by witnessing the entanglement produced starting from thermal states.

\section{Modified Newtonian Dynamics}

The basic idea of the MOND theory is the following: just like Newtonian gravity is an approximation of General Relativity when the gravitational field is not too strong, it might just be an approximation of an underlying theory when the accelerations are not too small.
The MOND theory re-investigates our understanding of gravity at tiny accelerations.
For $a \gtrsim a_0$ it postulates the Newtonian dynamics where the gravitational force falls as distance squared, but for $a \lesssim a_0$, the force plateaus to fall as the distance.
There are two ways of interpreting MONDian corrections. The first way is via modification of Newton's second law:
\begin{equation}
F =  m \ a \  \tilde \mu\qty(\frac{a}{a_0}),
\label{eq:Fm_MOND}
\end{equation}
where the exact form of function $ \tilde \mu$ is unknown and is not determined by the theory, but consistency with astronomical observations demands
\begin{equation}
 \tilde \mu\qty(\frac{a}{a_0}) =
\begin{cases}
a/a_0, \hspace{1cm} a \ll a_0,
\\
1, \hspace{1.6cm} a \gg a_0.
\end{cases}
\end{equation}
The interpolating function beyond these limits is weakly constrained~\cite{Famaey_2005_MOND_in_Milky_Way,Gentile_2011_Things_about_MOND}.
This modification applies to any force, not necessarily gravitational, and demands that the applied force is related to acquired acceleration with a proportionality factor that differs from the inertial mass.
This assertion has been experimentally tested with torsion pendula and agreement with Newtonian mechanics has been confirmed down to accelerations on the order $10^{-14}$ m/s$^2$~\cite{Gundlach2007}.

The other interpretation of MONDian corrections involves modifying Newton's law of universal gravitation:
\begin{equation}
\frac{1}{ \tilde \mu} \, G \frac{m M}{r^2} = a \, m.
\label{eq:grav_MOND}
\end{equation}
This time the second law of dynamics is held Newtonian, but the force is accordingly adjusted.
In principle, one could also consider modifying both laws with suitable interpolating functions, but this is beyond the scope of this thesis.


It so happens that the numerical value of the critical MOND acceleration is closely related to that of other cosmological quantities, e.g., $a_0 \approx cH_0/2\pi \approx (c^2/2\pi) \sqrt{\Lambda/3}$,
where $H_0 \approx 70$ (km/s)/Mpc is the Hubble's constant, and $\Lambda \approx 2 \times 10^{-35}$ s$^{-2}$ is the cosmological constant. These relations indicate that MOND could be derived from some underlying fundamental principles.
Bekenstein and Milgrom in 1984 proposed such a derivation and showed that the following Poisson-like equation generates MONDian correlations~\cite{Bekenstein_Lag4MOND}:
\begin{equation}
\nabla \cdot \qty[  \tilde  \mu \nabla \Phi ] = 4\pi G \rho,
\label{eq:Poission-likeMOND}
\end{equation}
where $\Phi$ is the gravitational potential and $\rho$ denotes mass density.
Note that $ \tilde \mu$ also involves gradient of the potential (acceleration).
Such a non-linearity gives rise to a peculiar `external field effect', which we shall discuss towards the end of this chapter.

\subsection{Gravitational field outside a spherical mass}

We shall now derive the potential of a spherical mass distribution in the MOND theory.
To start with let, us integrate the Poisson-like nonlinear equation [Eq.~\eqref{eq:Poission-likeMOND}] over the volume that contains the mass distribution. For a point located at a distance $r$ from the center of the sphere of mass $m$,
\begin{equation}
    \int d^3r \ \nabla \cdot \qty[  \tilde \mu \nabla \Phi ] = \int d^3r \ 4\pi G \rho = 4\pi G m.
\end{equation}
One can now use the divergence theorem to rewrite this as
\begin{equation}
   \oint \qty[   \tilde \mu \nabla \Phi ] \cdot \hat n \ d^2r = 4\pi G m, 
\end{equation}
Given that we assume spherical symmetry,
\begin{eqnarray}
   &&     \tilde  \mu \dv{\Phi}{r} \times 4 \pi r^2 = 4\pi G m, 
   \nonumber    \\
 \implies  &&      \tilde  \mu \dv{\Phi}{r}   = \frac{G m}{r^2}
 \nonumber  \\
 \implies    &&    \frac{a}{a_0} \times a = \frac{Gm}{r^2}	\hspace{1cm}: \qty{ a \ll a_0, \ \text{\&} \ a = \abs{d\Phi/dr} },
 \nonumber  \\
\implies    &&    a = \frac{\sqrt{Gma_0}}{r},
\end{eqnarray}
where $a$ is the gravitational field magnitude in the MOND approach.
The corresponding gravitational potential is given by 
$\Phi =  \sqrt{Gma_0} \ \ln(r)$,
where the constant of integration has been ignored as it is equivalent to an energy offset and contributes only to an irrelevant global phase factor in the quantum dynamics.
It can be easily seen that the derived functional form of the acceleration predicts the constancy of orbital speeds in spiral galaxies. Considering a huge mass $M$ concentrated at the center of a galaxy, one can equate the acceleration due to gravity with the centripetal acceleration at a distance to get
$v^4 = GMa_0$.

Though the MOND was designed to explain the flatness of rotation curves, it explains the Tully-Fisher relation in rotation-supported galaxies~\cite{Sanders_2015}, as well as the Jackson-Faber relation in pressure-supported galaxies~\cite{Famaey_2012}.
Such versatility has kept the MOND theory going despite its conflicts with various laws of fundamental physics.

\section{Gravitation between microspheres}

\begin{figure}[!t]
\centering
\includegraphics[width=0.5\linewidth]{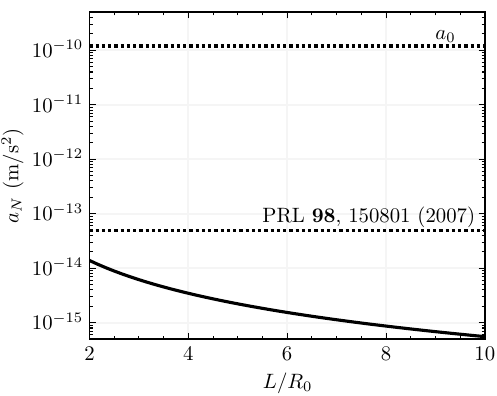}
\caption{Acceleration due to gravity between two Silica spheres of radius $R_0 = 75$ nm separated by a distance $L$~\cite{Sciene-Aspelmeyer}. The dotted line at $1.2 \times 10^{-10}$ m/s$^2$ is the MOND critical acceleration~\cite{Milogram_MOND}, and the one at $5 \times 10^{-14}$ m/s$^2$ is the minimum acceleration down to which there is no experimental evidence for any deviation from the law of inertia~\cite{Gundlach2007}.
}
\label{fig:accmond-aspelmeyer}
\end{figure}

In a recent development the Aspelmeyer group in Vienna has successfully cooled Silica microspheres with radius $\approx 75$ nm down to a temperature $\sim 10 \ \mu$K~\cite{Sciene-Aspelmeyer}.
As shown in Fig.~\ref{fig:accmond-aspelmeyer}, when such masses are separated by a distance that is a few times their radius, the internal acceleration due to gravity is multiple orders of magnitude smaller than the MOND critical limit.
We expect MONDian effects to dominate the gravitational coupling between these quantum particles.
This should noticeably alter the entanglement gain, and in this chapter we study the corresponding implications. The gravitational potential energy of identical particles of mass $m$, separated by a distance $L$, can be naively written as
\begin{equation}
V(r) = m\Phi = m\sqrt{Gma_0} \ \ln\qty( L + r  ),
\end{equation}
where $r$ now represents the relative displacement from their initial positions.
However, such a completion violates the law of equal and opposite action and reaction: the force on mass $m_2$ attracted by $m_1$ is $\sim m_2\sqrt{m_1}$ whereas that on mass $m_1$ attracted by $m_2$ is $\sim m_1\sqrt{m_2}$.
This issue is rectified if one solves the full nonlinear Poisson-like equation governing the bipartite dynamics in MONDian gravity.
In the non-relativistic limit of all MOND models, the gravitational potential energy of two identical particles of mass $m$ is actually given by~\cite{Milgrom_2014,MOND-differentfactor,TwoBodyForce-MOND}:
\begin{eqnarray}
	V &=& \frac{2}{3}  \sqrt{G a_0} \ \qty[ (m_1+m_2)^{3/2} - m_1^{3/2} - m_2^{3/2} ] \ \ln(L+r) 
\nonumber	\\
&=& \frac{2}{3}  \sqrt{G a_0} \ \qty[ (m+m)^{3/2} - m^{3/2} - m^{3/2} ] \ \ln(L+r) , \hspace{1cm} : \qty{m_1=m_2=m},
\nonumber	\\
&=&  \frac{4}{3} \qty( \sqrt{2} - 1 ) \ m \sqrt{G m a_0} \ \ln(L+r) .
	\label{eq:ForceIdenticalMasses}
\end{eqnarray}
This is very different from the usual Newtonian potential, and underlies the differences in observable quantities.
Given that the potential is central, we
truncate it up to the second-order in a binomial series $r/L$:
\begin{equation}
V(r) \approx  \frac{4}{3} \qty( \sqrt{2} - 1 ) \	m\sqrt{Gma_0} \ \qty(  \ln(L) +  \frac{r}{L} - \frac{r^2}{2L^2} ) .
\end{equation}
In accordance with the methodology developed in the previous chapter, the resultant Gaussian covariance matrix is (single-handedly) characterised by the parameter $\omega$. 
For the Newtonian gravity it was $\omega_N^2 = 4Gm/L^3$, and for MONDian potential expanded above we get
\begin{equation}
\omega_M^2 =  \frac{8}{3} \qty( \sqrt{2} - 1 ) \ \frac{\sqrt{Gma_0}}{L^2 } .
\end{equation}

\begin{figure}[!t]
\centering
\includegraphics[width=0.5\linewidth]{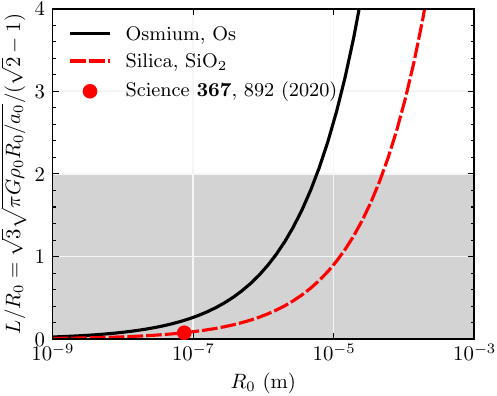}
\caption{RHS of Eq.~\eqref{eq:MOND_LR0} for Osmium and Silica. $R_0$ is the radius of a sphere with density $\rho_0$, $G$ is the Newton's constant, and $a_0$ is the MOND critical acceleration~\cite{Milogram_MOND}. The shaded area is physically inaccessible as two spheres overlap each other for $L \le 2R_0$.
In the regions above the curves MONDian entanglement is larger than the Newtonian entanglement.}
\label{fig:MOND_LR0}
\end{figure}

The entanglement increases with the force gradient, and hence with $\omega$ as well ($\omega^2 = -2\ev{F_2'}/m$).
Given the same initial state, the MONDian entanglement will be significantly larger than its Newtonian counterpart when $\omega_M \gg \omega_N$. 
These can be readily equated to see that this happens for
\begin{equation}
L \gg  \frac{3}{2(\sqrt{2}-1)} \sqrt{\frac{Gm}{a_0}}.
\end{equation}
The Newtonian acceleration due to gravity is given by $a_N = Gm/L^2$, and the above condition is (roughly) equivalent to $a_N \ll a_0/10$. We write the separation between the two masses in multiples of their radius and obtain
\begin{equation}
\frac{L}{R_0} \gg \frac{\sqrt{3}}{\sqrt{2}-1} \sqrt{ \frac{\pi G \rho_0 R_0}{a_0} } 
\label{eq:MOND_LR0}
\end{equation}
as the condition for MONDian entanglement to considerably exceed the Newtonian entanglement,
where $\rho_0 = 3m/4\pi R_0^3$ is the material density of the spheres used in the experiment. 
In Fig.~\ref{fig:MOND_LR0}, we show the RHS of Eq.~\eqref{eq:MOND_LR0} for Osmium  (the densest naturally occuring material) and Silica.
The configurations located well above the black and red lines are in the deep-MOND regime, which is the case most of the time as the region shaded in grey is physically inaccessible (the two masses touch each other for $L = 2R_0$).
In the following section we discuss how the gain of entanglement can serve as a tool to probe the modifications in gravity at small accelerations.

\section{Gravitational entanglement}

\begin{figure}[!t]
\centering
\includegraphics[width=\linewidth]{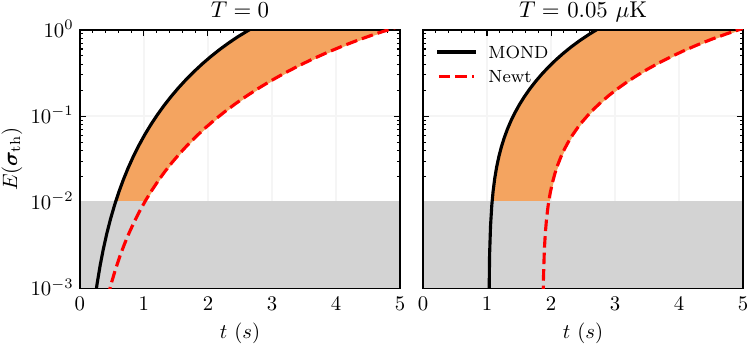}
\caption{Entanglement between two freely falling Osmium spheres of radius $R_0 = 250 $ nm separated by a distance of $L = 2.5R_0$.  The initial Gaussian state is prepared by cooling the two masses into \emph{thermal states} of identical harmonic traps of frequency $\omega_0 = 25$ kHz. $t$ is the time, and $E$ denotes the entanglement negativity beginning with the thermal state covariance matrix $\bm\sigma_\text{th}$ at a temperature $T$.
The region shaded in grey represents an entanglement negativity $E < 0.01,$ which is not accessible with current technologies~\cite{Palomaki-Science}. The orange shaded region represents all other modified gravity models having a force gradient between the Newtonian and the MONDian limits.}
\label{fig:entMOND_Os}
\end{figure}

In Fig.~\ref{fig:entMOND_Os} we show the entanglement between two freely falling Osmium spheres of radius $R_0 = 250 $ nm separated by a distance of $L = 2.5R_0$.
The initial Gaussian state is prepared by cooling the two masses in identical harmonic traps of frequency $\omega_0 = 25$ kHz. 
Note that the acceleration due to gravity is $\approx 2.5 \times 10^{-13}$ m/s$^2$, which is very well in the deep MOND regime.
The region shaded in grey represents $E < 0.01$, where the signal is too weak to be detected with current technologies~\cite{Palomaki-Science}.
The left panel is the ideal case of $T=0$, where it can be seen that the MONDian entanglement is much higher than what is accumulated by Newtonian gravity.
This is beneficial from an experimental point of view as the entanglement would also be detectable with noisy measurements.
The Newtonian entanglement is below detection capabilities for a short time window of $0.5 \lesssim t \lesssim 1.0$ seconds. The panel on the right shows a more realistic case when the experiment is performed at a finite temperature of $T = 0.05 \ \mu$K. There is no entanglement in the Newtonian theory for $1 \lesssim t \lesssim 2$ seconds, but the MONDian gravity generates a strong detectable signal already at $1$ second. Note that such an experiment does not require entanglement quantification but only a witness: any statistically significant detection of entanglement within a time window of $1 \lesssim t \lesssim 2$ seconds would indicate the presence of MONDian gravity.

\begin{figure}[!t]
\centering
\includegraphics[width=\linewidth]{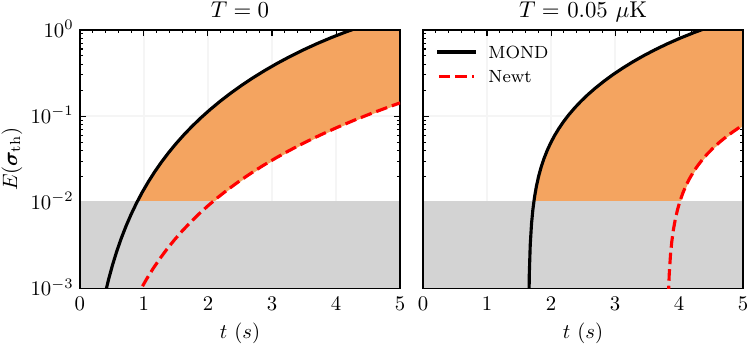}
\caption{Same as Fig.~\ref{fig:entMOND_Os} but for two freely falling Silica spheres with $R_0 = 500 $ nm, $L = 2.5R_0$, and $\omega_0 = 25$ kHz, i.e., the parameters from Ref.~\cite{Sciene-Aspelmeyer}. The region shaded in grey represents an entanglement negativity $E < 0.01,$ which is not accessible with current technologies~\cite{Palomaki-Science}. The orange shaded region represents all other modified gravity models having a force gradient between the Newtonian and the MONDian limits.}
\label{fig:entMOND_Si}
\end{figure}

In another example in Fig.~\ref{fig:entMOND_Si}, we show a similar configuration of two freely falling Silica spheres with $R_0 = 500 $ nm, $L = 2.5R_0$, and $\omega_0 = 25$ kHz. At $T=0$ the MONDian entanglement is detectable for $1 \lesssim t \lesssim 2$ seconds, but the Newtonian entanglement is below detection limits till $t = 2$ seconds. At $T = 0.05 \ \mu$K, the Newtonian entanglement vanishes for $t \lesssim 4$ seconds, and any entanglement detection within a time period of $2 \lesssim t \lesssim 4$ seconds would indicate the presence of MONDian gravity.
The idea is that given any configuration, the temperature can be easily adjusted such that there is a detectable entanglement exclusively in the MONDian gravity.

One can also imagine an experiment where the two masses are kept in harmonic traps. In such a case, the entanglement oscillates with an amplitude [see Eq.~\eqref{eq:NegHarmTrap_T0}]
\begin{equation}
E_0(T) \approx \max \Big[ 0, \frac{\omega^2}{2\ln(2) \omega_0^2} - \log_2(2\bar{n}+1)  \Big] ,
\end{equation}
where $\omega = \omega_N (\omega_M)$ for the Newtonian (MONDian) gravity, and $\bar{n} = \qty[\exp(\hbar \omega_0/k_BT)-1]^{-1}$ represents the average phonon number at temperature $T$.
The two terms on the right can be equated to see that the Newtonian entanglement vanishes completely at a critical temperature $T_0 = \hbar \omega_0 /k_B \ln\qty(\omega_0^2L/a_N)$,
where $a_N = Gm/L^2$ is the Newtonian acceleration.
The residual entanglement in the MONDian gravity oscillates with an amplitude
\begin{equation}
E_0^{(M)}(T_0) = \frac{2\sqrt{a_N a_0}}{\omega_0^2 L \ln(2)} \qty( \frac{2}{3}(\sqrt{2}-1) -  \sqrt{\frac{a_N}{a_0}} ) ,
\end{equation} 
Accordingly, for temperatures $T \gtrsim T_0$, any entanglement detection would certify the onset of MONDian effects.
Note that this experiment can be performed for longer times, but the entanglement accumulation is much less than in the freely falling configurations.

\section{The external field effect}

The MONDian law of gravity is non-linear in acceleration, violating the strong equivalence principle. Accordingly, the internal dynamics of a system cannot be decoupled from the external gravitational field it is placed in. This is known as the external field effect (EFE)~\cite{Milogram_MOND}. It is proposed that the MONDian effects will kick in only when the sum of all accelerations falls below the critical level $a_0$, i.e.,
\begin{equation}
\tilde\mu\qty(\frac{\abs{\vec{a}+\vec{a}_\text{ex}}}{a_0}) =
\begin{cases}
\abs{\vec{a}+\vec{a}_\text{ex}}/a_0, \hspace{1cm} \abs{\vec{a}+\vec{a}_\text{ex}} \ll a_0,
\\
1, \hspace{2.9cm} \abs{\vec{a}+\vec{a}_\text{ex}} \gg a_0 ,
\end{cases}
\end{equation}
where $\vec{a}_\text{ex}$ is the acceleration due to all external sources of gravitation.
For an experiment being performed on the surface of the Earth, $a_\text{ex} \approx$ $9.8$ m/s$^2$ due to the Earth, $6.1 \times 10^{-3}$ m/s$^2$ from the Sun, and $3.3 \times 10^{-5}$ m/s$^2$ from the moon.
Note that the EFE violates Einstein's equivalence principle~\cite{NORTON1985203} and has not been detected in the laboratory so far~\cite{EFE_NoObsEvidence}. If it exists, it will inhibit all MONDian effects on Earth, including the ones discussed here.

\section{Summary}

We have shown that the generation of position-momentum correlations between two nearby quantum masses can be used to probe the Modified Newtonian Dynamics (MOND). 
We derived the conditions under which the MONDian entanglement between two particles is much larger than its Newtonian counterpart. 
It turns out that two nearby quantum masses (similar to what is being used in the laboratory) always satisfy this criterion, thereby making the proposed experiment viable with near-future technologies.
A methodology is proposed where the temperature can be tuned such that a simple act of entanglement witnessing certifies the departures from Newtonian gravity at small accelerations.

%% file: science/ch6-Summary.tex
\chapter{Summary and Outlook}
\label{ch:chapter6}

\section{Summary and conclusions}

We have built a robust theoretical framework to study the entanglement dynamics of two nearby quantum masses prepared in natural Gaussian states. 
Even though this thesis is focused on gravity, all the methodologies developed are generic and versatile. They apply to arbitrary central interactions, even when many such forces are present side by side.

In particular, Chapter~\ref{ch:chapter1} describes our motivation and objectives. Chapter~\ref{ch:chapter2} introduces methods for the simulation of quantum mechanical time evolution.
We discussed the standard tridiagonal implementation of Cayley's operator, a known strategy that preserves the norm of quantum states.
The highly-accurate five-point stencil was utilized to discretize the problem onto an implicit-explicit pentadiagonal Crank-Nicolson scheme,
and the resultant solutions were demonstrated to have much higher accuracy than the standard tridiagonal ones. 
This will be useful where the potential is weak and precise simulations are required to maintain accuracy on long time scales. 
The codes have been made publicly available in the hope that they will find applications beyond the scope of this thesis.
We then developed a strategy to calculate the time evolution of (initially Gaussian) bipartite states interacting with central potentials.
This involved a transformation to the COM frame of reference where we found that only certain states (prepared in identical harmonic traps) transform as a disentangled product.
For central interactions the Hamiltonian separates into the COM and the relative degrees of freedom, which ensures that the product form is maintained at all times. Hence, the problem is fully decoupled into two fictitious particles evolving independently of each other.

Next, in Chapter~\ref{ch:chapter3}, we demonstrate the versatility of our methods by investigating the head-on collision in the Rutherford experiment, with the projectile described by (realistic) localised wave packet shot in the Coulomb potential sourced by a stationary target nucleus.
Various nonclassical effects emerge in the projectile's average trajectory, which are traced back to the convexity properties of the Coulomb potential. 
Jensen's inequality implies that the average force on a quantum wave packet is larger than what is experienced by a classical point particle.
Consequently, the quantum projectile cannot approach the target as closely as its classical counterpart.
We demonstrated that there exists a notion of a `quantum distance of closest approach' and provided lower and upper bounds with a combination of theoretical and numerical analysis.
We also investigated the quantum tunneling of localised wave packets and demonstrated that the tunneling probabilities are many orders of magnitude larger than what is obtained in the traditional plane-wave descriptions.
On the numerical side, the problems discussed in this chapter motivated us to develop a Dynamic Grid Allocation technique, which will find applications in evolution under long-range potentials.

In Chapter~\ref{ch:chapter4}, we developed tools to resolve the entanglement dynamics of two nearby quantum masses coupled gravitationally.
We proposed to push the masses towards each other, hoping that an ever-increasing gravitational interaction would automatically lead to a higher accumulated entanglement.
Starting with the traditional practice of a second-order truncation of the quantum Newtonian potential, we exactly solved the Ehrenfest's differential equations for the COM and the reduced mass. This resulted in an exact closed formulation for the covariance matrix.
We found that the covariance matrix, and the hence entanglement, is completely insensitive to any relative motion between the two spheres. This was indigestible in the beginning as it implies the same gain in correlations when two particles move towards (or, for that matter, away from) each other. 
Howsoever counter-intuitive, extensive numerical simulations and rigorous symbolic calculations confirmed this.
We thereafter investigated the non-Gaussian dynamics triggered by cubic and higher-order potentials.
The covariance matrix is unfaithful in non-Gaussian states, and the correlation dynamics could not be approximated with any existing framework.
We took a step back and wrote the bipartite wave as a product in the COM frame, where the COM is evolved analytically with Fourier techniques, and the reduced mass was evolved with the improved Cayley's propagator. This was followed by a Schmidt decomposition of the bipartite wave function with the help of the algorithms in Google TensorNetwork. 
Only in this non-Gaussian regime was the entanglement sensitive to the relative motion. 
With a foundational perspective we demonstrated that the force gradient across the reduced mass wave packet is the dominant driver of position-momentum correlations.
The entanglement was found to be amplified in the same way as the force gradient. This observation led us to the closed-form expressions for the non-Gaussian entanglement mediated by arbitrary central interactions expanded to any order.
In practice, it will be difficult to screen all interactions but gravity, e.g., Casimir could be present side by side.
With this in mind, we developed tools to  quantify the entanglement mediated by multiple central forces acting simultaneously.

Lastly, in Chapter~\ref{ch:chapter5}, we have shown that the generation of position-momentum correlations between two nearby quantum masses can be used to probe the Modified Newtonian Dynamics (MOND), a candidate explanation of dark matter effects, which proposes to modify Newton's second law and/or the gravitational force law for tiny accelerations.
We derived the conditions under which the MONDian entanglement between two mesospheres is much larger than its Newtonian counterpart. 
Two nearby quantum masses (similar to what is being used in the laboratory) always satisfy this criterion, which means that the proposed experiment is viable with near-future technologies.
Entanglement is easy to witness but difficult to quantify.
With this in mind, we demonstrated how the temperature can be fine-tuned such that a simple act of entanglement witnessing certifies departures from Newtonian gravity.

\section{Applications and future work}

\subsection{Projectile-target entanglement in Rutherford experiment}

Given that the methodologies for entanglement quantification are very generic, one can now step back and estimate the Coulomb-mediated entanglement in low-energy nuclear collisions~\cite{Ashutosh_EntNucColl}.
The parameter characterising the Gaussian covariance matrix for the potential truncated at the second order is given by
\begin{equation}
\omega^2 = \frac{4q_1q_2\alpha \hbar c}{e^2mL^3},
\end{equation}
where $q_1$ and $q_2$ are the charges on the projectile and the target, respectively, and $L$ is their initial separation. 
Note here the Coulomb force is repulsive, and hence in the covariance matrix in Eqs.~\eqref{eq:SolCovMat_FreeFall_Start}~--~\eqref{eq:SolCovMat_FreeFall_Finish}, one has to replace $\omega$ with $i\omega$ to arrive at
\begin{eqnarray}
	\bm{\sigma}_{00}   &=&
 \frac{\hbar}{4m\omega_0} \qty[ 2+\omega_0^2t^2 - \qty( 1-\frac{\omega_0^2}{\omega^2} )\sin^2(\omega t) ],
	\\
	\bm{\sigma}_{02}   &=& 
  \frac{\hbar}{4m\omega_0} \qty[ \omega_0^2t^2 + \qty(1-\frac{\omega_0^2}{\omega^2})\sin^2(\omega t) ],
	\\
\bm{\sigma}_{11} &=&
  \frac{m\hbar\omega_0}{4} \qty[ 2-\qty( 1-\frac{\omega^2}{\omega_0^2} )\sin^2(\omega t) ],
	\\
	\bm{\sigma}_{13} &=&  
  \frac{m\hbar\omega_0}{4} \qty( 1-\frac{\omega^2}{\omega_0^2} ) \sin^2(\omega t),
	\\
	\bm{\sigma}_{01} &=&
 	 \frac{\hbar}{8} \qty[  2\omega_0 t + \qty( \frac{\omega_0}{\omega}-\frac{\omega}{\omega_0} )\sin(2\omega t) ],
	\\
\bm{\sigma}_{03} &=&
  \frac{\hbar}{8} \qty[  2\omega_0 t - \qty( \frac{\omega_0}{\omega}-\frac{\omega}{\omega_0} )\sin(2\omega t) ].
 \end{eqnarray}
\begin{figure}
\centering
\includegraphics[width=0.6\linewidth]{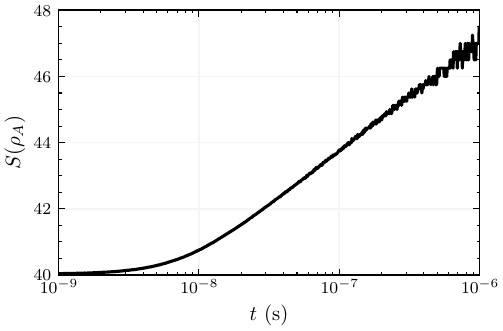}
\caption{Coulomb-mediated gain of entanglement between two bare ${}_{79}^{196}$Au nuclei separated by a distance of $1$ m and initially prepared in Gaussian states of width $1$ nm. $S(\rho_A)$ is the entanglement entropy, and $t$ denotes the time.}
\label{fig:entropy_nucleus}
\end{figure}
Considering an example of two bare nuclei of ${}_{79}^{196}$Au separated by a distance of $L = 1$ m, we get $\omega \sim 10^{20}$ s$^{-1}$. Accordingly, the entanglement has a highly-oscillatory component with a period of $\sim 10^{-20}$ s, which can be easily averaged out to get
\begin{eqnarray}
	\bm{\sigma}_{00}   &=&
 \frac{\hbar}{4m\omega_0} \qty[ 2+\omega_0^2t^2 - \frac{1}{2} \qty( 1-\frac{\omega_0^2}{\omega^2} ) ] ,
		\\
	\bm{\sigma}_{02}   &=& 
  \frac{\hbar}{4m\omega_0} \qty[ \omega_0^2t^2 + \frac{1}{2} \qty(1-\frac{\omega_0^2}{\omega^2}) ],
		\\
\bm{\sigma}_{11} &=&
  \frac{m\hbar\omega_0}{4} \qty[ 2- \frac{1}{2} \qty( 1-\frac{\omega^2}{\omega_0^2} ) ],
		\\
	\bm{\sigma}_{13} &=&  
  \frac{m\hbar\omega_0}{8} \qty( 1-\frac{\omega^2}{\omega_0^2} ) ,
		\\
	\bm{\sigma}_{01} &=& 
 	 \frac{1}{4} \hbar \omega_0 t.
\\
\bm{\sigma}_{03} &=& 
 	 \frac{1}{4} \hbar \omega_0 t.
\end{eqnarray}
In Fig.~\ref{fig:entropy_nucleus}, we assume $\sigma = 1$ nm and show a huge amount of Gaussian entanglement is generated in a microsecond.
The same can now be extrapolated for the momentum dependence (projectile and target colliding with each other) by using Eq.~\eqref{eq:ent_gradestimate_S}. The contribution of the cubic order term of the potential is quantified by $\epsilon_3 = - 3\ev{\hat r}/L$, where $\ev{\hat r}$ is to be calculated by solving the transcendental Eq.~\eqref{xcl}.

\subsection{Concrete scheme to measure the entanglement}

Gravity is weak, and hence it generates only a small amount of entanglement.
Accordingly, the detection step is a demanding part of the considered setup.
This thesis is focussed only on entanglement generation and not on its detection.
In future, we would like to develop a concrete scheme to measure all entries of the covariance matrix experimentally.
We aim to take cues and build upon various existing strategies, e.g.,
\begin{itemize}

\item
Palomaki \emph{et al}. Science \textbf{342}, 710 (2013)~\cite{Palomaki-Science}

The covariance matrix is here reconstructed for a mechanical oscillator coupled to a microwave field. 
The basic idea is to swap the state of the mechanical mode onto the field and measure the elements of the covariance matrix between the two field modes.
In principle, the same concept could work for the two masses considered in this thesis.

\item 
D’Angelo \emph{et al}. Journal of Modern Optics \textbf{53}, 16 (2006)~\cite{DAngelo-JoMO}

Another possibility would be to reconstruct the Wigner function of the quantum state. 
Here it is done for two-mode optical fields via homodyne measurements.

Our results can be easily re-formulated in this language. For a two-mode thermal Gaussian state represented by a vector operator $\hat u = \qty(\hat x_A, \hat p_A, \hat x_B, \hat p_B)^T$ and covariance matrix $\bm{\sigma}_\text{th} = (2\bar{n}+1)\bm\sigma$, the  Wigner function is given by~\cite{GaussianOperations-Brask}:
\begin{eqnarray}
W &=& \frac{1}{(2\pi)^2\sqrt{\text{Det}(\bm{\sigma}_\text{th})}} \ \exp\qty[ - \frac{1}{2} \qty(\hat u - \ev{\hat u})^T \ \bm{\sigma}_\text{th}^{-1} \ \qty(\hat u - \ev{\hat u})   ] 
\nonumber 	\\
&=& \frac{1}{(2\pi)^2\sqrt{(2\bar{n}+1)^4 \ \text{Det}(\bm{\sigma})}} \ \exp\qty[ - \frac{1}{2(2\bar{n}+1)} \qty(\hat u - \ev{\hat u})^T \ \bm{\sigma}^{-1} \ \qty(\hat u - \ev{\hat u}) ] .
\end{eqnarray}
Note that the purity of bipartite state at zero-temperature implies $\text{Det}(\bm{\sigma}) = (\hbar/2)^4$~\cite{Serafini_2003}, and hence the exact closed form for the two-mode Wigner function is
\begin{equation}
W = \frac{1}{\pi^2\hbar^2(2\bar{n}+1)^2} \exp\qty[ - \frac{1}{2(2\bar{n}+1)} \qty(\hat u - \ev{\hat u})^T \ \bm{\sigma}^{-1} \ \qty(\hat u - \ev{\hat u})  ] ,
\end{equation}
where the covariance matrix $\bm\sigma$ is  already calculated in Secs.~\ref{sec:covmat_freefall} and~\ref{sec:covmat_levitated}.

\item 
Krisnanda \emph{et al}. Physical Review D \textbf{107}, 086014 (2023)~\cite{Krisnanda2022_QNN} 

This methodology is inspired by quantum neural networks. The notable point is that the predicted entanglement sensitivity (for logarithmic negativity) is $10^{-4}$.

\end{itemize}

\subsection{Entanglement witnessing with minimum measurements}

For entanglement quantification, one needs to measure in LAB ten entries in the covariance matrix. 
However, for entanglement witnessing we need much less.
For a two-mode state of particles $A$ and $B$, there are only four correlated measurements between $x_A \pm x_B$ and $p_A \pm p_B$.
The entangled state will be unphysical under a partial transposition, and hence we only need to work out which of the four correlated measurements will subsequently violate Heisenberg's uncertainty relation. 
The same would be sufficient to witness the generation of entanglement in the considered setup.
Such a scheme would be useful from the perspective of Chapter~\ref{ch:chapter5}, where the whole idea of certifying MONDian effects is built around entanglement witnessing.

\subsection{Entanglement dynamics in quantum reference frames}

The Galilean principle of relativity inspired the development of Newtonian mechanics, which stimulated the industrial revolution during the 18th and 19th centuries.
The Lorentz transformations are at the heart of Einstein's special theory of relativity, which helped develop the highly-accurate Global Positioning System (GPS). 
In this thesis in Sec.~\ref{sec:GalileanTransformation}, we utilized the ideas of Galilean transformations to argue that a moving COM does not generate any correlation whatsoever. 
Moreover, this implies that the individual momenta of the two particles can be tweaked per experimental convenience as long as the relative motion is not affected. 
The lesson is that classical reference frame transformations have contributed a lot to our understanding of the world. They simplified various complexities in our theoretical and mathematical endeavors and revealed various natural phenomena that we harness today.
Given that we more or less understand how a quantum superposition looks from our point of view, it is long due that we work out how the world looks from the point of view of a delocalised quantum particle. Therefore, the logical step forward is the conceptual development of Quantum Reference Frame (QRF) transformations~\cite{DICKSON2004195,1984PhRvD..30..368A,Poulin_2007,Angelo_2011,RevModPhys.79.555}, followed by a re-formulation of physical laws that are invariant under these transformations~\cite{Giacomini2019}. 
Despite extensive research, a consistent QRF description is still under debate.

One of the interesting outcomes in the framework of Ref.~\cite{Giacomini2019} is that a separable state from our point of view can be entangled from the perspective of a quantum particle.
Assuming that two particles $A$ and $B$ are prepared in identical Gaussian wave packets, the initial tripartite wave function as seen from a classical localised frame $C$ is
\begin{equation}
\Psi^{(C)}  \sim    \delta(x_C) \  \exp\qty( - x_A^2 ) \  \exp\qty( - x_B^2 ) 
\end{equation}
We can jump onto particle $A$ by applying the unitary $\hat S_x = \mathcal{\hat P}_{AC} \exp(i \hat x_A \hat p_B) $, where $\hat P_{AC}$ is the parity and swap operator satisfying $\hat{\mathcal{P}}_{AC} \ \psi_A(x) = \psi_C(-x)$:
\begin{eqnarray}
\Psi^{(A)} = \hat S_x \Psi^{(C)}
&\sim& \mathcal{\hat P}_{AC} \ \exp\qty( i \hat x_A \hat p_B ) \  \delta(x_C) \  \exp\qty( - x_A^2 ) \  \exp\qty( - x_B^2 )  
\nonumber	\\
 &=& \mathcal{\hat P}_{AC} \ \delta(x_C)  \  \exp\qty( i \hat x_A \hat p_B )  \  \exp\qty( - x_A^2 )  \ \exp\qty( - x_B^2 ) 
\nonumber	\\
 &=& \mathcal{\hat P}_{AC} \  \delta(x_C)    \  \exp\qty( - x_A^2 )  \  \exp\qty[ - (x_B+x_A)^2 ] 
\nonumber	\\
 &=&   \delta(x_A)    \  \exp\qty( - x_C^2 )  \  \exp\qty[ -(x_B-x_C)^2 ] 
\nonumber	\\
 &\sim&   \delta(x_A)  \ \exp\qty( - x_B^2 ) \ \exp\qty( - x_C^2 )  \ \exp\qty( x_Bx_C ) .
\end{eqnarray}
Note the coupling term containing $x_B x_C$.
We plan to adopt to the QRFs and resolve the corresponding gravitational entanglement dynamics. 
Along the way, we aim to contribute to bring consistencies in QRF transformations with various thought experiments. In particular, one can consider the free evolution of a single particle Gaussian wave packet of width $\sigma$ corresponding to particle $A$. From the perspective of a localised frame $C$ we have
\begin{eqnarray}
\Psi^{(C)}  &=&    \delta(x_C) \ \frac{1}{\sqrt{\sigma\sqrt{2\pi}}} \   \exp\qty( - \frac{x_A^2}{4\sigma^2} ) ,
\\
\hat H^{(C)} &=& \frac{\hat p_A^2}{2m_A} .
\end{eqnarray}
The unitary for jumping onto the particle $A$ is now $\hat S_x  = \mathcal{\hat P}_{AC}$, which implies
\begin{eqnarray}
\Psi^{(A)} &=&  \hat S_x \Psi^{(C)} 
=   \delta(x_A) \ \frac{1}{\sqrt{\sigma\sqrt{2\pi}}} \  \exp\qty( - \frac{x_C^2}{4\sigma^2} ) , 
\\
\hat H^{(A)} &=&  \hat S_x \hat H^{(C)} \hat S_x^\dagger 
 = \frac{\hat p_C^2}{2m_C}. 
\end{eqnarray}
Recall that for a free Hamiltonian $\hat H = \hat p^2/2m$, the rate of wave packet expansion is characterized by a frequency $\omega_0 = \hbar/2m\sigma^2$.
Given the mass has changed during a QRF transformation between $A$ and $C$, the characteristic frequency also changes. 
The rate of wave packet expansion is now (unexpectedly) different in the two inertial frames of reference.
Proper research is required to bring consistency such that the mass of the classical frame $C$ is rendered irrelevant after an inverse transformation.

%% file: science/appendix.tex
\cleardoublepage
\appendix

\chapter{Statistical Transformation of a Two-Mode Gaussian State}
\label{appendix:MinUncertTransf}

Here we derive the condition for a two-mode Gaussian state in the LAB frame to transform into a two-mode Gaussian state in the COM frame. We stress that we are focused on the minimum uncertainty wavepackets, i.e., the ones which minimise the Heisenberg's uncertainty:
\begin{equation}
\bm{\Delta} x_A \bm{\Delta} p_A  
= \bm{\Delta} x_B \bm{\Delta} p_B  
= \frac{\hbar}{2}.
\end{equation}
The generalised inverse coordinate transformations to the COM frame read
\begin{eqnarray}
R = \frac{m_Ax_A+m_Bx_B}{m_A+m_B},
\hspace{1cm}
r = x_B-x_A,
\hspace{1cm}
P = p_A+p_B,
\hspace{1cm}
p = \frac{m_Ap_B-m_Bp_A}{m_A+m_B}.
\end{eqnarray}
If we assume that the position spreads as $\bm{\Delta} x_A  = \sigma_A$ and $\bm{\Delta} x_B = \sigma_B$, the spreads in momenta would be $\bm{\Delta}p_A = \hbar/2\sigma_A$, and $\bm{\Delta}p_B = \hbar/2\sigma_B$. Hence, the statistical uncertainties in the COM and the reduced mass are given by
\begin{eqnarray}
\bm{\Delta} R^2  &=&  \qty( \frac{m_A}{m_A+m_B} )^2 \bm{\Delta} x_A^2 +  \qty( \frac{m_B}{m_A+m_B} )^2 \bm{\Delta} x_B^2,	\nonumber	\\
&& =  \qty( \frac{m_A}{m_A+m_B} )^2 \sigma_A^2 +  \qty( \frac{m_B}{m_A+m_B} )^2  \sigma_B^2		\nonumber	\\
&& = \frac{ \qty( m_A^2 \sigma_A^2 +  m_B^2 \sigma_B^2 ) }{ (m_A+m_B)^2 },   
\\	\nonumber	\\
\bm{\Delta}r^2 &=& \bm{\Delta}x_B^2 + \bm{\Delta}x_A^2	 = \sigma_A^2 + \sigma_B^2,
\\	\nonumber	\\
\bm{\Delta} P^2 &=& \bm{\Delta} p_A^2 + \bm{\Delta} p_B^2
 = \frac{\hbar^2}{4\sigma_A^2} + \frac{\hbar^2}{4\sigma_B^2}	
=	\frac{\hbar^2}{4} \qty( \frac{\sigma_A^2+\sigma_B^2}{\sigma_A^2\sigma_B^2} ),
\\	\nonumber	\\
\bm{\Delta} p^2 &=& \qty( \frac{m_A}{m_A+m_B} )^2 \bm{\Delta} p_B^2 +  \qty( \frac{m_B}{m_A+m_B} )^2 \bm{\Delta} p_A^2	\nonumber	\\
&& = \qty( \frac{m_A}{m_A+m_B} )^2  \frac{\hbar^2}{4\sigma_B^2} +  \qty( \frac{m_B}{m_A+m_B} )^2  \frac{\hbar^2}{4\sigma_A^2} 	\nonumber	\\
&& = \frac{\hbar^2}{ 4 } \frac{ \qty( m_A^2 \sigma_A^2 +  m_B^2 \sigma_B^2 ) }{ (m_A+m_B)^2 \sigma_A^2\sigma_B^2 },
\end{eqnarray}
which imply that the Heisenberg uncertainties are
\begin{equation}
\bm{\Delta} R ^2\bm{\Delta} P^2 = \bm{\Delta} r ^2\bm{\Delta} p^2
=  \frac{\hbar^2}{ 4 } \frac{ \qty( m_A^2 \sigma_A^2 +  m_B^2 \sigma_B^2 ) \qty(\sigma_A^2+\sigma_B^2) }{ (m_A+m_B)^2 \sigma_A^2\sigma_B^2 } .
\end{equation}
For the COM and the reduced mass to be both in minimum uncertainty Gaussian wave packets, we must satisfy
\begin{eqnarray}
&& \qty( m_A^2 \sigma_A^2 +  m_B^2 \sigma_B^2 ) \qty(\sigma_A^2+\sigma_B^2)   =  (m_A+m_B)^2 \sigma_A^2\sigma_B^2	\nonumber	\\
\implies    &&    m_A^2 \sigma_A^4 +  m_B^2 \sigma_B^4 + (m_A^2+m_B^2) \sigma_A^2\sigma_B^2 = (m_A^2+m_B^2+2m_Am_B) \sigma_A^2\sigma_B^2	\nonumber	\\
\implies    &&    m_A^2 \sigma_A^4 +  m_B^2\sigma_B^4  - 2m_Am_B \sigma_A^2\sigma_B^2 = 0	\nonumber	\\
\implies    &&    (m_A\sigma_A^2)^2 + (m_B\sigma_B^2)^2 - 2(m_A\sigma_A^2)(m_B\sigma_B^2) = 0	\nonumber	\\
\implies    &&    (m_A\sigma_A^2-m_B\sigma_B^2)^2 = 0	\nonumber	\\
\implies    &&    m_A\sigma_A^2 = m_B\sigma_B^2,
\end{eqnarray}
which is the same expression we got algebraically in Chapter~\ref{ch:chapter2}.

\cleardoublepage
\chapter{Ehrenfest's Dynamics in COM Frame}
\label{appendix:Ehrenfest_COM_redmass}

The Ehrenfest's theorem relates the time derivative of the expectation value of an operator $ \hat A$ to the expectation of its commutator with the Hamiltonian $\hat H$~\cite{Ehrenfest1927}:
\begin{equation}
	\dv{t}\ev{ \hat A} = \frac{1}{i\hbar}\ev{\comm{ \hat A}{  \hat H}} + \ev{\pdv{ \hat A}{t}}.
\end{equation}
In this appendix, we derive the analytical solutions for the statistical moments of the COM and the reduced mass in the setup discussed in Chapter~\ref{ch:chapter4}, i.e., for the case of two identical particles of mass $m$ gravitating each other while in free fall:
\begin{eqnarray}
\hat{H} 
&=& \hat H_R + \hat H_r
\nonumber	\\
&=& \qty( \frac{\hat P^2}{4m} ) + \qty( \frac{\hat p^2}{m} - \frac{Gm^2}{L+\hat r} )
\nonumber	\\
&=& \qty( \frac{\hat P^2}{4m} ) + \qty( \frac{\hat p^2}{m} - \frac{1}{4} m \omega^2 \sum_{n=0}^{N} \frac{(-1)^n}{L^{n-2}} \hat r^n ),
\end{eqnarray}
where $\hat H_R$ and $\hat H_r$ are the Hamiltonians for the COM and the reduced mass, respectively, and $\omega^2 = 4Gm/L^3$ is assumed for later convenience.

 \section{Free evolution of the COM}

The COM Hamiltonian is
\begin{equation}
\hat H_R =  \frac{\hat P^2}{4m},
\end{equation} 
and hence the corresponding Ehrenfest's equations for the relevant moments are
\begin{eqnarray}
\dv{t} \ev{\hat R} &=&	\frac{1}{4mi\hbar} \ev{ \comm{\hat R}{\hat P^2} }
	=	\frac{1}{2m}\ev{\hat P},
\\	\nonumber	\\
\dv{t} \ev{\hat P} &=& \frac{1}{4mi\hbar} \ev{ \comm{\hat P}{\hat P^2} } 
= 0,	
\\	\nonumber	\\
\dv{t} \ev{\hat R\hat P+\hat P\hat R} &=&	 \frac{1}{4mi\hbar} \ev{ \comm{\hat R\hat P+\hat P\hat R}{\hat P^2} }	
  = \frac{1}{m} \ev{\hat P^2},
\\	\nonumber	\\
\dv{t} \ev{\hat R^2} &=&	\frac{1}{4mi\hbar} \ev{ \comm{\hat R^2}{\hat P^2} }
 = \frac{1}{2m}\ev{\hat R \hat P + \hat P \hat R},
\\	\nonumber	\\
\dv{t} \ev{\hat P^2} &=& \frac{1}{4mi\hbar} \ev{ \comm{\hat P^2}{\hat P^2} }	
 = 0.
\end{eqnarray}
The initial state (at $t=0$) is characterized by [see Fig.~\ref{fig:TMGS_special} or Eq.~\!\eqref{eq:COMwavefunction_t0}]
\begin{equation}
	\ev{\hat R} = 0, \	\ev{\hat P} = 0, \
 \ev{ \hat R\hat P+\hat P\hat R } = 0,	\
\ev{\hat R^2} = \frac{1}{2}\sigma^2, \ \ev{\hat P^2} = \frac{\hbar^2}{2\sigma^2},
\end{equation}
and the exact solutions to the Ehrenfest's differential equations imply
\begin{equation}
\ev{\hat R} = 0, \ \ev{\hat P} = 0, \ \ev{\hat R \hat P + \hat P \hat R} = \hbar\omega_0 t, \ \ev{\hat R^2} =  \frac{1}{2}\sigma^2 (1+\omega_0^2t^2), \ \ev{\hat P^2} = \frac{\hbar^2}{2\sigma^2}.
\end{equation}
Alternatively, one arrives at the exact same results by utilising the functional form of the time-dependent wave function~\cite{GaussEvolFreeSpace_SMBlinder}:
\begin{equation}
	\phi(R,t) = \frac{1}{\sqrt{\sigma(1+i\omega_0t)\sqrt{\pi}}} \exp\qty( -\frac{R^2}{2\sigma^2(1+i\omega_0t)} ),
	\label{eq:TDWF_COM}
\end{equation}
where $\omega_0 = \hbar/2m\sigma^2$ is the frequency of harmonic traps used to prepare the two particles in initial Gaussian states of width $\sigma$.

 \section{Evolution of the reduced mass}

The reduced mass Hamiltonian can be represented by a binomial series, Eq.~\eqref{eq:Ham_redmass}. For $N=2$, i.e., a quadratic Hamiltonian,
\begin{equation}
\hat{H}_r = \frac{ \hat{p}^2}{m} - \frac{1}{4} m \omega^2 \qty( L^2 - L\hat r + \hat r^2 ),
\end{equation}
the Ehrenfest's equations for the first two statistical moments are
\begin{eqnarray}
\dv{t} \ev{\hat r} &=& \frac{1}{mi\hbar} \ev{ \comm{\hat r}{\hat p^2} }	- \frac{m\omega^2}{4i\hbar} \qty( -L\ev{ \comm{\hat r}{\hat r} } +   \ev{ \comm{\hat r}{\hat r^2} } )  	\nonumber \\ 
	&&	=	\frac{2}{m}\ev{\hat p},
\\	\nonumber	\\
\dv{t} \ev{\hat p} &=& \frac{1}{mi\hbar} \ev{ \comm{\hat p}{\hat p^2} }	- \frac{m\omega^2}{4i\hbar} \qty( -L\ev{ \comm{\hat p}{\hat r} } +   \ev{ \comm{\hat p}{\hat r^2} } )  	\nonumber \\
	&& = -\frac{1}{4}m\omega^2L + \frac{1}{2}m\omega^2\ev{\hat r},	
\\	\nonumber	\\
\dv{t} \ev{\hat r\hat p+\hat p\hat r} &=&	 \frac{1}{mi\hbar} \ev{ \comm{\hat r\hat p+\hat p\hat r}{\hat p^2} } 	- \frac{m\omega^2}{4i\hbar} \qty( -L\ev{ \comm{\hat r\hat p+\hat p\hat r}{\hat r} } +   \ev{ \comm{\hat r\hat p+\hat p\hat r}{\hat r^2} } )  	\nonumber \\
	&&  = \frac{4}{m}\ev{\hat  p^2} -\frac{1}{2}m\omega^2L\ev{\hat r} + m\omega^2\ev{\hat r^2},
\\	\nonumber	\\
\dv{t} \ev{\hat r^2} &=&	\frac{1}{mi\hbar} \ev{ \comm{\hat r^2}{\hat p^2} }	- \frac{m\omega^2}{4i\hbar} \qty( -L\ev{ \comm{\hat r^2}{\hat r} } +   \ev{ \comm{\hat r^2}{\hat r^2} } )	\nonumber	\\
	&& = \frac{2}{m}
\ev{\hat r\hat p+\hat p \hat r},
\\	\nonumber	\\
\dv{t} \ev{\hat p^2} &=& \frac{1}{mi\hbar} \ev{ \comm{\hat p^2}{\hat p^2} } - \frac{m\omega^2}{4i\hbar} \qty( -L\ev{ \comm{\hat p^2}{\hat r} } +   \ev{ \comm{\hat p^2}{\hat r^2} } )	\nonumber	\\
	&& = \frac{1}{2}m\omega^2\ev{\hat r\hat p+\hat p \hat r} - \frac{1}{2}m\omega^2L\ev{\hat p}.
\end{eqnarray}
The initial state (at $t=0$) is characterized by [see Fig.~\ref{fig:TMGS_special} or Eq.~\!\eqref{eq:RMwavefunction_t0}]
\begin{equation}
\ev{\hat r} = 0, \ \ev{\hat p} = 0,	\
 \ev{ \hat r\hat p+\hat p\hat r} = 0,		\
\ev{\hat r^2} = 2\sigma^2, \ \ev{\hat p^2} = \frac{\hbar^2}{8\sigma^2},
\end{equation}
and the exact solutions for the Ehrenfest's differential equations imply
\begin{eqnarray}
	\ev{ \hat{r}}  &=&  \frac{1}{2}L \Big( 1 -\cosh(\omega t) \Big) -  \frac{2p_0}{m\omega} \sinh(\omega t),
\\	\nonumber	\\
	\ev{ \hat{p}}  &=& - p_0\cosh(\omega t) - \frac{1}{4} m\omega L \sinh(\omega t),
\\	\nonumber	\\
	\ev{ \hat{r} \hat{p} + \hat{p} \hat{r}}  &=&  Lp_0 \Big( \cosh(2\omega t) - \cosh(\omega t) \Big)	+ \frac{1}{8}m\omega L^2 \Big( \sinh(2\omega t) - 2\sinh(\omega t) \Big) 		\nonumber \\
	&&  + \frac{2}{m\omega} \qty( p_0^2+\frac{\hbar^2}{8\sigma^2}+ \frac{1}{2}m^2\omega^2\sigma^2 ) \sinh(2\omega t),
\\	\nonumber	\\
\ev{ \hat{r}^2 }  &=& 2\sigma^2 \Big( 1+\sinh[2](\omega t) \Big)		 + \frac{1}{8}L^2 \Big( 3 + \cosh(2\omega t)-4\cosh(\omega t) \Big) 	\nonumber \\
	&& + \frac{Lp_0}{m\omega} \Big( \sinh(2\omega t) - 2\sinh(\omega t) \Big) 	 + \frac{4}{m^2\omega^2} \qty( p_0^2+\frac{\hbar^2}{8\sigma^2} ) \sinh[2](\omega t),
\\	\nonumber	\\
\ev{ \hat{p}^2}  &=&  \qty(p_0^2+\frac{\hbar^2}{8\sigma^2} ) \Big( 1+\sinh[2](\omega t) \Big)    + \frac{1}{4}m\omega Lp_0\sinh(2\omega t)	\nonumber	\\
&& + \frac{1}{4}m^2\omega^2 \qty( 2\sigma^2+\frac{1}{4}L^2 ) \sinh[2](\omega t) .
	\label{eq:relmass_p2}
\end{eqnarray}

\cleardoublepage
\chapter{Quantification of Bipartite Entanglement}
\label{appendix:EntangFormalism} 

We have employed the formalism based on the covariance matrix to quantify entanglement gain via logarithmic negativity, and additionally used the density matrix to compute the von Neumann entropy of entanglement.

 \section{Bipartite covariance matrix}

The covariance matrix formalism is based on the first two statistical moments of a quantum state. Given a bipartite system $AB$ with $ \hat u = ( \hat x_A, \hat p_A, \hat x_B, \hat p_B)^T$, the covariance matrix is defined as~\cite{PRA_65.032314,PRA_70.022318,PRA_72.032334}:
\begin{equation}
	\bm{\sigma}_{jk}  = \frac{1}{2} \ev{ \hat u_j  \hat u_k +  \hat u_k \hat  u_j }  -  \ev{ \hat u_j }\ev{ \hat u_k}
\hspace{.5cm}  
\implies
\hspace{.5cm}  
 \bm{\sigma} = \mqty(
	\bm{\alpha} & \bm{\gamma} \\
	\bm{\gamma}^T & \bm{\beta} )  , 
\end{equation}
where $\bm{\alpha}(\bm{\beta}$) contains the local mode correlation for $A(B)$, and $\bm{\gamma}$ describes the intermodal correlation. In our setting the two masses are identical, which leads to the following inverse coordinate transformations between the LAB and the COM frames
\begin{equation}
	x_A(x_B) = R  -\!(\!+\!) \ \frac{r}{2} ,
\hspace{1cm}
	p_A(p_B) = \frac{P}{2}  -\!(\!+\!) \ p .
\end{equation}
Accordingly, the symmetric covariance matrix can be derived as
\begin{eqnarray}
\bm{\sigma}_{00} &=& \ev{\hat x_A^2} - \ev{\hat x_A}^2	
\nonumber	\\
&& = \ev{ \qty(\hat  R-\frac{\hat r}{2} )^2 } - \ev{\hat  R-\frac{\hat r}{2} }^2	\nonumber	\\
&& = \qty( \ev{\hat R^2} - \ev{\hat R}^2 ) + \frac{1}{4} \qty( \ev{\hat r^2} - \ev{\hat r}^2 ) 	\nonumber \\
&& = \bm{\Delta} R^2 + \frac{1}{4} \bm{\Delta} r^2,
\\	\nonumber	\\
\bm{\sigma}_{11} &=& \ev{\hat p_A^2} - \ev{\hat p_A}^2	
\nonumber	\\
&& =	\ev{ \qty( \frac{\hat P}{2}-\hat p )^2 } - \ev{ \frac{\hat P}{2}-\hat p }^2	\nonumber	\\
&& = \frac{1}{4}\qty( \ev{\hat P^2} - \ev{\hat P}^2 ) +  \qty( \ev{\hat p^2} - \ev{\hat p}^2 ) 	\nonumber \\
&& = \frac{1}{4}\bm{\Delta} P^2 + \bm{\Delta} p^2 ,
\\	\nonumber	\\
\bm{\sigma}_{02} &=& \frac{1}{2} \ev{\hat x_A \hat x_B + \hat x_B \hat x_A} - \ev{\hat x_A} \ev{\hat x_B}	\nonumber	\\
	&& = \frac{1}{2} \ev{ \qty(\hat R-\frac{\hat r}{2}) \qty(\hat R+\frac{\hat r}{2}) + \qty(\hat R+\frac{\hat r}{2}) \qty(\hat R-\frac{\hat r}{2}) } - \ev{\hat R-\frac{\hat r}{2}} \ev{\hat R+\frac{\hat r}{2}}	\nonumber	\\
	&& = \qty( \ev{\hat R^2} - \ev{\hat R}^2 ) - \frac{1}{4} \qty( \ev{\hat r^2} - \ev{\hat r}^2 ) 	\nonumber \\
	&& = \bm{\Delta} R^2 - \frac{1}{4} \bm{\Delta} r^2,
\\	\nonumber	\\
\bm{\sigma}_{13} &=&	\frac{1}{2}\ev{\hat p_A \hat p_B + \hat p_B \hat p_A} - \ev{\hat p_A} \ev{\hat p_B} 	\nonumber	\\
	&& =	\frac{1}{2} \ev{ \qty(\frac{\hat P}{2}-\hat p) \qty(\frac{\hat P}{2}+\hat p)  + \qty(\frac{\hat P}{2}+\hat p) \qty(\frac{\hat P}{2}-\hat p)  } - \ev{\frac{\hat P}{2}-\hat p} \ev{\frac{\hat P}{2}+\hat p}	\nonumber	\\
	&& = \frac{1}{4}\qty( \ev{\hat P^2} - \ev{\hat P}^2 ) -  \qty( \ev{\hat p^2} - \ev{\hat p}^2 ) 	\nonumber \\
	&& = \frac{1}{4}\bm{\Delta} P^2 - \bm{\Delta} p^2,
\\	\nonumber	\\
\bm{\sigma}_{01}  &=& \frac{1}{2}\ev{\hat x_A \hat p_A + \hat p_A \hat x_A} - \ev{\hat x_A} \ev{\hat p_A}
	\nonumber	\\
	&& =	\frac{1}{2} \ev{ \qty(\hat R-\frac{\hat r}{2}) \qty(\frac{\hat P}{2}-\hat p) + \qty(\frac{\hat P}{2}-\hat p) \qty(\hat R-\frac{\hat r}{2}) } - \ev{\hat R-\frac{\hat r}{2}} \ev{\frac{\hat P}{2}-\hat p}	\nonumber	\\
	&& = \qty( \frac{1}{4}\ev{\hat R \hat P + \hat P \hat R} - \frac{1}{2} \ev{\hat R} \ev{\hat P} ) + \qty( \frac{1}{4}  \ev{\hat r \hat p + \hat p \hat r}  -  \frac{1}{2} \ev{\hat r} \ev{\hat p} )	\nonumber	\\
	&& =  \frac{1}{2} \textbf{Cov}( {R}, {P})  + \frac{1}{2} \textbf{Cov}( {r}, {p}),
\\	\nonumber	\\
\bm{\sigma}_{03}  &=& \frac{1}{2}\ev{\hat x_A \hat p_B + \hat p_B \hat x_A} - \ev{\hat x_A} \ev{\hat p_B}	\nonumber	\\
	&& =	\frac{1}{2} \ev{ \qty(\hat R-\frac{\hat r}{2}) \qty(\frac{\hat P}{2}+\hat p) + \qty(\frac{\hat P}{2}+\hat p) \qty(\hat R-\frac{\hat r}{2}) } - \ev{\hat R-\frac{\hat r}{2}} \ev{\frac{\hat P}{2}+\hat p}	\nonumber	\\
	&& = \qty( \frac{1}{4}\ev{\hat R \hat P + \hat P \hat R} - \frac{1}{2} \ev{\hat R} \ev{\hat P} ) -  \qty( \frac{1}{4} \ev{\hat r \hat p + \hat p \hat r}  -  \frac{1}{2} \ev{\hat r} \ev{\hat p} )	\nonumber	\\
	&& =  \frac{1}{2} \textbf{Cov}( {R}, {P})  - \frac{1}{2} \textbf{Cov}( {r}, {p}) ,
\\	\nonumber	\\
\bm{\sigma}_{22} &=& \ev{\hat x_B^2} - \ev{\hat x_B}^2 
\equiv \ev{\hat x_A^2} - \ev{\hat x_A}^2 = \bm{\sigma}_{00} ,
\\  \nonumber   \\
\bm{\sigma}_{33} &=& \ev{\hat p_B^2} - \ev{\hat p_B}^2
\equiv \ev{\hat p_A^2} - \ev{\hat p_A}^2
= \bm{\sigma}_{11} ,
\\  \nonumber   \\
\bm{\sigma}_{23}  &=& \frac{1}{2}\ev{\hat x_B \hat p_B + \hat p_B \hat x_B} - \ev{\hat x_B} \ev{\hat p_B}	
\equiv \frac{1}{2}\ev{\hat x_A \hat p_A + \hat p_A \hat x_A} - \ev{\hat x_A} \ev{\hat p_A}	
= \bm{\sigma}_{01} ,
\\  \nonumber   \\
\bm{\sigma}_{12} &=& \frac{1}{2} \ev{\hat p_A \hat x_B + \hat x_B \hat p_A} - \ev{\hat p_A} \ev{\hat x_B}	
\equiv \frac{1}{2}\ev{\hat x_A \hat p_B + \hat p_B \hat x_A} - \ev{\hat x_A} \ev{\hat p_B}
= \bm{\sigma}_{03}  .
\end{eqnarray}
The same results are concisely written together in Chapter~\ref{ch:chapter4} in Eqs.~\eqref{eq:covmatak-varpos},~\eqref{eq:covmatak-varmom}, and~\eqref{eq:covmatak-covposmom}.

 \section{Logarithmic negativity}

The Negativity of the partially transposed density matrix is a necessary and sufficient condition for entanglement in two--mode Gaussian states~\cite{PRL_84.2726}. As a result of the partial transposition, the covariance matrix is transformed to $\tilde{\bm{\sigma}}$, which differs from $\bm{\sigma}$ by a sign-flip of $\text{Det} (\bm{\gamma})$~\cite{PRA_72.032334}.
The symplectic eigenvalues of $\tilde{\bm{\sigma}}$ are given by:
\begin{equation}
\tilde{\nu}_{\pm} = \sqrt{
\tilde{\Sigma}(\bm{\sigma}) \pm \sqrt{\tilde{\Sigma}^2(\bm{\sigma}) - 4 \ \text{Det}\qty( \bm{\sigma} ) } } \Bigg/ \sqrt{2} , 
\end{equation}
where $\tilde{\Sigma}(\bm{\sigma}) = \text{Det} (\bm{\alpha}) + \text{Det} ( \bm{\beta}) - 2 \ \text{Det} ( \bm{\gamma} )$. The gain of entanglement is quantified by the minimum symplectic eigenvalue through logarithmic negativity:
\begin{equation}
	E(\bm{\sigma}) = \max \Bigg[ 0, - \log_2\qty(  \frac{ \tilde{\nu}_{-} }{\hbar/2} ) \Bigg].
	\label{eq:E_from_covmat}
\end{equation}

 \section{Entropy of entanglement}

For a pure bipartite system described by a density matrix $\rho_{AB}$, the entanglement entropy is defined as the von Neumann entropy for any one of the subsystems, e.g., $S(\rho_A) = - \Tr\qty[ \rho_A\log_2(\rho_A) ]$, where $\rho_A = \Tr_B\qty( \rho_{AB} )$ is the reduced density matrix for subsystem $A$. In order to calculate $S(\rho_A)$ we start with the two-body wave function of Eq.~\!\eqref{eq:TBWF_LAB2COM_relation}:
\begin{equation}
	\Psi(x_A,x_B,t) = \phi \qty( \frac{x_A+x_B}{2},t  ) \ \psi \qty( x_B-x_A,t ),
\end{equation} 
where $\phi$ is derived analytically in Eq.~\!\eqref{eq:TDWF_COM}, and $\psi$ is calculated numerically by implementing the improved Cayley's propagator~\cite{Ankit_TDSE_Zenodo,Ankit_2022_TDSE}. 
Once this is available for a given time $t$, we perform a singular value decomposition~\cite{doi:10.1119/1.17904,thesis_pyqentangle}:
\begin{equation}
	\Psi(x_A,x_B,t) = \sum_{j} \sqrt{\lambda_j(t)} \ \chi^{(A)}_j(x_A,t) \ \chi^{(B)}_j(x_B,t),
\end{equation}
where $\qty{\chi^{(A)}_j}$ and $\qty{\chi^{(B)}_j}$ are orthonormal states in subsystems $A$ and $B$, respectively, and $\qty{\lambda_j}$ are the Schmidt coefficients. A numerical implementation utilizes the algorithms in Google TensorNetwork~\cite{arXiv:1905.01330,TensorNetwork-GitHub,pyqentangle-GitHub}. Note that we dynamically increase the number of Schmidt coefficients until the norm is preserved within an error of $10^{-7}$.
With this decomposition,
\begin{equation}
	S(\rho_A) = -\sum_{j} \lambda_j \log_2(\lambda_j).
\end{equation}
In the case of a Gaussian evolution, $S(\rho_A)$ is calculable using the covariance matrix~\cite{Serafini_2003}:
\begin{equation}
 S(\rho_A) = f\qty( \frac{1}{\hbar}\sqrt{\text{Det} (\bm{\alpha})} ),
	\label{eq:S_from_covmat}
\end{equation}
where
\begin{equation}
f(x) = \qty(x\!+\frac{1}{2})\log_2\qty(x\!+\frac{1}{2}) - \qty(x\!-\frac{1}{2})\log_2\qty(x\!-\frac{1}{2}) .
\end{equation}

\cleardoublepage
\chapter{Thermal Equilibrium of a Simple Harmonic Oscillator}
\label{appendix:ThermalStateCovMat}

Thermal states are the states of a system in equilibrium with a thermal reservoir:
\begin{equation}
\hat \rho_\text{th} = \sum_{n = 0}^{\infty} p_n \ketbra{n},
\end{equation}
where  $p_n$ is the probability of the eigenstate $\ket{n}$ of the system's energy operator. 
In this work, 
we consider the initial state is prepared by cooling the particles inside harmonic traps of frequency $\omega_0$ the trap frequency, and hence the Hamiltonian is
\begin{equation}
\hat H = \frac{\hat p^2}{2m} + \frac{1}{2}m\omega_0^2\hat x^2.
\end{equation}
The corresponding eigenstates are characterized by energies
 $\epsilon_n = \qty(n+\frac{1}{2})\hbar \omega_0$, with the corresponding eigenfunctions given by:
\begin{equation}
\psi_n(x) = \braket{x}{n} = \frac{1}{\sqrt{2^{n}n! \ \sigma\sqrt{2\pi}}}	\
 \exp\qty(-\frac{x^2}{4\sigma^2}) \
H_n\qty(\frac{x}{\sigma\sqrt{2}}),
\end{equation}
where $\sigma = \sqrt{\hbar/2m\omega_0}$ is the position spread in the ground state, and $H_n$ is the Hermite polynomial of $n^\text{th}$ order. In the language of second quantization, the Hamiltonian is written as
\begin{equation}
\hat H = \qty(\hat a^\dagger \hat a + \frac{1}{2}) \hbar \omega_0,
\end{equation}
where the ladder operators are defined by
\begin{equation}
\hat a  =  \sqrt{\frac{m\omega_0}{2\hbar}} \qty( \hat x + i \frac{\hat p}{m\omega_0} ),
\hspace{1cm}
\hat a^\dagger  =  \sqrt{\frac{m\omega_0}{2\hbar}} \qty( \hat x - i \frac{\hat p}{m\omega_0} ) .
\end{equation}
These two operators do not commute, instead $\comm{\hat a}{\hat a^\dagger} = \unitop$, and satisfy
\begin{equation}
\hat a \ket{n} = \sqrt{n}\ket{n-1},
\hspace{1cm}
\hat a^\dagger \ket{n} = \sqrt{n+1}\ket{n+1} .
\end{equation}
Consequently, $\hat a^\dagger \hat a$ is called the number operator: 
\begin{equation}
\expval{\hat a^\dagger \hat a}{n} = \bra{n}  \hat a^\dagger \sqrt{n}\ket{n-1}
= \sqrt{n} \bra{n}  \hat a^\dagger \ket{n-1}
= n.
\end{equation}

 \section{Partition function and phonon number}

The density matrix of the thermal state is $\hat \rho_\text{th} = e^{-\beta \hat H} / Z$, with the partition function given by
\begin{eqnarray}
Z = \Tr( e^{-\beta \hat H} )
&=& \sum_{n = 0}^{\infty} \expval{  e^{-\beta \hat H} }{ n }
\nonumber	\\
&=& e^{-\beta\hbar\omega_0/2} \ \sum_{n = 0}^{\infty} \expval{  e^{-\beta \hbar \omega_0 \hat a^\dagger \hat a } }{ n }
\nonumber	\\
&=& e^{-\beta\hbar\omega_0/2} \ \sum_{n = 0}^{\infty} \expval{  e^{-n\beta \hbar \omega_0  } }{ n }
\nonumber	\\
&=& e^{-\beta\hbar\omega_0/2} \ \sum_{n = 0}^{\infty}   e^{-n\beta \hbar \omega_0  } 
\nonumber	\\
&=&   e^{-\beta\hbar\omega_0/2} \times \frac{ 1 }{ 1 - e^{-\beta\hbar\omega_0} }
\nonumber	\\
&=& \frac{ e^{\beta\hbar\omega_0/2} }{ e^{\beta\hbar\omega_0} - 1 }.
\end{eqnarray}

The average phonon number is
\begin{eqnarray}
\bar{n} = \Tr( \hat a^\dagger \hat a \ \hat \rho_\text{th} )
&=&  \frac{1}{Z} e^{-\beta\hbar\omega_0/2} \ \sum_{n = 0}^{\infty}  \expval{ \hat a^\dagger \hat a e^{-\beta \hbar \omega_0 \hat a^\dagger \hat a }  }{n}
\nonumber	\\
&=&  \frac{1}{Z} e^{-\beta\hbar\omega_0/2} \ \sum_{n = 0}^{\infty}  \expval{ n e^{-n\beta \hbar \omega_0 }  }{n}
\nonumber	\\
&=&  \frac{1}{Z} e^{-\beta\hbar\omega_0/2} \ \sum_{n = 0}^{\infty}   n e^{-n\beta \hbar \omega_0 } 
\nonumber	\\
&=&  \frac{1}{Z} e^{-\beta\hbar\omega_0/2} \ e^{-\beta \hbar \omega_0 } \sum_{n = 1}^{\infty}   n e^{-(n-1)\beta \hbar \omega_0 } 
\nonumber	\\
&=&  \frac{1}{Z} e^{-\beta\hbar\omega_0/2} e^{-\beta\hbar\omega_0} \times \frac{1}{\qty(1-e^{-\beta\hbar\omega_0})^2}
\nonumber	\\
&=&  \frac{ 1 }{ e^{\beta\hbar\omega_0} - 1 }.
\end{eqnarray}
This implies $e^{\beta\hbar\omega_0} = (\bar{n}+1)/\bar{n}$, which allows us to express the density matrix as
\begin{eqnarray}
\hat \rho_\text{th} = \frac{1}{Z} e^{-\beta \hat H}
&=& \frac{ 1 - e^{-\beta\hbar\omega_0} }{ e^{-\beta\hbar\omega_0/2} } \times e^{-\beta\hbar\omega_0/2} e^{-\beta \hbar \omega_0 \hat a^\dagger \hat a }
\nonumber	\\
&& = \qty( 1 - e^{-\beta\hbar\omega_0} ) e^{-\beta \hbar \omega_0 \hat a^\dagger \hat a }
\nonumber	\\
&& \equiv \qty( 1 - e^{-\beta\hbar\omega_0} )  e^{-\beta \hbar \omega_0 \hat a^\dagger \hat a } \sum_{n = 0}^{\infty} \ketbra{n},	\hspace{1cm} : \qty{ \sum_{n = 0}^{\infty} \ketbra{n} = \unitop},
\nonumber	\\
&& = \qty( 1 - e^{-\beta\hbar\omega_0} ) \sum_{n = 0}^{\infty}  e^{-n \beta \hbar \omega_0}  \ketbra{n}
\nonumber	\\
&& = \sum_{n = 0}^{\infty}   \frac{\bar{n}^{n}}{(\bar{n}+1)^{n+1}}  \ketbra{n}.
\end{eqnarray}

 \section{Covariance matrix and entanglement negativity}

 \subsection{Ground state}

Consider the case of $T=0$, i.e., a particle prepared in the ground state $\ket{0}$. The position and momentum operators can be written in terms of the ladder operators as
\begin{equation}
\hat x = \sqrt{\frac{\hbar}{2m\omega_0}} \qty( \hat a^\dagger + \hat a ) \equiv \sigma \qty( \hat a^\dagger + \hat a ) ,
\hspace{1cm}
\hat p = i\sqrt{\frac{m\hbar\omega_0}{2}} \qty( \hat a^\dagger - \hat a ) \equiv  \frac{i\hbar}{2\sigma} \qty( \hat a^\dagger - \hat a )
\end{equation}
The harmonic oscillator is centered at the origin, i.e., $\ev{\hat x} = 0$,  $\ev{\hat p} = 0$. Hence, the variances and the correlation are
\begin{eqnarray}
\bm \Delta x^2 = \expval{\hat x^2}{0}
&=& \sigma^2 \expval{\qty( \hat a^\dagger + \hat a )\qty( \hat a^\dagger + \hat a )}{0}
\nonumber	\\
&=& \sigma^2 \expval{ \qty( 2\hat a^\dagger \hat a + 1  )}{0},	\hspace{1cm} : \qty{ \comm{\hat a}{\hat a^\dagger}  = \unitop },
\nonumber	\\	
&=& \sigma^2,
\\	\nonumber	\\
\bm \Delta p^2  = \expval{\hat p^2}{0}
&=& - \frac{\hbar^2}{4\sigma^2} \expval{\qty( \hat a^\dagger - \hat a )\qty( \hat a^\dagger - \hat a )}{0}
\nonumber	\\
&=&  \frac{\hbar^2}{4\sigma^2} \expval{\qty( 2\hat a^\dagger \hat a + 1 )}{0},		\hspace{1cm} : \qty{ \comm{\hat a}{\hat a^\dagger}  = \unitop },
\nonumber	\\
&=&  \frac{\hbar^2}{4\sigma^2},
\\	\nonumber	\\
\textbf{Cov}(x,p) = \frac{1}{2} \expval{\ev{\hat x \hat p + \hat p \hat x}}{0} 
&=& \frac{1}{2} \ev{\hat x \hat p}{0} - \frac{i\hbar}{2}, 	\hspace{2cm} : \qty{ \comm{\hat x}{\hat p} = i\hbar },
\nonumber	\\
&=&  \sigma \qty(\frac{i\hbar}{2\sigma}) 	\expval{\qty( \hat a^\dagger + \hat a )\qty( \hat a^\dagger - \hat a )}{0}		- \frac{i\hbar}{2} 
\nonumber	\\
&=& 	\frac{i\hbar}{2} 	\expval{ \comm{\hat a}{\hat a^\dagger} }{0}	- 	\frac{i\hbar}{2} 
\nonumber	\\
&=& 0 ,	\hspace{4cm} : \qty{ \comm{\hat a}{\hat a^\dagger}  = \unitop } .
\end{eqnarray}
If this system describes the local modes  in the bipartite covariance matrix of Appendix~\ref{appendix:EntangFormalism},
\begin{equation}
\bm{\alpha} =	
\begin{pmatrix}
\sigma^2	&	0	\\
0	& \hbar^2/4\sigma^2 
\end{pmatrix},
\hspace{.5cm}
\implies
\hspace{.5cm}
\bm{\sigma} = \begin{pmatrix}
\bm{\alpha}	& 0 \\
0	& \bm{\alpha}
\end{pmatrix}.
\end{equation}

 \subsection{Thermal state}

The variances and the correlation for the mixed thermal state are
\begin{eqnarray}
\bm \Delta x^2 = \Tr( \hat x^2 \ \hat  \rho_\text{th} )
&=& \sigma^2   \sum_{n = 0}^{\infty}  \frac{\bar{n}^{n}}{(\bar{n}+1)^{n+1}} \expval{ \qty( \hat a^\dagger + \hat a ) \qty( \hat a^\dagger + \hat a ) }{n}
\nonumber	\\
&=&  \sigma^2  \sum_{n = 0}^{\infty}  \frac{\bar{n}^{n}}{(\bar{n}+1)^{n+1}} \expval{ \qty( 2\hat a^\dagger \hat a + \unitop ) }{n}, \hspace{.5cm} : \qty{ \comm{\hat a}{\hat a^\dagger}  = \unitop },
\nonumber	\\
&=&  \sigma^2  \Tr[ \qty( 2\hat a^\dagger \hat a + \unitop ) \  \hat  \rho_\text{th} ]
\nonumber	\\
&=&  \qty( 2\bar{n} + 1 ) \sigma^2 ,
\\	\nonumber	\\
\bm \Delta p^2 = \Tr( \hat p^2 \ \hat  \rho_\text{th} )
&=& - \frac{\hbar^2}{4\sigma^2} \sum_{n = 0}^{\infty}  \frac{\bar{n}^{n}}{(\bar{n}+1)^{n+1}} \expval{ \qty( \hat a^\dagger - \hat a ) \qty( \hat a^\dagger - \hat a ) }{n}
\nonumber	\\
&=&  \frac{\hbar^2}{4\sigma^2}  \sum_{n = 0}^{\infty} \frac{\bar{n}^{n}}{(\bar{n}+1)^{n+1}} \expval{ \qty( 2\hat a^\dagger \hat a + \unitop ) }{n},
\hspace{.5cm} : \qty{ \comm{\hat a}{\hat a^\dagger}  = \unitop },
\nonumber	\\
&=& \frac{\hbar^2}{4\sigma^2}  \Tr[ \qty( 2\hat a^\dagger \hat a + \unitop ) \ \hat   \rho_\text{th} ]
\nonumber	\\
&=&  \qty( 2\bar{n} + 1 )  \frac{\hbar^2}{4\sigma^2}  ,
\\	\nonumber	\\
\textbf{Cov}(x,p) =  \frac{1}{2}  \Tr[ (\hat x \hat p + \hat p \hat x) \ \hat \rho_\text{th} ]
&=& \Tr( \hat x \hat p \ \hat  \rho_\text{th} ) - \frac{i\hbar}{2},	
\hspace{2cm}	: \qty{  \comm{\hat x}{\hat p} = i\hbar },
\nonumber	\\
&=&  \sigma \qty(\frac{i\hbar}{2\sigma})   \sum_{n = 0}^{\infty}  \frac{\bar{n}^{n}}{(\bar{n}+1)^{n+1}} \expval{ \qty( \hat a^\dagger + \hat a ) \qty( \hat a^\dagger - \hat a ) }{n}	-  \frac{i\hbar}{2} 
\nonumber	\\
&=& \frac{i\hbar}{2}   \sum_{n = 0}^{\infty}  \frac{\bar{n}^{n}}{(\bar{n}+1)^{n+1}} \expval{ \comm{\hat a}{\hat a^\dagger}  }{n}	- \frac{i\hbar}{2} 
\nonumber	\\
&=& \frac{i\hbar}{2} \Tr( \hat \rho_\text{th} ) - \frac{i\hbar}{2},  
\hspace{2cm} : \qty{ \comm{\hat a}{\hat a^\dagger}  = \unitop },
\nonumber	\\
&& = 0.
\end{eqnarray}
Hence, the corresponding local mode $\bm{\alpha}_\text{th}$ is
\begin{equation}
\bm{\alpha}_\text{th} 
=	
\begin{pmatrix}
\qty( 2\bar{n} + 1 ) \sigma^2	&	0	\\
0	&	\qty( 2\bar{n} + 1 ) \hbar^2/4\sigma^2
\end{pmatrix}
=	
\qty( 2\bar{n} + 1 ) \begin{pmatrix} \sigma^2	&	0	\\
0	& \hbar^2/4\sigma^2
\end{pmatrix}
=
\qty( 2\bar{n} + 1 ) \bm{\alpha} .
\end{equation}
Accordingly, the covariance matrix for thermal state with identical local modes is
\begin{equation}
\bm{\sigma}_\text{th}(0)
=
 \begin{pmatrix}
\bm{\alpha}_\text{th}	& 	0	\\
0	& \bm{\alpha}_\text{th}
\end{pmatrix}
=
\begin{pmatrix}
\qty( 2\bar{n} + 1 )\bm{\alpha}	& 	0	\\
0	& \qty( 2\bar{n} + 1 )\bm{\alpha}	
\end{pmatrix}
=
\qty( 2\bar{n} + 1 ) \begin{pmatrix}
\bm{\alpha}	& 	0	\\
0	& \bm{\alpha}	
\end{pmatrix}
=
\qty( 2\bar{n} + 1 ) \bm{\sigma}(0) .
\end{equation}
Compared to the ground state $\bm \sigma(0)$, this is only multiplied by a temperature dependent factor of $2\bar{n}+1$.
At low pressures the environmental impacts can be ignored, and this form is maintained approximately at all times~\cite{phdthesis-Tanjung}: $\bm{\sigma}_\text{th} \approx (2\bar{n}+1)\bm{\sigma}, \ \forall  t$,
where $\bm{\sigma}_\text{th}$ is the covariance matrix of a state at time $t$ that begins evolution as a thermal state and similarly for $\bm{\sigma}$.
The symplectic eigenvalues of the partially transposed matrix  $\tilde{\bm{\sigma}}_\text{th}$ are ~\cite{PRA_70.022318,PRA_65.032314}
\begin{equation}
	\tilde{\nu}_{\mp}^\text{(th)} = 
	\sqrt{ \tilde{\Sigma}(\bm{\sigma}_\text{th}) \mp \sqrt{\tilde{\Sigma}^2(\bm{\sigma}_\text{th}) - 4 \ \text{Det}\qty( \bm{\sigma}_\text{th} ) } } \Bigg/ \sqrt{2}.
\end{equation}
We can substitute $\text{Det} ( \bm{\sigma}_\text{th} ) = (2\bar{n}+1)^4 \ \text{Det}(\bm{\sigma})$ and $\tilde{\Sigma}(\bm{\sigma}_\text{th}) = (2\bar{n}+1)^2 \ \tilde{\Sigma}(\bm{\sigma})$ to see that $\tilde{\nu}_{\mp}^\text{(th)} = (2\bar{n}+1) \tilde{\nu}_{\mp}$.
Accordingly, the entanglement negativity of the state evolved from a thermal state is related to that evolved from the zero-temperature ground state by
\begin{eqnarray}
E(\bm{\sigma}_\text{th})  &=& 
\max \qty[ 0 ,  - \log_2 \qty( \frac{  \tilde{\nu}_{-}^\text{(th)} }{\hbar/2} ) ]
\nonumber   \\
&=& \max \qty[ 0 ,  - \log_2 \qty( \frac{  (2\bar{n}+1) \tilde{\nu}_{-} }{\hbar/2} ) ]
\nonumber   \\
&=& \max \qty[ 0 ,  - \log_2 \qty( \frac{  \tilde{\nu}_{-} }{\hbar/2} ) - \log_2 (2\bar{n}+1) ]
\nonumber   \\
&=& \max \Big[ 0 ,  E(\bm\sigma) - \log_2 (2\bar{n}+1) \Big] . 
\end{eqnarray}

 \section{Probability density functions}

In the position representation the thermal density matrix is given by
\begin{eqnarray}
\rho_\text{th}(x) = \expval{\hat \rho_\text{th}}{x}
&=& \sum_{n = 0}^{\infty} \frac{\bar{n}^{n}}{(\bar{n}+1)^{n+1}} \abs{\psi_n(x)}^2,	\hspace{1cm}	: \qty{ \braket{x}{n} = \psi_n(x) },
\nonumber	\\
&=&	\frac{1}{\sigma \sqrt{2\pi}} \sum_{n = 0}^{\infty} \frac{\bar{n}^{n}}{(\bar{n}+1)^{n+1}} \frac{1}{2^{n}n!} \ \exp\qty(-\frac{x^2}{2\sigma^2}) \ H_n^2\qty(\frac{x}{\sigma\sqrt{2}}).
\end{eqnarray}
We can now utilize the Mehler's formula,
\begin{equation}
\sum_{n = 0}^{\infty} \frac{x_0^n}{2^nn!} e^{-x^2} H_n^2(x) = \frac{1}{\sqrt{1-x_0^2}}	\
 \exp\qty[ - \qty( \frac{1-x_0}{1+x_0} ) x^2  ],
\end{equation}
to sum up the infinite series in the density matrix as
\begin{eqnarray}
\rho_\text{th}(x) &=& \frac{1}{\sigma(\bar{n}+1)\sqrt{2\pi}} \sum_{n = 0}^{\infty} \qty( \frac{\bar{n}}{\bar{n}+1} )^{n} \frac{1}{2^{n}n!} \ \exp\qty(-\frac{x^2}{2\sigma^2}) \ H_n^2\qty(\frac{x}{\sigma\sqrt{2}})
\nonumber	\\
&=& \frac{1}{\sigma(\bar{n}+1)\sqrt{2\pi}}  \frac{1}{ \sqrt{ 1 - \qty(\frac{\bar{n}}{\bar{n}+1})^2  } } \ \exp\qty[ -  \frac{ 1-\qty(\frac{\bar{n}}{\bar{n}+1}) }{ 1 + \qty(\frac{\bar{n}}{\bar{n}+1}) } \frac{x^2}{2\sigma^2}  ]
\nonumber	\\
&=& \frac{1}{\sigma\sqrt{2\bar{n}+1}\sqrt{2\pi}}  \ \exp\qty( -  \frac{x^2}{2(2\bar{n}+1)\sigma^2}  ).
\label{eq:ThermalGaussian_PosSpace}
\end{eqnarray}

In the momentum space we can use the Fourier eigenfunctions,
\begin{equation}
\tilde \psi(p) = \braket{p}{n} = \frac{(-i)^n}{ \sqrt{2^n n! \sqrt{2\pi} \ \hbar/2\sigma}} \ \exp(-\frac{p^2}{\hbar^2/\sigma^2})  \ H_n\qty( \frac{p}{\hbar/\sigma\sqrt{2}} ),
\end{equation}
and derive
\begin{eqnarray}
\tilde \rho_\text{th}(p) = \expval{\hat \rho_\text{th}}{p}
&=&   \sum_{n = 0}^{\infty}  \frac{\bar{n}^{n}}{(\bar{n}+1)^{n+1}}  \abs{\tilde \psi(p)}^2, \hspace{1cm}	: \qty{ \braket{p}{n} = \tilde \psi_n(p ) },
\nonumber	\\
&=& \frac{1}{(\bar{n}+1) \sqrt{2\pi} \ \hbar/2\sigma}  \sum_{n = 0}^{\infty} \qty( \frac{\bar{n}}{\bar{n}+1} )^{n}  \frac{1}{2^nn!} \ \exp(-\frac{p^2}{\hbar^2/2\sigma^2}) \ H_n^2\qty( \frac{p}{\hbar/\sigma\sqrt{2}} )
\nonumber	\\
&=& \frac{1}{\sqrt{2\bar{n}+1} \sqrt{2\pi} \ \hbar/2\sigma} \   \exp\qty( - \frac{p^2}{2(2\bar{n}+1)\hbar^2/4\sigma^2}  ).
\label{eq:ThermalGaussian_MomSpace}
\end{eqnarray}
This proves that thermal states are Gaussian, with their position and momentum variances are amplified by a factor of $2\bar{n}+1$ as compared to the ground state.

 \section{Wigner function}

In a more formal way, the Gaussianity of thermal states can be proved by showing Gaussianity of their corresponding Wigner function:
\begin{eqnarray}
W_\text{th}(x,p) &=& \frac{1}{\pi \hbar} \int_{-\infty}^{+\infty} dy \ \bra{x - \frac{y}{2}} \hat \rho_\text{th} \ket{x - \frac{y}{2}} e^{-ipy/\hbar}
\nonumber	\\
&=&  \sum_{n = 0}^{\infty} \frac{\bar{n}^{n}}{(\bar{n}+1)^{n+1}} \frac{1}{\pi \hbar} \int_{-\infty}^{+\infty} dy \ \braket{x-\frac{y}{2}}{n} \braket{n}{x+\frac{y}{2}} e^{-ipy/\hbar}
\nonumber	\\
&=&  \sum_{n = 0}^{\infty} \frac{\bar{n}^{n}}{(\bar{n}+1)^{n+1}} \frac{1}{\pi \hbar} \int_{-\infty}^{+\infty} dy \ \psi_n\qty(x-\frac{y}{2}) \psi_n^*\qty(x+\frac{y}{2}) e^{-ipy/\hbar}
\nonumber	\\
&=&  \sum_{n = 0}^{\infty} \frac{\bar{n}^{n}}{(\bar{n}+1)^{n+1}}	\ W_n(x,p),
\end{eqnarray}
where $W_n(x,p)$ is the Wigner function of the eigenstate $\psi_n$. This is given by
\begin{eqnarray}
W_n(x,p) &=& \frac{1}{\pi \hbar} \int_{-\infty}^{+\infty} dy \ \psi_n\qty(x-\frac{y}{2}) \ \psi_n^*\qty(x+\frac{y}{2}) \ e^{-ipy/\hbar}
\nonumber	\\
&=& \frac{1}{2^{n} n! \pi \hbar \sigma \sqrt{2\pi}} \int_{-\infty}^{+\infty} dy \ \exp\qty( -\frac{x^2+(y/2)^2}{2\sigma^2} ) 
\nonumber   \\
&& \hspace{3cm} \times 
H_n\qty(\frac{x-y/2}{\sigma\sqrt{2}}) \ 
H_n\qty(\frac{x+y/2}{\sigma\sqrt{2}}) \ e^{-ipy/\hbar}
\nonumber	\\
&\equiv&	\frac{(-1)^n}{\pi\hbar} \ \exp\qty[-\frac{1}{2}\qty( \frac{x^2}{\sigma^2} + \frac{p^2}{\hbar^2/4\sigma^2})] \ L_n\qty( \frac{x^2}{\sigma^2} + \frac{p^2}{\hbar^2/4\sigma^2}) ,	\hspace{.25cm}	: \qty{\text{\cite{book-QuantMechInPhaseSpace}}} , 
\end{eqnarray}
where $L_n$ is the Laguerre polynomial of $n^\text{th}$ order. Hence,
\begin{equation}
W_\text{th}(x,p) = \frac{1}{\pi\hbar(\bar{n}+1)}	\
 \exp\qty[-\frac{1}{2}\qty( \frac{x^2}{\sigma^2} + \frac{p^2}{\hbar^2/4\sigma^2})] \ \sum_{n = 0}^{\infty} \qty( -\frac{\bar{n}}{\bar{n}+1})^{n} \ L_n\qty( \frac{x^2}{\sigma^2} + \frac{p^2}{\hbar^2/4\sigma^2})  .
\end{equation}
One can use the generating function of the Laguerre polynomials,
\begin{equation}
\sum_{n=0}^\infty t^n L_n(z) = \frac{1}{1-t} \exp( - \frac{tz}{1-t} ),
\end{equation}
to rewrite the Wigner function as
\begin{eqnarray}
W_\text{th}(x,p) &=&	 \frac{1}{\pi\hbar(\bar{n}+1)}	\
 \exp\qty[-\frac{1}{2}\qty( \frac{x^2}{\sigma^2} + \frac{p^2}{\hbar^2/4\sigma^2})] 
\nonumber   \\
&& \hspace{1cm} \times \frac{1}{1 + \qty(\frac{\bar{n}}{\bar{n}+1})}	\
 \exp[ \frac{ \qty(\frac{\bar{n}}{\bar{n}+1}) }{1 + \qty( \frac{\bar{n}}{\bar{n}+1}) } \qty( \frac{x^2}{\sigma^2} + \frac{p^2}{\hbar^2/4\sigma^2}) ]
\nonumber	\\
&=& \frac{1}{\pi\hbar(2\bar{n}+1)} \ \exp[  \qty(- \frac{1}{2}+\frac{\bar{n}}{{2\bar{n}+1}})  \qty( \frac{x^2}{\sigma^2} + \frac{p^2}{\hbar^2/4\sigma^2}) ]
\nonumber	\\
&=&  \frac{1}{\pi\hbar(2\bar{n}+1)}   \ \exp[ - \frac{1}{2(2\bar{n}+1)} \qty( \frac{x^2}{\sigma^2} + \frac{p^2}{\hbar^2/4\sigma^2}) ]
\nonumber	\\
&\equiv&	\rho_\text{th}(x) 	\ 	\tilde \rho_\text{th}(p).
\end{eqnarray}
The same result can be derived much more easily by using the relation between the Wigner function and the covariance matrix~\cite{GaussianOperations-Brask}.
The absence of a $xp$ coupling term implies that the thermal state of a harmonic oscillator is an `uncorrelated' Gaussian.

%% file: Thesis.bbl
\begin{thebibliography}{100}

\bibitem{QuChem-rug01:001382226}
Ira~N. Levine.
\newblock ``{Quantum Chemistry}''.
\newblock Prentice-Hall, USA. ~(2000).
\newblock 5th edition.

\bibitem{QuBio-Lambert2013}
N.~Lambert, Yueh-Nan Chen, Yuan-Chung Cheng, Che-Ming Li, Guang-Yin Chen, and
  Franco Nori.
\newblock ``{Quantum Biology}''.
\newblock \href{https://dx.doi.org/10.1038/nphys2474}{Nature Physics {\bf 9},
  10}~(2013).

\bibitem{nielsen_chuang_2010}
Michael~A. Nielsen and Issac~L. Chuang.
\newblock ``{Quantum Computation and Quantum Information}''.
\newblock \href{https://dx.doi.org/10.1017/CBO9780511976667}{Cambridge
  University Press}. ~(2010).

\bibitem{MRI-Machine}
P.~C. {Lauterbur}.
\newblock ``{Image Formation by Induced Local Interactions: Examples Employing
  Nuclear Magnetic Resonance}''.
\newblock \href{https://dx.doi.org/10.1038/242190a0}{Nature {\bf 242},
  190}~(1973).

\bibitem{paper-EPR}
A.~Einstein, B.~Podolsky, and N.~Rosen.
\newblock ``{Can Quantum-Mechanical Description of Physical Reality Be
  Considered Complete?}''.
\newblock \href{https://dx.doi.org/10.1103/PhysRev.47.777}{Physical Review {\bf
  47}, 777}~(1935).

\bibitem{schrodinger_1935}
E.~Schr\"odinger.
\newblock ``{Discussion of Probability Relations between Separated Systems}''.
\newblock \href{https://dx.doi.org/10.1017/S0305004100013554}{Mathematical
  Proceedings of the Cambridge Philosophical Society {\bf 31}, 555}~(1935).

\bibitem{JohnBell1964}
J.~S. Bell.
\newblock ``{On the Einstein Podolsky Rosen paradox}''.
\newblock \href{https://dx.doi.org/10.1103/PhysicsPhysiqueFizika.1.195}{Physics
  Physique Fizika {\bf 1}, 195}~(1964).

\bibitem{Clauser1972}
S.~J. Freedman and J.~F. Clauser.
\newblock ``{Experimental Test of Local Hidden-Variable Theories}''.
\newblock \href{https://dx.doi.org/10.1103/PhysRevLett.28.938}{Physical Review
  Letters {\bf 28}, 938}~(1972).

\bibitem{QuantumComputing}
R.~P. Feynman.
\newblock ``{Simulating physics with computers}''.
\newblock \href{https://dx.doi.org/10.1007/BF02650179}{International Journal of
  Theoretical Physics {\bf 21}, 467}~(1982).

\bibitem{QuantumMetrology}
C.~M. Caves.
\newblock ``{Quantum limits on noise in linear amplifiers}''.
\newblock \href{https://dx.doi.org/10.1103/PhysRevD.26.1817}{Physical Review D
  {\bf 26}, 1817}~(1982).

\bibitem{QuantumCryptography}
A.~K. Ekert.
\newblock ``{Quantum cryptography based on Bell's theorem}''.
\newblock \href{https://dx.doi.org/10.1103/PhysRevLett.67.661}{Physical Review
  Letters {\bf 67}, 661}~(1991).

\bibitem{QuantumDenseCoding}
C.~H. Bennett and S.~J. Wiesner.
\newblock ``{Communication via one- and two-particle operators on
  Einstein-Podolsky-Rosen states}''.
\newblock \href{https://dx.doi.org/10.1103/PhysRevLett.69.2881}{Physical Review
  Letters {\bf 69}, 2881}~(1992).

\bibitem{QuantumTeleportation}
C.~H. Bennett, G.~Brassard, C.~Cr\'epeau, R.~Jozsa, A.~Peres, and William~K.
  Wootters.
\newblock ``{Teleporting an unknown quantum state via dual classical and
  Einstein-Podolsky-Rosen channels}''.
\newblock \href{https://dx.doi.org/10.1103/PhysRevLett.70.1895}{Physical Review
  Letters {\bf 70}, 1895}~(1993).

\bibitem{RevModPhys.81.865}
R.~Horodecki, P.~Horodecki, M.~Horodecki, and K.~Horodecki.
\newblock ``{Quantum entanglement}''.
\newblock \href{https://dx.doi.org/10.1103/RevModPhys.81.865}{Reviews of Modern
  Physics {\bf 81}, 865}~(2009).

\bibitem{StringTheory}
Michael~B. Green and John~H. Schwarz.
\newblock ``{Anomaly cancellations in supersymmetric $D = 10$ gauge theory and
  superstring theory}''.
\newblock \href{https://dx.doi.org/10.1016/0370-2693(84)91565-X}{Physics
  Letters B {\bf 149}, 117}~(1984).

\bibitem{LoopQuantumGravity}
A.~Ashtekar.
\newblock ``{New Variables for Classical and Quantum Gravity}''.
\newblock \href{https://dx.doi.org/10.1103/PhysRevLett.57.2244}{Physical Review
  Letters {\bf 57}, 2244}~(1986).

\bibitem{TwistorTheory}
R.~Penrose and M.~A.~H. MacCallum.
\newblock ``{Twistor theory: An approach to the quantisation of fields and
  space-time}''.
\newblock \href{https://dx.doi.org/10.1016/0370-1573(73)90008-2}{Physics
  Reports {\bf 6}, 241}~(1973).

\bibitem{CanonicalQuantumGravity}
B.~S. {DeWitt}.
\newblock ``{Quantum Theory of Gravity. I. The Canonical Theory}''.
\newblock \href{https://dx.doi.org/10.1103/PhysRev.160.1113}{Physical Review
  {\bf 160}, 1113}~(1967).

\bibitem{PhysRevLett.4.337}
R.~V. Pound and G.~A. Rebka.
\newblock ``{Apparent Weight of Photons}''.
\newblock \href{https://dx.doi.org/10.1103/PhysRevLett.4.337}{Physical Review
  Letters {\bf 4}, 337}~(1960).

\bibitem{doi:10.1126/science.177.4044.168}
J.~C. Hafele and R.~E. Keating.
\newblock ``{Around-the-World Atomic Clocks: Observed Relativistic Time
  Gains}''.
\newblock \href{https://dx.doi.org/10.1126/science.177.4044.168}{Science {\bf
  177}, 168}~(1972).

\bibitem{PhysRevLett.34.1472}
R.~Colella, A.~W. Overhauser, and S.~A. Werner.
\newblock ``{Observation of Gravitationally Induced Quantum Interference}''.
\newblock \href{https://dx.doi.org/10.1103/PhysRevLett.34.1472}{Physical Review
  Letters {\bf 34}, 1472}~(1975).

\bibitem{Nesvizhevsky2002}
V.~V. Nesvizhevsky, H.~G. B{\"o}rner, A.~K. Petukhov, H.~Abele, S.~Bae{\ss}ler,
  F.~J. Rue{\ss}, T.~St{\"o}ferle, A.~Westphal, A.~M. Gagarski, G.~A. Petrov,
  and A.~V. Strelkov.
\newblock ``{Quantum states of neutrons in the Earth's gravitational field}''.
\newblock \href{https://dx.doi.org/10.1038/415297a}{Nature {\bf 415},
  297}~(2002).

\bibitem{Peters1999}
A.~Peters, K.~Yeow Chung, and S.~Chu.
\newblock ``{Measurement of gravitational acceleration by dropping atoms}''.
\newblock \href{https://dx.doi.org/10.1038/23655}{Nature {\bf 400},
  849}~(1999).

\bibitem{PhysRevLett.118.183602}
P.~Asenbaum, C.~Overstreet, T.~Kovachy, D.~D. Brown, J.~M. Hogan, and M.~A.
  Kasevich.
\newblock ``{Phase Shift in an Atom Interferometer due to Spacetime Curvature
  across its Wave Function}''.
\newblock \href{https://dx.doi.org/10.1103/PhysRevLett.118.183602}{Physical
  Review Letters {\bf 118}, 183602}~(2017).

\bibitem{PhysRevLett.119.120402}
T.~Krisnanda, M.~Zuppardo, M.~Paternostro, and T.~Paterek.
\newblock ``{Revealing Nonclassicality of Inaccessible Objects}''.
\newblock \href{https://dx.doi.org/10.1103/PhysRevLett.119.120402}{Physical
  Review Letters {\bf 119}, 120402}~(2017).

\bibitem{PhysRevLett.119.240401}
S.~Bose, A.~Mazumdar, G.~W. Morley, H.~Ulbricht, M.~Toro\ifmmode~\check{s}\else
  \v{s}\fi{}, M.~Paternostro, A.~A. Geraci, P.~F. Barker, M.~S. Kim, and
  G.~Milburn.
\newblock ``{Spin Entanglement Witness for Quantum Gravity}''.
\newblock \href{https://dx.doi.org/10.1103/PhysRevLett.119.240401}{Physical
  Review Letters {\bf 119}, 240401}~(2017).

\bibitem{PhysRevLett.119.240402}
C.~Marletto and V.~Vedral.
\newblock ``{Gravitationally Induced Entanglement between Two Massive Particles
  is Sufficient Evidence of Quantum Effects in Gravity}''.
\newblock \href{https://dx.doi.org/10.1103/PhysRevLett.119.240402}{Physical
  Review Letters {\bf 119}, 240402}~(2017).

\bibitem{PhysRevA.98.043811}
A.~Al~Balushi, W.~Cong, and R.~B. Mann.
\newblock ``{Optomechanical quantum Cavendish experiment}''.
\newblock \href{https://dx.doi.org/10.1103/PhysRevA.98.043811}{Physical Review
  A {\bf 98}, 043811}~(2018).

\bibitem{npjQI_6.12}
T.~Krisnanda, G.~Y.~Tham, M.~Paternostro, and T.~Paterek.
\newblock ``{Observable quantum entanglement due to gravity}''.
\newblock \href{https://dx.doi.org/10.1038/s41534-020-0243-y}{npj Quantum
  Information {\bf 6}, 12}~(2020).

\bibitem{JOPB_23.235501}
S.~Qvarfort, S.~Bose, and A.~Serafini.
\newblock ``{Mesoscopic entanglement through central{\textendash}potential
  interactions}''.
\newblock \href{https://dx.doi.org/10.1088/1361-6455/abbe8d}{Journal of Physics
  B: Atomic, Molecular and Optical Physics {\bf 53}, 235501}~(2020).

\bibitem{Rijavec_2021}
Simone R., Matteo C., A.~Bassi, V.~Vedral, and C.~Marletto.
\newblock ``{Decoherence effects in non-classicality tests of gravity}''.
\newblock \href{https://dx.doi.org/10.1088/1367-2630/abf3eb}{New Journal of
  Physics {\bf 23}, 043040}~(2021).

\bibitem{PhysRevLett.128.143601}
K.~Kustura, C.~Gonzalez-Ballestero, Andr\'es de los~R\'{\i}os Sommer, N.~Meyer,
  R.~Quidant, and O.~Romero-Isart.
\newblock ``{Mechanical Squeezing via Unstable Dynamics in a Microcavity}''.
\newblock \href{https://dx.doi.org/10.1103/PhysRevLett.128.143601}{Physical
  Review Letters {\bf 128}, 143601}~(2022).

\bibitem{PhysRevA.102.062807}
T.~W. van~de Kamp, R.~J. Marshman, S.~Bose, and A.~Mazumdar.
\newblock ``{Quantum gravity witness via entanglement of masses: Casimir
  screening}''.
\newblock \href{https://dx.doi.org/10.1103/PhysRevA.102.062807}{Physical Review
  A {\bf 102}, 062807}~(2020).

\bibitem{PhysRevResearch.4.023087}
R.~J. Marshman, A.~Mazumdar, R.~Folman, and S.~Bose.
\newblock ``{Constructing nano-object quantum superpositions with a
  Stern-Gerlach interferometer}''.
\newblock \href{https://dx.doi.org/10.1103/PhysRevResearch.4.023087}{Physical
  Review Research {\bf 4}, 023087}~(2022).

\bibitem{PhysRevLett.128.110401}
J.~S. Pedernales, K.~Streltsov, and M.~B. Plenio.
\newblock ``{Enhancing Gravitational Interaction between Quantum Systems by a
  Massive Mediator}''.
\newblock \href{https://dx.doi.org/10.1103/PhysRevLett.128.110401}{Physical
  Review Letters {\bf 128}, 110401}~(2022).

\bibitem{Sidajaya_2022}
P.~Sidajaya, W.~Cong, and V.~Scarani.
\newblock ``{Possibility of detecting the gravity of an object frozen in a
  spatial superposition by the Zeno effect}''.
\newblock \href{https://dx.doi.org/10.1103/physreva.106.042217}{Physical Review
  A {\bf 106}, 042217}~(2022).

\bibitem{PhysRevA.105.032411}
M.~Schut, J.~Tilly, R.~J. Marshman, S.~Bose, and A.~Mazumdar.
\newblock ``{Improving resilience of quantum-gravity-induced entanglement of
  masses to decoherence using three superpositions}''.
\newblock \href{https://dx.doi.org/10.1103/PhysRevA.105.032411}{Physical Review
  A {\bf 105}, 032411}~(2022).

\bibitem{phdthesis-Tanjung}
T.~Krisnanda.
\newblock ``{Distribution of quantum entanglement: Principles and
  applications}''.
\newblock PhD thesis.
\newblock Nanyang Technological University, Singapore.
\newblock ~(2020).

\bibitem{RMP_86.1391}
M.~Aspelmeyer, T.~J. Kippenberg, and F.~Marquardt.
\newblock ``{Cavity optomechanics}''.
\newblock \href{https://dx.doi.org/10.1103/RevModPhys.86.1391}{Reviews of
  Modern Physics {\bf 86}, 1391}~(2014).

\bibitem{Teufel2011}
J.~D. Teufel, T.~Donner, Dale Li, J.~W. Harlow, M.~S. Allman, K.~Cicak, A.~J.
  Sirois, J.~D. Whittaker, K.~W. Lehnert, and R.~W. Simmonds.
\newblock ``{Sideband cooling of micromechanical motion to the quantum ground
  state}''.
\newblock \href{https://dx.doi.org/10.1038/nature10261}{Nature {\bf 475},
  359}~(2011).

\bibitem{Nature.478.89}
J.~Chan, T.~P.~M. Alegre, A.~H. Safavi-Naeini, J.~T. Hill, A.~Krause,
  S.~Gr\"oblacher, M.~Aspelmeyer, and O.~Painter.
\newblock ``{Laser cooling of a nanomechanical oscillator into its quantum
  ground state}''.
\newblock \href{https://dx.doi.org/10.1038/nature10461}{Nature {\bf 478},
  89}~(2011).

\bibitem{NJP_11.073032}
B.~Abbott et~al.
\newblock ``{Observation of a kilogram-scale oscillator near its quantum ground
  state}''.
\newblock \href{https://dx.doi.org/10.1088/1367-2630/11/7/073032}{New Journal
  of Physics {\bf 11}, 073032}~(2009).

\bibitem{Palomaki-Science}
T.~A. Palomaki, J.~D. Teufel, R.~W. Simmonds, and K.~W. Lehnert.
\newblock ``{Entangling Mechanical Motion with Microwave Fields}''.
\newblock \href{https://dx.doi.org/10.1126/science.1244563}{Science {\bf 342},
  710}~(2013).

\bibitem{Nature.556.473}
R.~Riedinger, A.~Wallucks, I.~Marinkovi\'c, C.~L\"oschnauer, M.~Aspelmeyer,
  S.~Hong, and S.~Gr\"oblacher.
\newblock ``{Remote quantum entanglement between two micromechanical
  oscillators}''.
\newblock \href{https://dx.doi.org/10.1038/s41586-018-0036-z}{Nature {\bf 556},
  473}~(2018).

\bibitem{PhysRevLett.121.220404}
I.~Marinkovi\ifmmode~\acute{c}\else \'{c}\fi{}, A.~Wallucks, R.~Riedinger,
  S.~Hong, M.~Aspelmeyer, and S.~Gr\"oblacher.
\newblock ``{Optomechanical Bell Test}''.
\newblock \href{https://dx.doi.org/10.1103/PhysRevLett.121.220404}{Physical
  Review Letters {\bf 121}, 220404}~(2018).

\bibitem{GaussEvolFreeSpace_SMBlinder}
S.~M. Blinder.
\newblock ``{Evolution of a Gaussian Wavepacket}''.
\newblock \href{https://dx.doi.org/10.1119/1.1974961}{American Journal of
  Physics {\bf 36}, 525}~(1968).

\bibitem{GaussEvolHarmOsc_Tsuru}
H.~Tsuru.
\newblock ``{Wave Packet Motion in Harmonic Potential}''.
\newblock \href{https://dx.doi.org/10.1143/JPSJ.60.3657}{Journal of the
  Physical Society of Japan {\bf 60}, 3657}~(1991).

\bibitem{FEIT1982412}
M.~D. Feit, J.~A. Fleck, and A.~Steiger.
\newblock ``{Solution of the Schr\"odinger equation by a spectral method}''.
\newblock \href{https://dx.doi.org/10.1016/0021-9991(82)90091-2}{Journal of
  Computational Physics {\bf 47}, 412}~(1982).

\bibitem{Park1986}
T.~J. Park and J.~C. Light.
\newblock ``{Unitary quantum time evolution by iterative Lanczos reduction}''.
\newblock \href{https://dx.doi.org/10.1063/1.451548}{The Journal of Chemical
  Physics {\bf 85}, 5870--5876}~(1986).

\bibitem{BANDRAUK1991428}
A.~D. Bandrauk and H.~Shen.
\newblock ``{Improved exponential split operator method for solving the
  time-dependent Schrödinger equation}''.
\newblock \href{https://dx.doi.org/10.1016/0009-2614(91)90232-X}{Chemical
  Physics Letters {\bf 176}, 428}~(1991).

\bibitem{Muller1999}
H.~G. Muller.
\newblock ``{An efficient propagation scheme for the time-dependent
  Schr\"odinger equation in the velocity gauge}''.
\newblock Laser Physics {\bf 9}, 138~(1999).

\bibitem{Nurhuda1999}
M.~Nurhuda and F.~H.~M. Faisal.
\newblock ``{Numerical solution of time-dependent Schr\"odinger equation for
  multiphoton processes: A matrix iterative method}''.
\newblock \href{https://dx.doi.org/10.1103/PhysRevA.60.3125}{Physical Review A
  {\bf 60}, 3125}~(1999).

\bibitem{Watanabe2000}
N.~Watanabe and M.~Tsukada.
\newblock ``{Fast and Stable Method for Simulating Quantum Electron
  Dynamics}''.
\newblock \href{https://dx.doi.org/10.1143/PTPS.138.115}{Progress of
  Theoretical Physics Supplement {\bf 138}, 115}~(2000).

\bibitem{book_FiniteDiff_JWThomas}
J.~W. Thomas.
\newblock ``{Numerical Partial Differential Equations: Finite Difference
  Methods}''.
\newblock Springer-Verlag, USA. ~(1995).
\newblock 1st edition.

\bibitem{Ankit_TDSE_Zenodo}
{Ankit Kumar}.
\newblock ``{
  \href{https://doi.org/10.5281/zenodo.7275667}{https://doi.org/10.5281/zenodo.7275667}
  }''.

\bibitem{Ankit_TDSE_GitHub}
{Ankit Kumar}.
\newblock
  ``\href{https://github.com/vyason/Cayley-TDSE}{https://github.com/vyason/Cayley-TDSE}''.

\bibitem{Ankit_2022_TDSE}
A.~{Kumar} and P.~{Arumugam}.
\newblock ``{An accurate pentadiagonal matrix solution for the time-dependent
  Schr{\"o}dinger equation}''~(2022).
\newblock  \href{http://arxiv.org/abs/2205.13467}{arXiv:2205.13467}.

\bibitem{KOSLOFF198335}
D.~Kosloff and R.~Kosloff.
\newblock ``{A fourier method solution for the time dependent Schrödinger
  equation as a tool in molecular dynamics}''.
\newblock \href{https://dx.doi.org/10.1016/0021-9991(83)90015-3}{Journal of
  Computational Physics {\bf 52}, 35}~(1983).

\bibitem{Ankit_2022_Gravity}
A.~Kumar, T.~Krisnanda, P.~Arumugam, and T.~Paterek.
\newblock ``{Continuous-{V}ariable {E}ntanglement through {C}entral {F}orces:
  {A}pplication to {G}ravity between {Q}uantum {M}asses}''.
\newblock \href{https://dx.doi.org/10.22331/q-2023-05-15-1008}{{Quantum} {\bf
  7}, 1008}~(2023).

\bibitem{Ankit_2021_Quantum}
A.~Kumar, T.~Krisnanda, P.~Arumugam, and T.~Paterek.
\newblock ``{Nonclassical trajectories in head-on collisions}''.
\newblock \href{https://dx.doi.org/10.22331/q-2021-07-19-506}{{Quantum} {\bf
  5}, 506}~(2021).

\bibitem{CoP_PPuschnig}
P.~Puschnig.
\newblock ``{Computerorientierte Physik}''~(2016).

\bibitem{RutherfordExp_Students}
H.~Geiger, E.~Marsden, and E.~Rutherford.
\newblock ``{On a diffuse reflection of the $\alpha$-particles}''.
\newblock \href{https://dx.doi.org/10.1098/rspa.1909.0054}{Proceedings of the
  Royal Society of London. Series A {\bf 82}, 495}~(1909).

\bibitem{RutherfordExp_Prof}
E.~Rutherford.
\newblock ``{LXXIX. The scattering of $\alpha$ and $\beta$ particles by matter
  and the structure of the atom}''.
\newblock \href{https://dx.doi.org/10.1080/14786440508637080}{The London,
  Edinburgh, and Dublin Philosophical Magazine and Journal of Science {\bf 21},
  669}~(1911).

\bibitem{book_CJJoachain}
C.~J. Joachain.
\newblock ``{Quantum Collision Theory}''.
\newblock North-Holland Publishing Company, USA. ~(1975).
\newblock 3rd edition.

\bibitem{Ehrenfest1927}
P.~Ehrenfest.
\newblock ``{Bemerkung {\"u}ber die angen{\"a}herte G{\"u}ltigkeit der
  klassischen Mechanik innerhalb der Quantenmechanik}''.
\newblock \href{https://dx.doi.org/10.1007/BF01329203}{Zeitschrift f{\"u}r
  Physik {\bf 45}, 455}~(1927).

\bibitem{book-QMech-BCHall}
B.~C. Hall.
\newblock ``Quantum theory for mathematicians''.
\newblock Springer-Verlag, USA. ~(2013).
\newblock 1st edition.

\bibitem{book-QMech-MaxJammer}
M.~Jammer and E.~Merzbacher.
\newblock ``{The Conceptual Development of Quantum Mechanics}''.
\newblock \href{https://dx.doi.org/10.1063/1.3034186}{Physics Today {\bf 20},
  102}~(1967).

\bibitem{ZP-54.656}
R.~d.~E. Atkinson and F.~G. Houtermans.
\newblock ``{Zur Frage der Aufbaum{\"o}glichkeit der Elemente in Sternen}''.
\newblock \href{https://dx.doi.org/10.1007/BF01341595}{Zeitschrift f{\"u}r
  Physik {\bf 54}, 656}~(1929).

\bibitem{Nature-106.14}
A.~S. Eddington.
\newblock ``{The Internal Constitution of the Stars}''.
\newblock \href{https://dx.doi.org/10.1038/106014a0}{Nature {\bf 106},
  14}~(1920).

\bibitem{PR-40.621}
L.~A. MacColl.
\newblock ``{Note on the Transmission and Reflection of Wave Packets by
  Potential Barriers}''.
\newblock \href{https://dx.doi.org/10.1103/PhysRev.40.621}{Physical Review {\bf
  40}, 621}~(1932).

\bibitem{PLA-220.41}
V.~V. Dodonov, A.~B. Klimov, and V.~I. Man'ko.
\newblock ``{Low energy wave packet tunneling from a parabolic potential well
  through a high potential barrier}''.
\newblock \href{https://dx.doi.org/10.1016/0375-9601(96)00482-3}{Physics
  Letters A {\bf 220}, 41}~(1996).

\bibitem{PLA-225.303}
B.~B. Kadomtsev and M.~B. Kadomtsev.
\newblock ``{Wavefunctions of gas atoms}''.
\newblock \href{https://dx.doi.org/10.1016/S0375-9601(96)00804-3}{Physics
  Letters A {\bf 225}, 303}~(1997).

\bibitem{JPA-37.2423}
M.~A. Andreata and V.~V. Dodonov.
\newblock ``{Tunnelling of narrow Gaussian packets through delta potentials}''.
\newblock \href{https://dx.doi.org/10.1088/0305-4470/37/6/031}{Journal of
  Physics A: Mathematical and General {\bf 37}, 2423}~(2004).

\bibitem{PLA-378.1071}
A.~V. Dodonov and V.~V. Dodonov.
\newblock ``{Tunneling of slow quantum packets through the high Coulomb
  barrier}''.
\newblock \href{https://dx.doi.org/10.1016/j.physleta.2014.02.016}{Physics
  Letters A {\bf 378}, 1071}~(2014).

\bibitem{book_QM_Eisberg}
R.~Eisberg and R.~Resnick.
\newblock ``{Quantum Physics of Atoms, Molecules, Solids, Nuclei, and
  Particles}''.
\newblock John Wiley \& Sons, USA. ~(1985).
\newblock 2nd edition.

\bibitem{EJP-20.29}
M.~A. Doncheski and R.~W. Robinett.
\newblock ``{Anatomy of a quantum `bounce'}''.
\newblock \href{https://dx.doi.org/10.1088/0143-0807/20/1/009}{European Journal
  of Physics {\bf 20}, 29}~(1999).

\bibitem{AJP-66.252}
M.~Andrews.
\newblock ``{Wave packets bouncing off walls}''.
\newblock \href{https://dx.doi.org/10.1119/1.18854}{American Journal of Physics
  {\bf 66}, 252}~(1998).

\bibitem{AJP-35.177}
A.~Goldberg, H.~M. Schey, and J.~L. Schwartz.
\newblock ``{Computer-Generated Motion Pictures of One-Dimensional
  Quantum-Mechanical Transmission and Reflection Phenomena}''.
\newblock \href{https://dx.doi.org/10.1119/1.1973991}{American Journal of
  Physics {\bf 35}, 177}~(1967).

\bibitem{PhyScr-71.136}
M.~Belloni, M.~A. Doncheski, and R.~W. Robinett.
\newblock ``{Exact Results for {\textasciigrave}Bouncing{\textquotesingle}
  Gaussian Wave Packets}''.
\newblock \href{https://dx.doi.org/10.1238/Physica.Regular.071a00136}{Physica
  Scripta {\bf 71}, 136}~(2005).

\bibitem{ZP-39.828}
H.~A. Kramers.
\newblock ``{Wellenmechanik und halbzahlige Quantisierung}''.
\newblock \href{https://dx.doi.org/10.1007/BF01451751}{Zeitschrift f\"{u}r
  Physik {\bf 39}, 828}~(1926).

\bibitem{ZP-38.518}
G.~Wentzel.
\newblock ``{Eine Verallgemeinerung der Quantenbedingungen f\"{u}r die Zwecke
  der Wellenmechanik}''.
\newblock \href{https://dx.doi.org/10.1007/BF01397171}{Zeitschrift f\"{u}r
  Physik {\bf 38}, 518}~(1926).

\bibitem{AJP-81.405}
M.~Selmke and F.~Cichos.
\newblock ``{Photonic Rutherford scattering: A classical and quantum mechanical
  analogy in ray and wave optics}''.
\newblock \href{https://dx.doi.org/10.1119/1.4798259}{American Journal of
  Physics {\bf 81}, 405}~(2013).

\bibitem{ActaPhyPolB-33.2059}
W.~\.{Z}akowicz.
\newblock ``{Graphical Examples of Geometrical and Wave Optics}''.
\newblock Acta Physica Polonica B {\bf 33}, 2059~(2002).

\bibitem{ActaPhyPolA-101.369}
W.~\.{Z}akowicz.
\newblock ``{On the Extinction Paradox}''.
\newblock \href{https://dx.doi.org/10.12693/APhysPolA.101.369}{Acta Physica
  Polonica A {\bf 101}, 369}~(2002).

\bibitem{NP-78.409}
P.~J.~A. Buttle and L.~J.~B. Goldfarb.
\newblock ``{Neutron transfer in heavy ion reactions}''.
\newblock \href{https://dx.doi.org/10.1016/0029-5582(66)90617-1}{Nuclear
  Physics {\bf 78}, 409}~(1966).

\bibitem{NPA-148.529}
W.~von Oertzen.
\newblock ``{On the interaction induced by the exchange of nucleons between two
  identical nuclear cores}''.
\newblock \href{https://dx.doi.org/10.1016/0375-9474(70)90646-9}{Nuclear
  Physics A {\bf 148}, 529}~(1970).

\bibitem{PR-155.29}
B.~Imanishi and W.~von Oertzen.
\newblock ``{Molecular orbitals of nucleons in nucleus-nucleus collisions}''.
\newblock \href{https://dx.doi.org/10.1016/0370-1573(87)90101-3}{Physics
  Reports {\bf 155}, 29}~(1987).

\bibitem{PRC-61.054610}
J.~M. Sparenberg, D.~Baye, and B.~Imanishi.
\newblock ``{Coupled-reaction-channel calculations of the
  ${}^{16}\mathrm{O}{+}^{17}\mathrm{O}$ and
  ${}^{16}\mathrm{O}{+}^{17}\mathrm{F}$ charge-symmetric systems}''.
\newblock \href{https://dx.doi.org/10.1103/PhysRevC.61.054610}{Physical Review
  C {\bf 61}, 054610}~(2000).

\bibitem{NP-47.652}
S.~Edwards.
\newblock ``{Exchange terms in direct nuclear reaction theories}''.
\newblock \href{https://dx.doi.org/10.1016/0029-5582(63)90911-8}{Nuclear
  Physics {\bf 47}, 652}~(1963).

\bibitem{PTP-122.1055}
K.~Ogata, M.~Kan, and M.~Kamimura.
\newblock ``{Quantum Three-Body Calculation of the Nonresonant Triple-$\alpha$
  Reaction Rate at Low Temperatures}''.
\newblock \href{https://dx.doi.org/10.1143/PTP.122.1055}{Progress of
  Theoretical Physics {\bf 122}, 1055}~(2009).

\bibitem{PLB-772.1}
P.~Descouvemont.
\newblock ``{Four-body continuum effects in $^{11}$Be + d scattering}''.
\newblock \href{https://dx.doi.org/10.1016/j.physletb.2017.06.024}{Physics
  Letters B {\bf 772}, 1}~(2017).

\bibitem{JPCC-122.14606}
M.~Liao, R.~Grenier, Quy-Dong To, M.~P. de~Lara-Castells, and C.~L\'{e}onard.
\newblock ``{Helium and Argon Interactions with Gold Surfaces: Ab
  Initio-Assisted Determination of the He–Au Pairwise Potential and Its
  Application to Accommodation Coefficient Determination}''.
\newblock \href{https://dx.doi.org/10.1021/acs.jpcc.8b03555}{Journal of
  Physical Chemistry C {\bf 122}, 14606}~(2018).

\bibitem{MicMach-11.319}
S.~M. Nejad, S.~Nedea, A.~Frijns, and D.~Smeulders.
\newblock ``{The Influence of Gas–Wall and Gas–Gas Interactions on the
  Accommodation Coefficients for Rarefied Gases: A Molecular Dynamics Study}''.
\newblock \href{https://dx.doi.org/10.3390/mi11030319}{Micromachines {\bf 11},
  319}~(2020).

\bibitem{JPCA-119.6897}
R.~{Grenier}, Quy-Dong {To}, M.~P.~de {Lara-Castells}, and C.~{L{\'e}onard}.
\newblock ``{Argon Interaction with Gold Surfaces:Ab Initio-Assisted
  Determination of Pair Ar-Au Potentials for Molecular Dynamics Simulations}''.
\newblock \href{https://dx.doi.org/10.1021/acs.jpca.5b03769}{Journal of
  Physical Chemistry A {\bf 119}, 6897}~(2015).

\bibitem{AJP-52.60}
I.~Galbraith, Y.~S. Ching, and E.~Abraham.
\newblock ``{Two‐dimensional time‐dependent quantum‐mechanical scattering
  event}''.
\newblock \href{https://dx.doi.org/10.1119/1.13811}{American Journal of Physics
  {\bf 52}, 60}~(1984).

\bibitem{AJP-68.1113}
J.~J.~V. {Maestri}, R.~H. {Landau}, and M.~J. {P{\'a}ez}.
\newblock ``{Two-particle Schr{\"o}dinger equation animations of wave
  packet-wave packet scattering}''.
\newblock \href{https://dx.doi.org/10.1119/1.1286310}{American Journal of
  Physics {\bf 68}, 1113}~(2000).

\bibitem{FBS-56.727}
C.~A. Bertulani.
\newblock ``{Tunneling of atoms, nuclei and molecules}''.
\newblock \href{https://dx.doi.org/10.1007/s00601-015-0990-z}{Few-Body Systems
  {\bf 56}, 727}~(2015).

\bibitem{PRC-73.054608}
S.~Bacca and H.~Feldmeier.
\newblock ``{Resonant tunneling in a schematic model}''.
\newblock \href{https://dx.doi.org/10.1103/PhysRevC.73.054608}{Physical Review
  C {\bf 73}, 054608}~(2006).

\bibitem{CMAME-193.1733}
H.~S. Park and W.~K. Liu.
\newblock ``{An introduction and tutorial on multiple-scale analysis in
  solids}''.
\newblock \href{https://dx.doi.org/10.1016/j.cma.2003.12.054}{Computer Methods
  in Applied Mechanics and Engineering {\bf 193}, 1733}~(2004).

\bibitem{Science_372.6548}
C.~Whittle et~al.
\newblock ``{Approaching the motional ground state of a 10-kg object}''.
\newblock \href{https://dx.doi.org/10.1126/science.abh2634}{Science {\bf 372},
  1333}~(2021).

\bibitem{NatPhot.15.817}
N.~Fiaschi, B.~Hensen, A.~Wallucks, R.~Benevides, J.~Li, T.~P.~M. Alegre, and
  S.~Gr\"oblacher.
\newblock ``{Optomechanical quantum teleportation}''.
\newblock \href{https://dx.doi.org/10.1038/s41566-021-00866-z}{Nature Photonics
  {\bf 15}, 817}~(2021).

\bibitem{Datta_2021}
A.~Datta and H.~Miao.
\newblock ``{Signatures of the quantum nature of gravity in the differential
  motion of two masses}''.
\newblock \href{https://dx.doi.org/10.1088/2058-9565/ac1adf}{Quantum Sci.
  Technol. {\bf 6}, 045014}~(2021).

\bibitem{Roccati2022}
F.~Roccati, B.~Militello, E.~Fiordilino, R.~Iaria, L.~Burderi, T.~Di~Salvo, and
  F.~Ciccarello.
\newblock ``{Quantum correlations beyond entanglement in a classical-channel
  model of gravity}''.
\newblock \href{https://dx.doi.org/10.1038/s41598-022-22212-1}{Scientific
  Reports {\bf 12}, 17641}~(2022).

\bibitem{PRA_65.032314}
G.~Vidal and R.~F. Werner.
\newblock ``{Computable measure of entanglement}''.
\newblock \href{https://dx.doi.org/10.1103/PhysRevA.65.032314}{Physical Review
  A {\bf 65}, 032314}~(2002).

\bibitem{PRA_70.022318}
G.~Adesso, A.~Serafini, and F.~Illuminati.
\newblock ``{Extremal entanglement and mixedness in continuous variable
  systems}''.
\newblock \href{https://dx.doi.org/10.1103/PhysRevA.70.022318}{Physical Review
  A {\bf 70}, 022318}~(2004).

\bibitem{PRA_72.032334}
G.~Adesso and F.~Illuminati.
\newblock ``{Gaussian measures of entanglement versus negativities: Ordering of
  two-mode Gaussian states}''.
\newblock \href{https://dx.doi.org/10.1103/PhysRevA.72.032334}{Physical Review
  A {\bf 72}, 032334}~(2005).

\bibitem{JOPB_33.4447}
B.~M. Garraway.
\newblock ``{Extended Gaussian wavepacket dynamics}''.
\newblock \href{https://dx.doi.org/10.1088/0953-4075/33/20/318}{Journal of
  Physics B: Atomic, Molecular and Optical Physics {\bf 33}, 4447}~(2000).

\bibitem{RevModPhys.84.621}
C.~Weedbrook, S.~Pirandola, R.~Garc\'{\i}a-Patr\'on, N.~J. Cerf, T.~C. Ralph,
  J.~H. Shapiro, and S.~Lloyd.
\newblock ``{Gaussian quantum information}''.
\newblock \href{https://dx.doi.org/10.1103/RevModPhys.84.621}{Reviews of Modern
  Physics {\bf 84}, 621}~(2012).

\bibitem{book_decoherence_Maximilian}
M.~Schlosshauer.
\newblock ``{Decoherence and the Quantum-To-Classical Transition}''.
\newblock \href{https://dx.doi.org/10.1007/978-3-540-35775-9}{Springer-Verlag,
  Germany}. ~(2007).
\newblock 1st edition.

\bibitem{book_NumericalRecipies}
W.~H. Press, S.~A. Teukolsky, H.~A. Bethe, W.~T. Vetterling, and B.~P.
  Flannery.
\newblock ``{Numerical Recipes: The Art of Scientific Computing}''.
\newblock Cambridge University Press, USA. ~(2007).
\newblock 3rd edition.

\bibitem{PhysRevA.101.063804}
H.~Miao, D.~Martynov, H.~Yang, and A.~Datta.
\newblock ``{Quantum correlations of light mediated by gravity}''.
\newblock \href{https://dx.doi.org/10.1103/PhysRevA.101.063804}{Physical Review
  A {\bf 101}, 063804}~(2020).

\bibitem{PhysRevD.105.106028}
S.~Bose, A.~Mazumdar, M.~Schut, and M.~Toro\ifmmode~\check{s}\else \v{s}\fi{}.
\newblock ``{Mechanism for the quantum natured gravitons to entangle masses}''.
\newblock \href{https://dx.doi.org/10.1103/PhysRevD.105.106028}{Physical Review
  D {\bf 105}, 106028}~(2022).

\bibitem{PhysRevA.101.052110}
R.~J. Marshman, A.~Mazumdar, and S.~Bose.
\newblock ``{Locality and entanglement in table-top testing of the quantum
  nature of linearized gravity}''.
\newblock \href{https://dx.doi.org/10.1103/PhysRevA.101.052110}{Physical Review
  A {\bf 101}, 052110}~(2020).

\bibitem{PhysRevD.105.026011}
D.~Miki, A.~Matsumura, and K.~Yamamoto.
\newblock ``{Non-Gaussian entanglement in gravitating masses: The role of
  cumulants}''.
\newblock \href{https://dx.doi.org/10.1103/PhysRevD.105.026011}{Physical Review
  D {\bf 105}, 026011}~(2022).

\bibitem{Krisnanda2022_QNN}
T.~Krisnanda, T.~Paterek, M.~Paternostro, and Timothy C.~H. Liew.
\newblock ``Quantum neuromorphic approach to efficient sensing of
  gravity-induced entanglement''.
\newblock \href{https://dx.doi.org/10.1103/PhysRevD.107.086014}{Physical Review
  D {\bf 107}, 086014}~(2023).

\bibitem{bscthesis-GuoYao}
G.~Y. Tham.
\newblock ``Gravity-mediated gain of quantum entanglement''.
\newblock Bachelor's thesis.
\newblock Nanyang Technological University, Singapore.
\newblock ~(2019).

\bibitem{PhysRevLett.99.170403}
T.~Emig, N.~Graham, R.~L. Jaffe, and M.~Kardar.
\newblock ``{Casimir Forces between Arbitrary Compact Objects}''.
\newblock \href{https://dx.doi.org/10.1103/PhysRevLett.99.170403}{Physical
  Review Letters {\bf 99}, 170403}~(2007).

\bibitem{NASA-PlanetSheet}
{ NASA }.
\newblock
  ``{\href{https://nssdc.gsfc.nasa.gov/planetary/factsheet/planet_table_ratio.html}{{https://nssdc.gsfc.nasa.gov/planetary/factsheet/planet\_table\_ratio.html}}
  }''.

\bibitem{sanders_2010}
Robert~H. Sanders.
\newblock ``{The Dark Matter Problem: A Historical Perspective}''.
\newblock \href{https://dx.doi.org/10.1017/CBO9781139192309}{Cambridge
  University Press}. ~(2010).

\bibitem{Babock_AndromedaGalaxy}
Horace~W. {Babcock}.
\newblock ``{The rotation of the Andromeda Nebula}''.
\newblock \href{https://dx.doi.org/10.5479/ADS/bib/1939LicOB.19.41B}{Lick
  Observatory Bulletin {\bf 498}, 41}~(1939).

\bibitem{Zwicky2009}
F.~Zwicky.
\newblock ``{Republication of: The redshift of extragalactic nebulae}''.
\newblock \href{https://dx.doi.org/10.1007/s10714-008-0707-4}{General
  Relativity and Gravitation {\bf 41}, 207}~(2009).

\bibitem{DarkMatter-RMP}
G.~Bertone and D.~Hooper.
\newblock ``{History of dark matter}''.
\newblock \href{https://dx.doi.org/10.1103/RevModPhys.90.045002}{Reviews of
  Modern Physics {\bf 90}, 045002}~(2018).

\bibitem{Milogram_MOND}
M.~Milgrom.
\newblock ``{A modification of the newtonian dynamics as a possible alternative
  to the hidden mass hypothesis}''.
\newblock \href{https://dx.doi.org/10.1086/161130}{The Astrophysical Journal
  {\bf 270}, 365}~(1983).

\bibitem{Famaey_2005_MOND_in_Milky_Way}
B.~Famaey and J.~Binney.
\newblock ``{Modified Newtonian dynamics in the Milky Way}''.
\newblock \href{https://dx.doi.org/10.1111/j.1365-2966.2005.09474.x}{Monthly
  Notices of the Royal Astronomical Society {\bf 363}, 603}~(2005).

\bibitem{Gentile_2011_Things_about_MOND}
G.~Gentile, B.~Famaey, and W.~J.~G. de~Blok.
\newblock ``{Things about {MOND}}''.
\newblock \href{https://dx.doi.org/10.1051/0004-6361/201015283}{Astronomy \&
  Astrophysics {\bf 527}, A76}~(2011).

\bibitem{Bekenstein_Lag4MOND}
J.~D. {Bekenstein} and M.~{Milgrom}.
\newblock ``{Does the missing mass problem signal the breakdown of Newtonian
  gravity?}''.
\newblock \href{https://dx.doi.org/10.1086/162570}{The Astrophysical Journal
  {\bf 286}, 7}~(1984).

\bibitem{Bekenstein-TeVes}
J.~D. Bekenstein.
\newblock ``{Relativistic gravitation theory for the modified Newtonian
  dynamics paradigm}''.
\newblock \href{https://dx.doi.org/10.1103/PhysRevD.70.083509}{Physical Review
  D {\bf 70}, 083509}~(2004).

\bibitem{Zlosnik-TeVes}
C.~Skordis and T.~Z\l{}o\ifmmode~\acute{s}\else \'{s}\fi{}nik.
\newblock ``{New Relativistic Theory for Modified Newtonian Dynamics}''.
\newblock \href{https://dx.doi.org/10.1103/PhysRevLett.127.161302}{Physical
  Review Letters {\bf 127}, 161302}~(2021).

\bibitem{Sciene-Aspelmeyer}
U.~Deli\'c, M.~Reisenbauer, K.~Dare, D.~Grass, V.~Vuleti\'c, N.~Kiesel, and
  M.~Aspelmeyer.
\newblock ``{Cooling of a levitated nanoparticle to the motional quantum ground
  state}''.
\newblock \href{https://dx.doi.org/10.1126/science.aba3993}{Science {\bf 367},
  892}~(2020).

\bibitem{Gundlach2007}
J.~H. Gundlach, S.~Schlamminger, C.~D. Spitzer, K.-Y. Choi, B.~A. Woodahl,
  J.~J. Coy, and E.~Fischbach.
\newblock ``{Laboratory Test of Newton's Second Law for Small Accelerations}''.
\newblock \href{https://dx.doi.org/10.1103/PhysRevLett.98.150801}{Physical
  Review Letters {\bf 98}, 150801}~(2007).

\bibitem{Sanders_2015}
R.~H. Sanders.
\newblock ``{A historical perspective on modified Newtonian dynamics}''.
\newblock \href{https://dx.doi.org/10.1139/cjp-2014-0206}{Canadian Journal of
  Physics {\bf 93}, 126}~(2015).

\bibitem{Famaey_2012}
B.~Famaey and S.~S. McGaugh.
\newblock ``{Modified Newtonian Dynamics ({MOND}): Observational Phenomenology
  and Relativistic Extensions}''.
\newblock \href{https://dx.doi.org/10.12942/lrr-2012-10}{Living Reviews in
  Relativity {\bf 15}, 10}~(2012).

\bibitem{Milgrom_2014}
M.~Milgrom.
\newblock ``{General virial theorem for modified-gravity {MOND}}''.
\newblock \href{https://dx.doi.org/10.1103/physrevd.89.024016}{Physical Review
  D {\bf 89}, 024016}~(2014).

\bibitem{MOND-differentfactor}
H.~Zhao, B.~Li, and O.~Bienaym\'e.
\newblock ``{Modified Kepler's law, escape speed, and two-body problem in
  modified Newtonian dynamics-like theories}''.
\newblock \href{https://dx.doi.org/10.1103/PhysRevD.82.103001}{Physical Review
  D {\bf 82}, 103001}~(2010).

\bibitem{TwoBodyForce-MOND}
M.~{Milgrom}.
\newblock ``{Solutions for the Modified Newtonian Dynamics Field Equation}''.
\newblock \href{https://dx.doi.org/10.1086/164021}{The Astrophysical Journal
  {\bf 302}, 617}~(1986).

\bibitem{NORTON1985203}
J.~Norton.
\newblock ``{What was Einstein's principle of equivalence?}''.
\newblock \href{https://dx.doi.org/10.1016/0039-3681(85)90002-0}{Studies in
  History and Philosophy of Science Part A {\bf 16}, 203}~(1985).

\bibitem{EFE_NoObsEvidence}
Kyu-Hyun {Chae}, F.~{Lelli}, H.~{Desmond}, S.~S. {McGaugh}, P.~{Li}, and J.~M.
  {Schombert}.
\newblock ``{Testing the Strong Equivalence Principle: Detection of the
  External Field Effect in Rotationally Supported Galaxies}''.
\newblock \href{https://dx.doi.org/10.3847/1538-4357/abbb96}{The Astrophysical
  Journal {\bf 904}, 51}~(2020).

\bibitem{Ashutosh_EntNucColl}
A.~Singh, A.~Kumar, and P.~Arumugam.
\newblock ``{Coulomb-mediated entanglement between two nuclei in free fall}''.
\newblock Proceedings of the DAE Symposium on Nuclear Physics {\bf 66},
  693~(2022).

\bibitem{DAngelo-JoMO}
M.~D’angelo, A.~Zavatta, V.~Parigi, and M.~Bellini.
\newblock ``{Remotely prepared single-photon time-encoded ebits: homodyne
  tomography characterization}''.
\newblock \href{https://dx.doi.org/10.1080/09500340600895342}{Journal of Modern
  Optics {\bf 53}, 2259}~(2006).

\bibitem{GaussianOperations-Brask}
J.~B. Brask.
\newblock ``{Gaussian states and operations -- a quick reference}''~(2021).
\newblock  \href{http://arxiv.org/abs/2102.05748}{arXiv:2102.05748}.

\bibitem{Serafini_2003}
A.~Serafini, F.~Illuminati, and S.~De~Siena.
\newblock ``{Symplectic invariants, entropic measures and correlations of
  Gaussian states}''.
\newblock \href{https://dx.doi.org/10.1088/0953-4075/37/2/l02}{Journal of
  Physics B: Atomic, Molecular and Optical Physics {\bf 37}, L21}~(2003).

\bibitem{DICKSON2004195}
M.~Dickson.
\newblock ``{A view from nowhere: quantum reference frames and uncertainty}''.
\newblock \href{https://dx.doi.org/10.1016/j.shpsb.2003.12.003}{Studies in
  History and Philosophy of Science Part B: Studies in History and Philosophy
  of Modern Physics {\bf 35}, 195}~(2004).

\bibitem{1984PhRvD..30..368A}
Y.~{Aharonov} and T.~{Kaufherr}.
\newblock ``{Quantum frames of reference}''.
\newblock \href{https://dx.doi.org/10.1103/PhysRevD.30.368}{Physical Review D
  {\bf 30}, 368}~(1984).

\bibitem{Poulin_2007}
D.~Poulin and J.~Yard.
\newblock ``{Dynamics of a quantum reference frame}''.
\newblock \href{https://dx.doi.org/10.1088/1367-2630/9/5/156}{New Journal of
  Physics {\bf 9}, 156}~(2007).

\bibitem{Angelo_2011}
R.~M. Angelo, N.~Brunner, S.~Popescu, A.~J. Short, and P.~Skrzypczyk.
\newblock ``{Physics within a quantum reference frame}''.
\newblock \href{https://dx.doi.org/10.1088/1751-8113/44/14/145304}{Journal of
  Physics A: Mathematical and Theoretical {\bf 44}, 145304}~(2011).

\bibitem{RevModPhys.79.555}
Stephen~D. Bartlett, Terry Rudolph, and Robert~W. Spekkens.
\newblock ``{Reference frames, superselection rules, and quantum
  information}''.
\newblock \href{https://dx.doi.org/10.1103/RevModPhys.79.555}{Reviews of Modern
  Physics {\bf 79}, 555}~(2007).

\bibitem{Giacomini2019}
F.~Giacomini, E.~Castro-Ruiz, and {\v{C}}.~Brukner.
\newblock ``{Quantum mechanics and the covariance of physical laws in quantum
  reference frames}''.
\newblock \href{https://dx.doi.org/10.1038/s41467-018-08155-0}{Nature
  Communications {\bf 10}, 494}~(2019).

\bibitem{PRL_84.2726}
R.~Simon.
\newblock ``{Peres-Horodecki Separability Criterion for Continuous Variable
  Systems}''.
\newblock \href{https://dx.doi.org/10.1103/PhysRevLett.84.2726}{Physical Review
  Letters {\bf 84}, 2726}~(2000).

\bibitem{doi:10.1119/1.17904}
A.~Ekert and P.~L. Knight.
\newblock ``{Entangled quantum systems and the Schmidt decomposition}''.
\newblock \href{https://dx.doi.org/10.1119/1.17904}{American Journal of Physics
  {\bf 63}, 415}~(1995).

\bibitem{thesis_pyqentangle}
{Kwan-Yuet Ho}.
\newblock ``{Quantum Entanglement in Continuous System}''.
\newblock \href{https://dx.doi.org/10.13140/RG.2.2.24338.45764}{PhD thesis}.
\newblock The Chinese University of Hong Kong, China.
\newblock ~(2004).

\bibitem{arXiv:1905.01330}
C.~Roberts, A.~Milsted, M.~Ganahl, A.~Zalcman, B.~Fontaine, Y.~Zou, J.~Hidary,
  G.~Vidal, and S.~Leichenauer.
\newblock ``{TensorNetwork: A Library for Physics and Machine
  Learning}''~(2019).
\newblock  \href{http://arxiv.org/abs/1905.01330}{arXiv:1905.01330}.

\bibitem{TensorNetwork-GitHub}
{Google}.
\newblock
  ``\href{https://github.com/google/TensorNetwork}{https://github.com/google/TensorNetwork}''.

\bibitem{pyqentangle-GitHub}
{Kwan-Yuet ``Stephen'' Ho}.
\newblock
  ``\href{https://github.com/stephenhky/pyqentangle}{{https://github.com/stephenhky/pyqentangle}}''.

\bibitem{book-QuantMechInPhaseSpace}
T.~Curtright, D.~Fairlie, and C.~Zachos.
\newblock ``{A Concise Treatise on Quantum Mechanics in Phase Space}''.
\newblock World Scientific, Singapore. ~(2014).

\end{thebibliography}
